\RequirePackage[2020-02-02]{latexrelease}

\documentclass[conference]{IEEEtran}

\usepackage{subcaption}

\usepackage{cite}
\usepackage{url}
\usepackage{amsmath,amssymb,amsfonts}
\usepackage{algorithmic}
\usepackage{graphicx}
\usepackage{textcomp}
\usepackage{xcolor}
\usepackage{booktabs}
\usepackage{float}
\usepackage{multirow}
\usepackage[hidelinks]{hyperref}

\usepackage{listings,lstautogobble}

\usepackage{color, colortbl}
\definecolor{ListBGColor}{rgb}{0.9,0.9,0.9}
\definecolor{ListCommentColor}{rgb}{0.4, 0.27, 0.173}
\makeatletter
\newcommand\labelline[1]{%
  \def\@currentlabel{\thelstnumber}\label{#1}}
\newcommand\lineref[1]{
  \@ifundefined{r@#1}{0}{\ref{#1}}
}
\makeatother

\usepackage{booktabs, makecell, tabularx}
\usepackage{rotating}
\usepackage[export]{adjustbox}
\input{tex/macros}

\begin{document}

\title{Better GPU Hash Tables}

\newcommand{\ece}{Dept.\ of Elect.\ and Comp.\ Engineering}

\author{
\IEEEauthorblockN{Muhammad A. Awad}
\IEEEauthorblockA{\emph{\ece} \\
  \emph{UC Davis}\\
  Davis, CA, USA \\
  mawad@ucdavis.edu}
\and
\IEEEauthorblockN{Saman Ashkiani}
\IEEEauthorblockA{\emph{Google} \\
  Mountain View, CA, USA \\
  sa.ashkiani@gmail.com}
\and
\IEEEauthorblockN{Serban D. Porumbescu}
\IEEEauthorblockA{\emph{\ece} \\
  \emph{UC Davis}\\
  Davis, CA, USA \\
  sdporumbescu@ucdavis.edu}
\linebreakand
\IEEEauthorblockN{Mart{\'{i}}n Farach-Colton}
\IEEEauthorblockA{\emph{Dept.\ of Computer Science} \\
  \emph{Rutgers University}\\
  Piscataway, NJ, USA \\
  farach@cs.rutgers.edu}
\and
\IEEEauthorblockN{John D. Owens}
\IEEEauthorblockA{\emph{\ece} \\
  \emph{UC Davis}\\
  Davis, CA, USA \\
  jowens@ece.ucdavis.edu\\
  ORCID: 0000-0001-6582-8237}
}

\maketitle

\begin{abstract}
  We revisit the problem of building static hash tables on the GPU and design and build three bucketed hash tables that use different probing schemes. Our implementations are lock-free and offer efficient memory access patterns; thus, only the probing scheme is the factor affecting the performance of the hash table's different operations. Our results show that a bucketed cuckoo hash table that uses three hash functions (BCHT) outperforms alternative methods that use power-of-two choices, iceberg hashing, and a cuckoo hash table that uses a bucket size one. At high load factors as high as 0.99, BCHT enjoys an average probe count of 1.43 during insertion. Using three hash functions only, positive and negative queries require at most 1.39 and  2.8 average probes per key, respectively.
\end{abstract}

\section{Introduction}
\label{sec:intro}
\label{sec:introduction}

Hash tables are fundamental data structures that answer membership queries (``is this key present'') and lookups (``what is the value associated with this key''). Computer systems from small to large implement hash tables with a  variety of design goals and capabilities. The primary design tradeoff in any of these implementations is between three metrics:

\begin{LaTeXdescription}
  \item[Query performance] is measured in queries per second and reflects the throughput of membership queries or key-value lookups into the hash table. Queries can either be positive (exist in the hash table) or negative (do not exist in the hash table).
  \item[Build performance] is measured in keys or key-value pairs per second and indicates the throughput of building a hash table from an unordered list of keys or key-value pairs.
  \item[Memory efficiency] expresses how well the implementation uses its allocated memory. In particular, the metric of \emph{load factor} is the ratio between the amount of storage used to store keys or key-value pairs and the total amount of storage allocated for the entire hash table.
\end{LaTeXdescription}

While the design space of all hash table implementations is quite large, it is useful to have a variety of hash table implementations that make different tradeoff choices because different applications may have different performance requirements. In this work, we focus on a specific segment of this space: a single (NVIDIA) GPU as the platform; a dataset that fits into the GPU's main memory; and a static hash table that is constructed once on the GPU from a list of keys or key-value pairs and then never modified. Within this segment, we implement and analyze different hash table designs and recommend designs for different use cases.

The designs of the hash tables we consider are influenced by four main choices:
\begin{LaTeXdescription}
  \item[Probing scheme.] Probing schemes offer different techniques to resolve collision between multiple keys mapping to the same bucket in the hash table. These schemes include linear probing, quadratic probing, double hashing, and cuckoo hashing. Compared to double hashing and cuckoo hashing, linear and quadratic probing schemes are prone to clustering (when colliding keys form a cluster).

  \item[Bucket size.] Our strategies place items into buckets in the hash table; we choose the size of these buckets. Using larger buckets helps to reduce the variance of the load between the buckets. As the bucket size increases, the achievable load factor for the hash table increases. The choice of the bucket size is influenced by the hardware constraints; for instance, we can choose the bucket size to be the same as the cache line size.

  \item[Probe complexity.] A hash table can place keys in different buckets; each of these buckets is associated with a different hash function. Increasing the number of hash functions improves the achievable load factor but also increases the number of buckets we need to inspect while inserting or querying a key.

  \item[Placement strategy.] Since there are potentially multiple buckets where a given key might end up, a hash table with multiple hash functions needs to decide on a placement strategy. One choice is to always insert in the first hash function's bucket, avoiding any placement strategy overhead. A different choice is to place a key into the least loaded bucket out of the possible choices. This adds the additional overhead of inspecting these possible buckets.
\end{LaTeXdescription}

In this work, we focus on cuckoo hash tables and explore these different choices. We implement and evaluate three families of GPU-optimized hash tables:
\begin{itemize}
  \item Bucketed cuckoo hash table (BCHT): a cuckoo hashing strategy where the bucket size can be one (corresponding to CUDPP's implementation (1CHT), or larger than one.
  \item Bucketed power-of-two-choices hash table (BP2HT): where a key chooses among its two (or more) possible buckets by choosing the least-loaded bucket.
  \item Iceberg hash table (IHT): where we optimize for the common case when insertion succeeds without collision. Using a primary hash function, we attempt to fill buckets up to a threshold, then when a bucket reaches the threshold, two additional hash functions offer two alternative locations for insertion. We describe this process in more detail in Section~\ref{sec:iceberg}.
\end{itemize}

Our contributions are:
\begin{itemize}
  \item The design and implementation of GPU-optimized hash tables that offer different tradeoffs and properties.
  \item Analysis of the performance of each technique and identification of their performance limiters.
  \item Hash table recommendations for different use cases.
\end{itemize}

\paragraph*{High-level recommendations}
We find that when the main constraint is memory, and we require the hash table to achieve a high load factor, bucketed cuckoo hash with a bucket size of 16 and using three hash functions provides the highest load factors (up to 0.98). BCHT also achieves the highest query and insertion throughput. When stability is the main property required in the hash table, the iceberg hash table using a bucket size of 16 (IHT) provides a high insertion throughput but only achieves up to 0.82 load factor. Using a bucket size of 32 allows IHT to achieve higher load factors (up to 0.91) but with a lower insertion and query throughput. When both stability and high query throughput is required, BP2HT with a bucket size of 16 is the best choice, but it only achieves up to around 0.82 load factor. Similar to IHT, increasing the bucket size improves the achievable load factor (up to 0.91) at the cost of both query and insertion throughput. All the previous recommendations guarantee a success rate of at least 99\% in building the table. Table~\ref{tab:recommendations} shows a summary of the recommendations.

\begin{table}
  \centering
  \begin{tabular}{lccc}
    \toprule
              & Load factor  & Insertion    & Query         \\ \midrule
    Insertion & BCHT, $b=16$ & ---          & ---           \\
    Query     & BCHT, $b=16$ & BCHT, $b=16$ & ---           \\
    Stability & IHT, $b=32$  & IHT, $b=16$  & BP2HT, $b=16$ \\
    \bottomrule
  \end{tabular}
  \caption{Hash table recommendations. Each cell in the table represents the recommended hash table for its row and column metrics. For example, to achieve both the highest insertion and query throughput, BCHT with $b=16$ is our recommended hash table.}
  \label{tab:recommendations}
\end{table}
\section{Background: GPU Hash Tables and Cuckoo Hashing}

Cuckoo hashing~\cite{Pagh:2001:CUH} differs from open-addressing hash schemes in that an item may only be inserted into one out of two possible locations, making queries fast. If old keys occupy both of the locations, then we have a collision. We resolve the collision by performing the ``cuckooing'' operation in which the new key pushes one of the old keys to a new location. Fotakis et al.~\shortcite{Fotakis:2005:SEH} generalized cuckoo hashing and introduced $d$-ary cuckoo hashing, where the number of possible locations per key is $d$ (in this paper, we use $h$ to refer to the possible number of locations). Using more than one hash function, they achieved 0.97 and 0.99 load factors for 4 and 5 hash functions, respectively. Panigrahy~\shortcite{Panigrahy:2005:EHW} generalized the bucket size $b$ to achieve higher load factors introducing a ``bucketized'' cuckoo hash table (BCHT) while using $b=2$. Later, Erlingsson et al.~\shortcite{Erlingsson:2006:ACA} combined both generalizations and experimented with a cuckoo hash table parameterized using both $h$ and $b$. The problem of cuckoo hashing is typically viewed as a graph problem where buckets are vertices and the mapping between a key to a bucket (or more) is a hyperedge. For any combination of bucket size and number of hash functions, there exists a hyperedge density threshold such that below this threshold the graph is \emph{orientable}, in other words, we can build the cuckoo hash table (See Walzer~\cite{Walzer:2020:RHF}). As the number of hash functions or bucket size increases, the threshold increases. Reaching a load factor of 100\% is possible when the number of hash functions and the bucket size is (5, 5) and (4, 6). Using only two hash functions would be enough to achieve more than 99\% load factor, but it requires using a load-balancing strategy when inserting into one of these buckets.

The first hash table that could be built on the GPU~\cite{Alcantara:2009:RPH} used a cuckoo hashing hash table formulation, and was succeeded by a simpler cuckoo hashing implementation~\cite{Alcantara:2011:BAE} that stored the entire hash table as a single table in global memory. Its implementation is in the CUDPP library~\cite{Harris:2017:CUDPP} and is commonly used as a performance comparison in the GPU literature. CUDPP's implementation uses $h=4$ and $b=1$. Subsequent work has explored both different performance tradeoffs and a broader hash-table feature set; rather than summarize this work here, we direct the reader to the comprehensive 2020 survey by Lessley and Childs~\cite{Lessley:2020:DHT}.

Mega-KV's 8-entry, two-hash-function BCHT targets a hybrid CPU-GPU in-memory key-value store~\cite{Zhang:2015:MAC}. Mega-KV divides the hash table into several partitions, and a single thread block processes each partition. The differences between our BCHT with $b=8$ and Mega-KV are that we use three hash functions instead of only two and a single hash table instead of multiple partitions. Using three hash functions allows our BCHT to achieve higher load factors and insertion throughput. However, the drawback of using an additional hash function is that the query throughout becomes lower (1.38x at load factor 0.96) when all queries are negative. As we shall see in our results (Section~\ref{sec:results}), using a bucket of size 16 outperforms alternate other bucket sizes.

Horton tables~\cite{Breslow:2016:HTF} optimize for negative queries in the hash table. They add ``remap tables'' to BCHTs to reduce the BCHT's expected number of memory accesses. Horton hashes a key into a primary bucket using a single hash function, while the secondary bucket is chosen between 7 secondary hash functions by choosing the least loaded bucket. They achieve load factors of 95\% and an average probe count of 1.18 and 1.06 for positive and negative queries, respectively. Construction, however, is performed on the CPU\@. Our work focuses on performing both construction and query operations on the GPU\@. Neither Mega-KV nor Horton compare their performance to a traditional $b=1$ cuckoo hash table on the GPU, but both achieve results that are faster than CUDPP for both positive and negative queries.


WarpDrive~\cite{Junger:2018:WMP} and the follow-up in WarpCore~\cite{Junger:2018:WAL} introduced an open-addressing static hash table that uses a double hashing scheme. Using CUDA's cooperative groups\footnote{\url{https://developer.nvidia.com/blog/cooperative-groups/}}, they experiment with different probe-window sizes. Their window size optimal performance is when the window size $|g| \in \{2, 4, 8\}$. They achieve up to 2.84x and 1.3x faster insertion and query performance than CUDPP\@. The main difference between our approaches and WarpCore's is the probing scheme choice. Compared to WarpCore, our bucketed cuckoo hash table (BCHT with $b=16$) achieves similar insertion performance and slightly faster ($\leq1.2$x) positive query performance. For negative queries, BCHT achieves speedups of \{2.3x, 1.6x\} compared to WarpCore with probing windows of \{4, 8\}. BCHT negative query performance significantly outperforms WarpCore at high load factors, reaching up to 6.4x speedups\footnote{\label{comparison_note}Averaged results over different load factors for inserting and querying 50M unique keys.}. While performance may be comparable for some operations, WarpCore's unacceptably low performance for negative queries motivates our recommendation of BCHT\@.

GPU implementations of cuckoo hashing do not typically target dynamic scenarios where new items are inserted into or deleted from an existing hash table. However, Li et al.~\cite{Li:2021:DDH} recently proposed a dynamic cuckoo hash table implementation (DyCuckoo)\@. To avoid rebuilding the entire hash table from scratch, they divide the table into multiple subtables. Each subtable has a lower and upper bound on its load factor, and when resizing is required, they double the size of the smallest subtable (or halve the smallest subtable). A key can only be hashed to two subtables; thus, a query only needs to perform at most two lookups.

In static scenarios, DyCuckoo achieves similar performance to Mega-KV and WarpDrive. Compared to our BCHT with $b=16$, insertion is 1.1x faster than DyCuckoo. Positive queries in BCHT achieve higher speedups with an average speedup\footnoteref{comparison_note} of 1.2x. For negative queries, BCHT achieves significant speedups \{2.1x, 1.1x\} at load factors between \{0.6, 0.93\}. However, at higher load factors, DyCuckoo starts to outperform our BCHT, reaching a speedup of 1.1x at a load factor of 0.97. DyCuckoo outperforms our BCHT since BCHT performs up to three probes when the key does not exist on the hash table; on the other hand, DyCuckoo only requires two lookups in the worst case. We do not address dynamic updates in this work.

\subsection{High-Level GPU Cuckoo Hashing Implementation Strategy}

In the CUDPP implementation, as in most subsequent work, queries are evaluated in parallel across CUDA threads (each queried key is assigned to a different thread) but the computation of each query is serial within a thread. The thread checks each bucket $H_i(k)$, returning ``found'' if the key is stored in one of those buckets and ``not found'' otherwise. Because each individual query is serialized within a thread, we can stop searching (with no additional memory accesses) once we find an item.

Just as with queries, insertions are parallelized across threads, but each individual insertion process (possibly involving multiple exchanges) is serial within a thread. In other words, an insertion process begins with an item to place and may do multiple exchanges (cuckooing), ending when an item is finally placed into an empty slot. Each exchange operation must be atomic to guarantee proper operation, so insertions into the same bucket serialize, but the ability to simultaneously insert into different buckets allows a large amount of concurrency. Because the exchange is atomic, the practical size of a key or a key-value pair is limited to the atomic-exchange size supported by the hardware. NVIDIA's new CUDA C++ Standard Library \texttt{libcu++},\footnote{\url{https://nvidia.github.io/libcudacxx/}} with initial release in October 2020, promises generalized atomic support on any datatype, but its performance characteristics are not yet known. For arbitrary datatypes, it is unlikely to match the performance of current atomic exchanges.

Implementations may choose to query buckets serially or in parallel\footnote{Parallel queries have an overall lower latency than serial queries but potentially also increase the number of memory accesses.} and also may choose a bucket-insertion and bucket-querying strategy that together minimize the expected number of memory accesses. As an example, if we always choose hash functions for insertion in the order $H_0, H_1, \ldots$, and we later query serially using the same order, we can stop querying if we ever find an empty slot, because we are not supporting deletions.

\pp{Performance characteristics of GPU cuckoo hashing}
In the best (and common) case for queries, a key is found in its first searched bucket, meaning that the only memory access necessary to find a key is a single random read into the global memory where the hash table is stored. For insertions, the ideal case is when the first location checked is empty, and thus no cuckooing is necessary.  In this case, the only memory access necessary to insert a key into the table is an atomic exchange into global memory. As the table becomes full, however, the average number of exchanges rises quickly.  In addition, such swaps increase query costs because they move items out of their primary slot into their secondary slot. The costs of the ideal case for queries and insertions are low, perhaps minimal. If the common case approximates the ideal case, then from a throughput perspective on the GPU, cuckoo hashing implementations are hard to beat.
In their 2011 paper, Alcantara et al.\ indicate that memory performance is the bottleneck~\cite{Alcantara:2011:BAE}: ``\ldots\ we can run up to 1000 instructions per memory update and remain memory bound. Consequently, arithmetic utilization and thread divergence have only marginal impact on hash performance.'' As more recent GPUs offer an increasing amount of compute per memory operation (a higher ``compute intensity''), we expect this conclusion will continue to be even more true with future architectures for both queries and insertions.

The above assumptions about the dominant costs for an insertion (atomic exchanges) or queries (global reads) assume that the address translation between virtual and physical memory is free. This is not accurate in the presence of TLB misses. On NVIDIA's Volta microarchitecture, for instance, Jia et al.~\cite{Jia:2018:DTN} suggest that Volta's L2 TLB stores only enough entries to cover 8~GB of memory, so we would expect a performance drop due to TLB misses as the cuckoo hash table grows larger than this size. We do not address this issue in this work.
\section{Algorithms}
\label{sec:algorithms}

In this work, we describe the implementation of three different hash table approaches: bucketed cuckoo hash table (BCHT); bucketed power-of-two-choices hash table (BP2HT); and iceberg hash table (IHT)\@. BCHT is a generalization of CUDPP's cuckoo hash table implementation (1CHT), where BCHT's bucket size can be greater than one. In this section, we discuss the insertion and query algorithms for each of these approaches. A hash table is represented using a fixed size array split into a number of buckets $m$ of size $b$ (i.e., a bucket contains $b$ locations for $b$ pairs). The array has a capacity $C$ of key-value pairs where $C = m \times b$. A hash table containing $n$ keys has a load factor of $n / C$.

\begin{figure}
  \centering
  \begin{subfigure}{\columnwidth}
    \includegraphics[page=1, width=\textwidth]{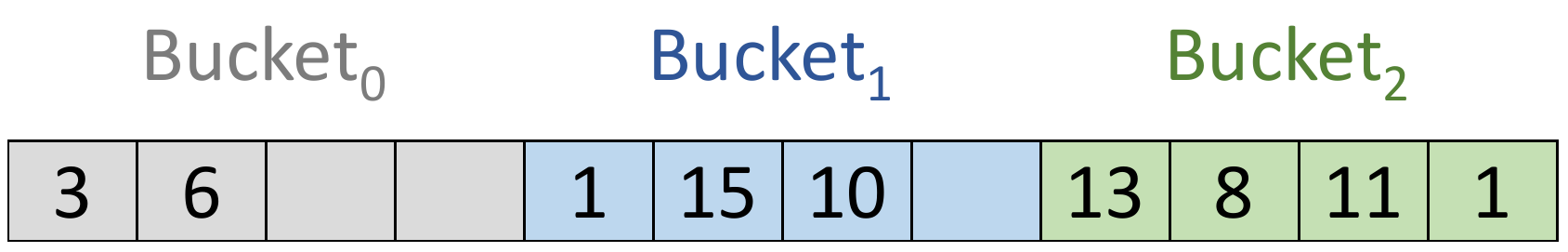}
    \caption{Initial hash table with three buckets and $b =4$.}
  \end{subfigure}
  \begin{subfigure}{\columnwidth}
    \includegraphics[page=3, width=\textwidth]{fig/tables_cropped.pdf}
    \caption{In BCHT the key is hashed into a full bucket $h_0(7) = 2$. We exchange the new key with a random one from the full bucket. The swapped out key is moved to the next bucket according to the three hash functions $h_0(11) = 0$, $h_1(11) = 2$, and $h_2(11) = 1$.}
  \end{subfigure}
  \begin{subfigure}{\columnwidth}
    \includegraphics[page=2, width=\textwidth]{fig/tables_cropped.pdf}
    \caption{In BP2HT we compute the two hash functions $h_0(7) = 0$ and $h_1(7) = 1$, then we insert the key in the least loaded bucket.}
  \end{subfigure}
  \begin{subfigure}{\columnwidth}
    \includegraphics[page=2, width=\textwidth]{fig/tables_cropped.pdf}
    \caption{In IHT with a threshold $t=3$ the key is first mapped to its primary bucket $h_p(7) = 1$. Since the bucket exceeds the threshold we compute the two secondary hash functions: $h_{s0}(2) = 0$, and $h_{s1}(2) = 0$. The new key is inserted into the bucket with lower load.}
  \end{subfigure}
  \caption{Inserting a key $=7$ in the different hash tables.}
\label{fig:algos_fig}
\end{figure}

\subsection{Bucketed cuckoo hash table (BCHT or 1CHT)}

We first begin by specifying $h$ hash functions $H_0, H_1 \ldots H_{h-1}$. We then allocate a table of $m$ buckets, each with $b$ valid locations in each bucket. For simplicity, we only discuss storing keys (values are typically handled by bundling them with their associated key).

A number of hash functions $h$ maps a key $k$ to the buckets $H_0(k) \ldots H_{h-1}(k)$. Thus, the lookup process only requires checking these buckets. Only if the key is found in one of these buckets is the key present in the hash table. We use a constant number of hash functions, so queries have an $O(1)$ complexity.

Construction is (slightly) more complex and demonstrates the unique idea behind cuckoo hashing. We initialize all buckets of the hash table to empty and then insert each key into the hash table. For key $k$, we begin with the first hash function $H_0$ and attempt to insert $k$ into bucket $H_0(k)$. If the bucket is empty, we insert the key and are done. If it is full, we \emph{exchange} our item $k$ with $\hat{k}$, a random key already in $H_0(k)$. It is this exchange that is the characteristic feature of cuckoo hashing, and we say that $k$ \emph{cuckoos} $\hat{k}$. The key $k$ is now stored in the hash table at $H_0(k)$ and we now must insert $\hat{k}$. If $\hat{k}$ was previously stored with $H_i$, we (typically) begin this new insertion with the next sequential hash function, $H_{(i+1) \bmod h}$.\footnote{We can determine which of our $h$ hash functions was used to store a particular key at a particular address $H_i(k$) by hashing our key with all $h$ of our hash functions and seeing which of those hash functions yields the hash table address $H_i(k)$, breaking ties by favoring the lowest index hash function.} We finish when we successfully store our key into an empty bucket or reach a maximum number of exchanges, in which case we have failed to construct our hash table. The theoretical analysis of cuckoo hashing requires a fairly low load factor in order to guarantee that the construction fails with low probability, but in practice, it can achieve high load factors. An advantage of using an increasing order when picking a hash function is that during queries, if a bucket contains an empty key, the query can exit early.

Another variation of bucketed cuckoo hash tables uses a load-balancing scheme when picking a bucket out of the possible $h$ buckets. Instead of always inserting a key into the first bucket as we discussed previously, we insert the key into the least loaded bucket out of all the possible buckets. A load-balancing technique improves the achievable load factor at the cost of the additional memory transfers required for evaluating the load of all possible $h$ buckets. Since insertion could pick any bucket out of the possible buckets, during queries early-exit strategies cannot be used. This results in an overhead similar to the insertion's overhead (unless the key is found in an early bucket). We will next discuss load balancing (power of two) as a technique on its own.

\subsection{The power of two (or more) choices (P2HT)}
Consider a cuckoo hash table with two hash functions. When we insert a new key, we evaluate one hash function and insert that key  into its corresponding bucket. The other hash function will only be used if our key is ejected and reinserted.

Instead, what if we evaluated \emph{both} hash functions before deciding where to insert? Armed with this information, we could, and would want to, choose the less loaded bucket. We would conjecture that we could achieve a higher load factor with this strategy at the cost of more memory accesses (since we now must check multiple buckets on each insertion, and we no longer have a preferred bucket for queries).

This technique, termed ``power of two choices'', is a powerful one: ``even a small amount of choice can lead to drastically different results in load balancing''~\cite{Mitzenmacher:2001:TPO}. Given $h$ choices, the maximum load for $n$ keys now decreases from $O( \log n / \log \log n)$ to $O(\log \log n / \log h)$~\cite{Azar:1999:BA}.

Unlike cuckoo hashing, power-of-two-choices hashing is \emph{stable}: once we place a key, that key never moves. Stability is useful in applications where we require a (stable) pointer to a particular key in the hash table. Similar to a bucketed cuckoo hash table, we can build bucketed power-of-two-choices (BP2HT) where the bucket size is greater than one.

\subsection{Iceberg Hashing}
\label{sec:iceberg}
\label{sec:iht}
In a stable hash table, larger buckets increase query and build throughput, but do not allow us to reach high load factors. Power-of-two-choices hashing can improve the load factor, but it increases the amount of work we must do for queries and insertions. While we can directly combine the two techniques, doing so does not yield both the performance gains and memory efficiency gains that we target.

What we like about larger buckets is the common-case behavior: we most likely find the key we are searching for, or succeed with an insertion, in the first bucket. What we like about power-of-two-choices is that it gives us better memory utilization for the extraordinary cases. For these cases, we are willing to pay the extra cost of the multiple choices.

Iceberg hashing~\cite{Anon:2021:DBA} combines the common-case and extraordinary-case behaviors\footnote{Separating a compute task into common and extraordinary components has been previously explored in other GPU application domains, e.g., sparse-matrix dense-vector multiplication~\cite{Bell:2009:ISM}.} in a novel way. First, we establish the intuition. Consider filling up most of the hash table cheaply by placing keys in the first bucket where they fit, then fill the rest of the table by using the approach that optimizes placement (i.e., power-of-two-choices). The term ``iceberg hashing'' reflects the two different strategies: the common case (the most prevalent, corresponding to the largest part of the iceberg under the water), and the extraordinary case (the small part of the iceberg above the water).

Like power-of-two-choices hashing, iceberg hashing is stable.

\subsubsection{Iceberg hashing algorithm}

\paragraph{Preliminaries}
Choose a threshold $t$ for buckets. $t$ marks the boundary between the aforementioned ``common'' (buckets have fewer than $t$ keys) and ``extraordinary'' (at least $t$) cases. If each bucket has on average $b_\text{avg}$ keys, then make $t$ a little larger than $b_\text{avg}$.

Choose 3 hash functions ($H_p, H_{s0}, H_{s1}$). $H_p$ is the hash function for the ``primary'' bucket; the two $H_s$es are for ``secondary'' buckets.

\paragraph{Insertion}

To insert a key $k$:

\begin{itemize}
  \item Compute its primary hash function $H_p(k)$. If the resulting bucket has fewer than $t$ keys, insert the item there and return.
  \item If the resulting bucket is too full, compute \emph{both} secondary hash functions $H_{s0}(k)$ and $H_{s1}(k)$. Insert the key into the less full of those two buckets.
  \item If all of these buckets are full, our build has failed. Choose another set of hash functions or enlarge the table and try again. We expect this scenario will only occur with a very low probability if the threshold and the bucket size are properly chosen.
\end{itemize}

\paragraph{Query}

To look up a key $k$:

\begin{itemize}
  \item First check location $H_p(k)$. Return the key if found.
  \item If the key is not found in $H_p(k)$, check both locations $H_{s0}(k)$ and $H_{s1}(k)$. Return the item if found.
  \item If the key is not found in any of those locations, it is not in the hash table.
\end{itemize}

\section{Common Design Decisions}
\label{sec:design_decisions}

When implementing the previous different hash table strategies, we face some common design decisions across them: how many items we allocate per bucket, how many hash functions we use in the hash table, and how we decide to place items when we have multiple choices. We now discuss these design decisions and the intuition behind them.

\subsection{Bucket size}

\pp{Effect of bucket size on a cuckoo hash table.} Consider a hash table that uses two hash functions, $h=2$, and where the bucket size is one, $b=1$. In this case, an item is in its first location with a probability of $1/2$. Now consider  a larger bucket of $b=16$. In this case, making some simplifying assumptions, we boost the probability of finding the item in the first location to $1-1/2^{16}$. Thus, with these parameters, we would be increasing the computation time by about $10\times$ (in the $b=1$ case, we would perform an expected 1.5 comparisons per lookup, whereas, in the hypothetical $b=16$, we would perform slightly more than 16 comparisons).  These calculations are only rough but do illustrate that larger buckets can reduce memory use at the cost of increasing computation, which should result in an overall improvement in system throughput.

Increasing the bucket size can also reduce the expected amount of cuckooing.  To see this, notice that if $b=1$, only one item can be inserted into any particular bin, while all other threads will need to cuckoo to other bins.  When $b>1$, more threads can succeed, so there will be fewer cuckoo chains.

\pp{Effect of bucket size on the memory access pattern.} Not only does using larger bucket sizes improves the achievable load factor, it also satisfies the GPU optimal memory access pattern. In general, we would like adjacent threads to access adjacent memory locations achieving a \emph{coalesced} memory access. With 4-byte keys and 4-byte values and a GPU cache-line size of 128 bytes, we expect the optimal bucket size from the memory-system perspective to be 16.

We conclude that bucketizing cuckoo hashing is a good idea, in terms of decreasing expensive memory accesses in exchange for inexpensive computation. Increasing the bucket size improves both the achievable load factor and the number of probes per key. However, increasing the bucket size beyond the cache-line size reduces the achievable insertion and query throughput.

\subsection{Number of hash functions}

The number of hash functions, corresponding to the number of possible locations for any item, has both positive and negative effects on hash table performance.

\begin{itemize}
  \item Increasing the number of hash functions enables achieving high load factors. For instance, in a standard cuckoo hash table, using two hash functions can only achieve load factors up to only 0.5. Adding a single additional possible location increases this achievable load factor up to 0.92~\cite{Dietzfelbinger:2010:GMM}. However, the benefits of adding more locations are marginal beyond three hash functions. The reason behind these achievable higher load factors is that adding more possible locations reduces the possibility of ending up in a cycle while moving keys between buckets when a collision happens as well as decreasing the number of probes required for insertion.
  \item However, increasing the number of hash functions also increases the work necessary to query a hash table, particularly when a queried item is not present in the hash table. In this case, having $h$ hash functions requires $h$ probes into the hash table before concluding the queried item is not present.
\end{itemize}

In a bucketed cuckoo hash table, when the bucket size is large, using two hash functions is enough to achieve high load factors, but this requires balancing the load between the two possible buckets. A simpler approach where we insert a key into the first bucket using only two hash functions reduces the achievable load factor, but adding a single hash function to this simple approach allows achieving higher load factors. Moreover, using a simple insertion approach improves the insertion throughput (over the load-balanced technique), but it reduces the query throughput. Specifically, when the number of positive queries is low, using more hash functions reduces the performance of queries, since a query requires loading a number of buckets equal to the number of hash functions. In general, we opt to use 3 hash functions and a simple insertion strategy; however, if the primary goal is to improve the query rate, using a load-balanced insertion strategy and only 2 hash functions is a better choice.

\subsection{Placement strategies}
While increasing the number of hash functions improves achievable load factors, it also increases the number of choices when placing a item. The simplest choice is to always pick the first bucket and only perform cuckooing when it is full. A more sophisticated approach would be to pick the least loaded bucket (i.e., power of two hashing). This increases the cost of the insertion since it requires loading all possible buckets before deciding on which bucket to perform the insertion into. Later we will show that the simple placement strategy is more efficient than a load-balancing placement strategy. Iceberg hashing is a third strategy for placement.

\section{GPU Implementation Details}

This section focuses on how we implement our hash tables on the GPU\@. Between the different hash tables is one common implementation detail: supporting buckets of arbitrary size. We first discuss the implementation details for a hash table with the bucket size specified by the programmer, then we discuss the specific insertion and query implementation details for each technique separately. Our implementation targets 32-bit keys and values. We use the largest number an unsigned integer can hold as our sentinel key (or value). For example, an empty key (or value) would be \texttt{std::numeric\_limits<uint32\_t>::max()}.

\subsection{Supporting Arbitrary Bucket Sizes}
\label{sec:larger_buckets}

In a bucketed hash table GPU implementation, we balance two considerations: larger buckets reduce the number of bucket accesses required for queries and insertions, but larger buckets are also more complex to search and manage. Nonetheless, a larger bucket is a sensible idea for a GPU implementation for the following two reasons:

\begin{itemize}
  \item The cost of accessing a single word in a GPU cache line, to first order, is the same as accessing the entire cache line. Thus, if a bucket size is no larger than a cache line, we can access the entire bucket with the same memory cost as accessing a single item.
  \item Even while we incur more work per query or insertion with a larger bucket size, we are still memory-bound, so the additional work does not affect our performance. Moreover, with CUDA's cooperative groups abstraction, we can assign a group of threads to each query or insertion and use that group to access an entire bucket, communicating efficiently between threads in the group.
\end{itemize}

The challenge with buckets of size $b>1$ is to efficiently implement operations that involve the entire bucket, e.g., search for a particular item in a bucket; find an empty slot in a bucket; choose an item to remove from a bucket. One way to implement this would be to maintain a data structure (metadata) for each bucket that, for instance, stored a bitmask of full/empty slots in the bucket or an index of the next empty item. We choose instead to compute any necessary information on the fly when it is required (without the use of any metadata). Why? First, any data structure would use space in the bucket and reduce our load factor. Second, we not only have ample compute capability, but our computation here, because of warp-wide (cooperative-group-wide) intrinsics, is quite efficient.

\pp{GPU Implementation details}
Our implementation uses cooperative groups to achieve flexible control over the bucket size while taking advantage of the tile-wide communication intrinsics between the threads in the tile. The cooperative groups API, specifically the \texttt{tiled\_partition} explicit groups, offers tile-wide functions such as \texttt{ballot}, which evaluates a predicate and returns an integer where a bit is set for each thread predicate evaluation; and \texttt{shfl}, which broadcasts a variable across all the tile's threads.

\begin{figure}[ht]
  \centering
  \begin{lstlisting}[caption={Tile-wide cooperative insertion.},captionpos=b, label=lst:coop_insert]
bool cooperative_insert(bool to_insert, pair_type pair, pair_type* table){
  //Construct the work tile
  cg::thread_block thb = cg::this_thread_block();(*@ \labelline{lst:coop_insert:thread_block} @*)
  auto tile = cg::tiled_partition<bucket_size>(thb);(*@ \labelline{lst:coop_insert:tile} @*)
  auto thread_rank = tile.thread_rank();(*@ \labelline{lst:coop_insert:rank} @*)
  bool success = true;
  //Perform the insertions
  while(uint32_t work_queue = tile.ballot(to_insert)){(*@ \labelline{lst:coop_insert:ballot} @*)
    auto cur_lane = __ffs(work_queue) - 1;(*@ \labelline{lst:coop_insert:cur_lane} @*)
    auto cur_pair = tile.shfl(pair, cur_lane);(*@ \labelline{lst:coop_insert:cur_pair} @*)
    auto cur_result = insert(tile, cur_pair, table);(*@ \labelline{lst:coop_insert:do_insert} @*)
    if(tile.thread_rank() == cur_lane){(*@ \labelline{lst:coop_insert:finalize} @*)
      to_insert = false;(*@ \labelline{lst:coop_insert:pop} @*)
      success = cur_result;(*@ \labelline{lst:coop_insert:write_result} @*)
    }
  }
  return success;
}
\end{lstlisting}
\end{figure}

\begin{figure}[ht]
  \centering
  \begin{lstlisting}[caption={Tile-sized bucket implementation.},captionpos=b, label=lst:larger_bucket]
struct bucket{
  // Constructor to load the key-value pair of the bucket
  bucket(pair_type* ptr, const tile_type& tile) : ptr_(ptr), tile_(tile){(*@ \labelline{lst:larger_bucket:construct} @*)
    lane_pair_ = ptr[tile_.thread_rank()];
  }
  // Compute the load of the bucket
  int compute_load() const {(*@ \labelline{lst:larger_bucket:load} @*)
    auto load_bitmap = tile_.ballot(lane_pair_.key != EMPTY_KEY);
    return __popc(load_bitmap);
  }
  // Find the value associated with a key
  value_type find_key_value(const key_type key) {(*@ \labelline{lst:larger_bucket:find} @*)
    bool key_exist = (key == lane_pair_.key);
    int key_lane = __ffs(tile.ballot(key_exist));
    if(key_lane == 0) return EMPTY_KEY;
    return tile_.shfl(lane_pair_.value, key_lane - 1);
  }
  // Perform an insertion using compare-and-swap
  pair_type cas_at_location(const pair_type& pair, const int location) {(*@ \labelline{lst:larger_bucket:cas} @*)
    pair_type old_pair;
    if(tile_.thread_rank() == elected_lane_){
      old_pair = atomicCAS(ptr_ + location, EMPTY_PAIR, pair);
    }
    return tile_.shfl(old_pair, elected_lane_);
  }
  // Perform an exchange operation
  pair_type exch_at_location(const pair_type& pair, const int location) {(*@ \labelline{lst:larger_bucket:exch} @*)
    pair_type old_pair;
    if(tile_.thread_rank() == elected_lane_){
      old_pair = atomicExch(ptr_ + location, pair);
    }
    return tile_.shfl(old_pair, elected_lane_);
  }
  private:
  pair_type* ptr_;
  pair_type lane_pair_;
  const tile_type tile_;
  const int elected_lane_ = 0;
}
\end{lstlisting}
\end{figure}

We first discuss the implementation details for performing insertion in a tile-cooperative fashion. Performing insertion in a cooperative fashion allows serializing all operations within a tile and then each serial operation can read a bucket in a coalesced fashion (a generalized version of Ashkiani's warp-cooperative work sharing strategy~\cite{Ashkiani:2018:ADH}). Listing~\ref{lst:coop_insert} shows the construction of a tile in a cooperative insertion function. In line~\ref{lst:coop_insert:thread_block} we first construct a thread block that will contain our tiles. Then, we can use the thread block to construct a tile with a size equal to the bucket size (known at compile time) in line~\ref{lst:coop_insert:tile}. Now we can utilize the tile and perform the various operations to either query the rank of the thread in the tile (line~\ref{lst:coop_insert:rank}) or evaluate a predicate to build a queue of all the threads that will perform an operation later (line~\ref{lst:coop_insert:ballot}). Line~\ref{lst:coop_insert:cur_lane} shows finding the next item in the queue, using this information, the pair associated with the lane is broadcasted to all threads in the tile in line~\ref{lst:coop_insert:cur_pair}. In line~\ref{lst:coop_insert:do_insert}, the pair is inserted into the hash table. Finally, in line~\ref{lst:coop_insert:finalize} only the current lane removes the item from the queue by setting \texttt{to\_insert} to \texttt{false} (line~\ref{lst:coop_insert:pop}) and writes back its result to its own register in line~\ref{lst:coop_insert:write_result}. Performing a query is similar with the only differences being that the result contains the value of the key (or a sentinel key if it was not found) and calling the find operation instead of the insertion.

Now we discuss the common operation on a bucket shown in Listing~\ref{lst:larger_bucket}. In line~\ref{lst:larger_bucket:construct} we construct the bucket structure using a pointer to the bucket and the cooperative-group tile. To compute the load of the bucket (line~\ref{lst:larger_bucket:load}) we perform a \texttt{ballot} operation to populate an unsigned integer with bits (for each thread in the tile) that are set whenever the bucket's key is not a sentinel key. Using the populated integer, we can count the number of valid keys in the bucket using the population count (\texttt{\_\_popc}) intrinsic. Additionally, we can use the load to find the next available insertion location in the bucket.

To perform a query (line~\ref{lst:larger_bucket:find}), we first compare the query key with the keys present in the bucket followed up by a \texttt{ballot} instruction to compact the comparison result into a single unsigned integer. By finding the first-set bit in the the result (using an \texttt{\_\_ffs}) we can determine the location of the key. Note that the index of the least-significant bit is 1 while a result of zero indicates that the key is not present. Finally, we broadcast the result to all threads in the tile.

One way to perform insertion is to swap an empty sentinel pair with a new pair (line~\ref{lst:larger_bucket:cas}) using the compare-and-swap (CAS) instruction. Only one thread performs the CAS operation and attempts to swap an empty sentinel pair with the new pair, then we broadcast the swapped-out pair to all threads in the tile. Only when the swapped out pair is indeed an empty pair we can conclude that the insertion succeeded. Another way to perform insertion is to exchange an old pair with the new one (line~\ref{lst:larger_bucket:exch}). Similarly, only one thread performs the exchange operation then we broadcast the exchanged pair to all threads in the tile.
\subsection{Bucketed cuckoo hash table}
\label{sec:impl:bcht}

\pp{Insertion} Listing~\ref{lst:bucketed_cuckoo_insertion} shows the implementation details for insertion in a bucketed cuckoo hash table. We begin by computing the first bucket we will attempt to insert into (line~\ref{lst:bucketed_cuckoo_insertion:hf0}). Then, we compute the bucket's load and if the bucket is full and we did not reach the maximum number of cuckoo chains we exchange our new pair with a random pair from the bucket (line~\ref{lst:bucketed_cuckoo_insertion:exch}) then we increment the counter of the number of swaps we performed (line~\ref{lst:bucketed_cuckoo_insertion:increment_counter}). To generate a random number, we use Marsaglia's Xorshf~\cite{Marsaglia:2003:XR} initialized once per tile. We dicussed the case when the first bucket was full and we needed to perform cuckooing; now if the bucket has an empty spot, we try to insert the pair into it. We perform an atomic compare-and-swap (line~\ref{lst:bucketed_cuckoo_insertion:cas}) then we compare the swapped out pair to the sentinel empty pair and terminate if insertion succeeded (lines~\ref{lst:bucketed_cuckoo_insertion:success}). If insertion does not succeed, we attempt the insertion again until we either succeed or reach the maximum number of cuckoo chains (line~\ref{lst:bucketed_cuckoo_insertion:failed}).

\begin{figure}[ht]
  \centering
  \begin{lstlisting}[caption={Bucketed cuckoo hash table insertion.},captionpos=b, label=lst:bucketed_cuckoo_insertion]
bool insert(const tile_type tile, const pair_type pair, pair_type* table){
 auto bucket_id = hf0(pair.key) % num_buckets;  (*@ \labelline{lst:bucketed_cuckoo_insertion:hf0} @*)
 uint32_t cuckoo_counter = 0;
 random_number_generator rng(0, bucket_size);
 bool success = false;
 do{
  auto cur_bucket = bucket(hash_table + bucket_id * bucket_size, tile);
  auto load = cur_bucket.compute_load();(*@ \labelline{lst:bucketed_cuckoo_insertion:load} @*)
  if(load == bucket_size){(*@ \labelline{lst:bucketed_cuckoo_insertion:full} @*)
   if(cuckoo_counter == max_cuckoo_chains) return false;(*@ \labelline{lst:bucketed_cuckoo_insertion:failed} @*)
   pair = cur_bucket.exch_at_location(pair, rng());(*@ \labelline{lst:bucketed_cuckoo_insertion:exch} @*)
   auto prev_bucket = bucket;
   auto bucket0 = hf0(pair.key) % num_buckets;(*@ \labelline{lst:bucketed_cuckoo_insertion:new_hf0} @*)
   auto bucket1 = hf1(pair.key) % num_buckets;(*@ \labelline{lst:bucketed_cuckoo_insertion:new_hf1} @*)
   auto bucket2 = hf2(pair.key) % num_buckets;(*@ \labelline{lst:bucketed_cuckoo_insertion:new_hf2} @*)
   bucket = bucket0;(*@ \labelline{lst:bucketed_cuckoo_insertion:new_bucket} @*)
   bucket = prev_bucket == bucket1 ? bucket2 : bucket;(*@ \labelline{lst:bucketed_cuckoo_insertion:prev_1} @*)
   bucket = prev_bucket == bucket0 ? bucket1 : bucket;(*@ \labelline{lst:bucketed_cuckoo_insertion:prev_0} @*)
   cuckoo_counter++;(*@ \labelline{lst:bucketed_cuckoo_insertion:increment_counter} @*)
  }else{
    auto old_pair = cur_bucket.insert_at_location(pair, load);(*@ \labelline{lst:bucketed_cuckoo_insertion:cas} @*)
    success = old_pair == EMPTY_PAIR;(*@ \labelline{lst:bucketed_cuckoo_insertion:success} @*)
   }
 }while(!success);
 return false;
}
\end{lstlisting}
\end{figure}

\pp{Find} Listing~\ref{lst:bucketed_cuckoo_query} shows our implementation for the find operation in a bucketed cuckoo hash table. To evaluate the query we need to inspect the 3 buckets associated with the three hash functions. We perform these three bucket inspections serially within the tile. Each time we evaluate the hash function (lines~\ref{lst:bucketed_cuckoo_query:hf0},~\ref{lst:bucketed_cuckoo_query:hf1}, or~\ref{lst:bucketed_cuckoo_query:hf2}). Then we load the bucket (line~\ref{lst:bucketed_cuckoo_query:load}). We look up the key's value (if it exists) in the bucket (line~\ref{lst:bucketed_cuckoo_query:exist}) and we also check if the bucket has an empty key (line~\ref{lst:bucketed_cuckoo_query:early_exist}). We either terminate after we inspect all the three hash functions or when one of the buckets has an empty key.

\begin{figure}[ht]
  \centering
  \begin{lstlisting}[caption={Bucketed cuckoo hash table find.},captionpos=b, label=lst:bucketed_cuckoo_query]
value_type find(const key_type key, pair_type* table){
 value_type result = EMPTY_VALUE;
 const int num_hash_functions = 3;
 for(int i = 0; i < num_hash_functions && result == EMPTY_VALUE; i++){
    uint32_t bucket_id;
    if(i == 0) bucket_id = hf0(key) % num_buckets;(*@ \labelline{lst:bucketed_cuckoo_query:hf0} @*)
    else if (i == 1) bucket_id = hf1(key) % num_buckets;(*@ \labelline{lst:bucketed_cuckoo_query:hf1} @*)
    else bucket_id = hf2(key) % num_buckets;(*@ \labelline{lst:bucketed_cuckoo_query:hf2} @*)
    auto cur_bucket = bucket(hash_table + bucket_id * bucket_size, tile);(*@ \labelline{lst:bucketed_cuckoo_query:load} @*)
    auto result = cur_bucket.find_key_value(key);(*@ \labelline{lst:bucketed_cuckoo_query:exist} @*)
    if(result == EMPTY_VALUE && i != num_hash_functions){
        if(cur_bucket.compute_load() != bucket_size) return result;(*@ \labelline{lst:bucketed_cuckoo_query:early_exist} @*)
    }
 }
 return result;
}
\end{lstlisting}
\end{figure}

\subsection{Placement strategy: The power of two (or more) choices}
\label{sec:placement_strategies}
\label{sec:placement_strategies:p2}

\pp{Insertion} We now describe the implementation details for the power-of-two-choices insertion as shown in Listing~\ref{lst:power_of_two_insertion}. We first evaluate the buckets associated with the two hash functions (i.e., the two choices) (lines~\ref{lst:power_of_two_insertion:hf0} and~\ref{lst:power_of_two_insertion:hf1}). After computing the loads, if both buckets are full, then insertion fails (line~\ref{lst:power_of_two_insertion:failure}). Otherwise, we find the bucket with lower load (breaking ties by favoring the first hash function) then perform the insertion into that bucket (lines~\ref{lst:power_of_two_insertion:min_load1},~\ref{lst:power_of_two_insertion:min_load2} and~\ref{lst:power_of_two_insertion:try_insert}). Finally, we attempt an atomic compare-and-swap operation to swap the last empty pair with the new pair, and if the operation succeeds, insertion succeeds. Otherwise, we repeat the process until we insert the pair or the two possible locations become full.

\begin{figure}[ht]
\centering
\begin{lstlisting}[caption={Power of two insertion.},captionpos=b, label=lst:power_of_two_insertion]
bool insert(const tile_type tile, const pair_type pair){
  auto bucket0_id = hf0(pair.key) % num_buckets;  (*@ \labelline{lst:power_of_two_insertion:hf0} @*)
  auto bucket1_id = hf1(pair.key) % num_buckets;  (*@ \labelline{lst:power_of_two_insertion:hf1} @*)
  bool keep_trying = true;
  do{
    auto bucket0 = bucket(hash_table + bucket0_id * bucket_size, tile);
    auto bucket1 = bucket(hash_table + bucket0_id * bucket_size, tile);
    auto load0 = bucket0.compute_load();(*@ \labelline{lst:power_of_two_insertion:b0_load} @*)
    auto load1 = bucket1.compute_load();(*@ \labelline{lst:power_of_two_insertion:b1_load} @*)
    if(load0 == bucket_size && load1 == bucket_size) return false;(*@ \labelline{lst:power_of_two_insertion:failure} @*)
    auto bucket_choice = load0 <= load1 ? bucket0 : bucket1;(*@ \labelline{lst:power_of_two_insertion:min_load1} @*)
    auto load_choice = load0 <= load1 ? load0 : load1;(*@ \labelline{lst:power_of_two_insertion:min_load2} @*)
    keep_trying = bucket_choice.cas_at_location(pair, load_choice, 0) != EMPTY_PAIR;(*@ \labelline{lst:power_of_two_insertion:try_insert} @*)
  }while(keep_trying);
  return true;
}
\end{lstlisting}
\end{figure}

\pp{Find} Find in BP2HT is similar to the one in BCHT (Listing~\ref{lst:bucketed_cuckoo_query}) with only two differences: 1) we only need to inspect two buckets, and 2) we can only terminate when we inspect both buckets (i.e., no early exit as in line~\ref{lst:bucketed_cuckoo_query:early_exist}).

\subsection{Placement strategy: Iceberg hashing}
\label{sec:placement_strategies:iceberg}

\pp{Insertion} Listing~\ref{lst:iceberg_insertion} shows the implementation for insertion in an Iceberg hash table. We first begin by computing the primary bucket and its load (lines~\ref{lst:iceberg_insertion:hfp}, and~\ref{lst:iceberg_insertion:p_load}). Now the load is known, if it is larger than the threshold we check the two secondary hash functions and their buckets. We calculate the buckets' indices and their load (lines~\ref{lst:iceberg_insertion:hf0} to~\ref{lst:iceberg_insertion:hf1_load}).  Note that if the two secondary buckets are full we use the primary bucket, otherwise we find the least loaded bucket and use it for the insertion (lines~\ref{lst:iceberg_insertion:p2_load} and~\ref{lst:iceberg_insertion:p2_location}). If after inspecting all three possible buckets, there is no empty space for insertion, insertion fails (line~\ref{lst:iceberg_insertion:exist}). Finally, we try to perform the insertion (line~\ref{lst:iceberg_insertion:do_insert}) and if the returned pair is the sentinel pair the insertion succeeded. Otherwise, we repeat the process until insertion succeeds or all three possible locations are full.

\begin{figure}[ht]
\centering
\begin{lstlisting}[caption={Iceberg insertion.},captionpos=b, label=lst:iceberg_insertion]
bool insert(const tile_type tile, const pair_type pair, pair_type* table){
  auto primary_bucket_id = hfp(pair.key) % num_buckets;  (*@ \labelline{lst:iceberg_insertion:hfp} @*)
  while(true){
    auto bucket = bucket(table + primary_bucket_id * bucket_size, tile);
    auto load = compute_load();  (*@ \labelline{lst:iceberg_insertion:p_load} @*)
    if(primary_bucket_load > threshold){
      auto bucket0_id = hf0(key) % num_buckets;(*@ \labelline{lst:iceberg_insertion:hf0} @*)
      auto bucket1_id = hf1(key) % num_buckets;(*@ \labelline{lst:iceberg_insertion:hf1} @*)
      auto bucket_0 = bucket(table + bucket0_id * bucket_size, tile);
      auto bucket_1 = bucket(table + bucket_id * bucket_size, tile);
      auto load0 = bucket0.compute_load();(*@ \labelline{lst:iceberg_insertion:hf0_load} @*)
      auto load1 = bucket1.compute_load();(*@ \labelline{lst:iceberg_insertion:hf1_load} @*)
      if(load0 != bucket_size && load1 != bucket_size){
        bucket = load0 <= load1 ? bucket0 : bucket1;(*@ \labelline{lst:iceberg_insertion:p2_load} @*)
        load = load0 <= load1 ? load0 : load1;(*@ \labelline{lst:iceberg_insertion:p2_location} @*)
      }
    }
    if(insertion_location == bucket_size) return false;(*@ \labelline{lst:iceberg_insertion:exist} @*)
    if(bucket.cas_at_location(pair, load) == EMPTY_PAIR){(*@ \labelline{lst:iceberg_insertion:do_insert} @*)
      return true;
    }
  }
  return false;
}
\end{lstlisting}
\end{figure}

\pp{Find} Similar to BP2HT's find operation, find in IHT cannot perform early exits, instead requiring inspecting all of the three buckets (the primary and the two secondary buckets) unless the query terminates after finding its key's value in an early bucket.

\section{High-Level Hash Table Recommendations}
\label{sec:recommendations}

\begin{table}
    \centering
    \begin{tabular}{ccccc}
      \toprule
        Method & Load factor & Insertion Probes & Query Probes & Stability\\ \midrule
        1CHT   & 0.88        & $\approx 2.8$    & up to 4     & no       \\
        BCHT   & 0.98        & $\approx 1.8$    & up to 3     & no       \\
        BP2HT  & 0.92        & 2                & up to 2     & yes      \\
        IHT    & 0.92        & 1 or 3           & up to 3     & yes      \\
      \bottomrule
    \end{tabular}
    \caption{Properties for different hash tables. 1CHT uses 4 hash functions while BCHT uses only 3 hash functions. BP2HT uses two choices (i.e., two hash functions) while IHT uses a primary hash function and two secondary hash functions.}
    \label{tab:hashtables_properties}
\end{table}

The designer of a hash table may prioritize peak achievable load factor, insertion rate, query rate, or some combination of those. Table~\ref{tab:hashtables_properties} summarizes the performance of each technique given each different priority. We discuss next our preferred choice and intuition for each of these priorities.

\subsection{High load factor}
\paragraph{Choice} BCHT achieves the highest load factors, followed by IHT and BP2HT then 1CHT\@.
\paragraph{Why?} BCHT can move keys between buckets until it succeeds or reaches the maximum allowable cuckoo chain length, but IHT only has 3 possible locations for insertion while BP2HT has only two possible locations. 1CHT has 4 locations but each location can only hold a single entry. Increasing the number of hash functions for any of these techniques yields higher load factors, but decreases the throughput of insertion and query operations.

\subsection{Insertion throughput}
\paragraph{Choice} BCHT achieves the highest throughput, followed by IHT, BP2HT then 1CHT\@.
\paragraph{Why?} 1CHT's memory access pattern is not optimal, since it only accesses one key (and its value) from a 128-byte cache line, utilizing only 1/16 of the cache line and bandwidth. Moreover, its small bucket size increases the average probes per key. In contrast, bucketed techniques utilize the whole cache line. Among the bucketed techniques, BP2HT transfers more memory since it requires loading both of the buckets before deciding to which bucket to write. Similarly, IHT inspects the primary bucket and if its capacity exceeds the threshold, it inspects the two secondary buckets. At low load factors, fewer secondary buckets are inspected, and it requires similar memory behavior as BCHT, but at high load factors, inspecting secondary buckets is more likely and the required memory transfers are higher than BCHT\@.
\subsection{Query throughput}
\paragraph{Choice} BCHT achieves the highest throughput, followed by IHT, BP2HT, then 1CHT\@.
\paragraph{Why?} Similar to insertion, the cache-line-sized memory access of BCHT allows it to exceed the performance of 1CHT\@. IHT favors insertion in the primary bucket over the two secondary buckets, which makes it inspect fewer buckets than BP2HT, but more than BCHT\@. However, if the queries do not exist in the hash table, IHT falls behind BP2HT since it will always need to read 3 buckets compared to 2 in BP2HT\@. 1CHT can achieve better performance compared to a large-bucket BP2HT and IHT (for example, a bucket size of 32). The cost of loading these additional cache lines exceeds the ones that 1CHT requires (1 cache line for each hash function, i.e., 4 cache lines).

\subsection{Stability}
\paragraph{Choice} IHT and BP2HT are the only hash tables that support stability.
\paragraph{Why?} Either IHT or BP2HT writes keys to a number of buckets, depending on the number of hash functions used, and after a key is placed in a bucket, the key is never moved. On the other hand, 1CHT and BCHT have the freedom to move keys between buckets, which means that they are not stable. The tradeoff here is between stability and the highest achievable load factors.
\section{Results}
\label{sec:results}

Now we analyze and compare in detail the three hash table configurations that we describe in this paper and CUDPP's 1CHT\@. Our figures of merit are query rate, build rate, and load factor. We evaluate all different hash table implementations\footnote{Our implementation is available at \url{https://github.com/owensgroup/BGHT}.} on an NVIDIA TITAN~V (Volta) GPU with 12~GB DRAM and an Intel Xeon E5-2637 CPU\@. Our code is complied with CUDA~11.1. The GPU has a theoretical achievable DRAM bandwidth of 652.8~GiB/s. All results are averaged over 10 successful experiments with a maximum of 50 failures unless stated otherwise. All keys (and values) are 32-bit randomly generated unique keys using \texttt{std::mt19937}, a random number engine based on the Mersenne Twister algorithm, then transformed using \texttt{std::uniform\_int\_distribution} to uniformly distributed unsigned integers from 0 up to the maximum unsigned integer. We use the hash function $h(k; a, b) = ((ak + b)  \mod{p}) \mod{L}$, where $a$ and $b$ are randomly generated integers and $p$ is a random prime number (we use $p = 4294967291$) and $L$ is the number of buckets in the hash table.

We use the following variations of hash tables in our evaluation:
\begin{LaTeXdescription}
  \item[1CHT] A cuckoo hash table with bucket size of one, faithful to Alcantara et al.~\cite{Alcantara:2011:BAE} and their CUDPP implementation~\cite{Harris:2017:CUDPP}. We use four hash functions like CUDPP's default configuration.
  \item[BCHT] Bucketed cuckoo hash table where we increase $b$, the number of items per bucket, to larger values.
  \item[BP2HT] Bucketed power-of-two-choices cuckoo hash table with $b>1$.
  \item[IHT] Iceberg hash table, parameterized by the number of items per bucket $b$ and the threshold $t$.
\end{LaTeXdescription}

\subsection{Configuring bucketized and iceberg parameters}

We first begin our performance analysis by configuring the optimal parameters for each hash table. For each technique, we evaluate the query and insertion throughput for two scenarios: when the number of keys is constant with varying load factors; and a constant load factor and a different number of keys. We also evaluate the query performance for three different fractions of positive queries (fraction of lookup keys that exist in the hash table): 100\%, 50\%, and 0\%. When the number of keys is fixed, we use 50 million keys as our input size, and when we vary the load factor, we choose the two load factors 0.8 and 0.9. Since our implementation uses cooperative groups as its main building block, the bucket size is always a power-of-two number that varies up to the warp size (i.e., 32). We avoid using larger bucket sizes (we expect they will not be performance-competitive) as the functionality of cooperative groups is more limited and it has the additional cost of using shared memory for communicating between threads in the tile when performing tile-wide operations.

\begin{table}
  \centering
  \begin{tabular}{cccccc}
    \toprule
    \multirow{2}{*}{Fixed parameter} & \multirow{2}{*}{$b$} & \multirow{2}{*}{Insert} & \multicolumn{3}{c}{\% of queries present in hash table}                                       \\[0.2em]\cmidrule(l){4-6}
                                     &                      &                         & 100\%                                                   & 50\%             & 0\%              \\ \midrule
    \multirow{4}{*}{50M keys}        & 1                    & 735.84                  & 2235                                                    & 1884.83          & 1496.21          \\
                                     & 8                    & 1150.60                 & 3396.57                                                 & 2645.33          & 2127.42          \\
                                     & 16                   & \textbf{1273.86}        & \textbf{3642.61}                                        & \textbf{2840.57} & \textbf{2313.17} \\
                                     & 32                   & 1140.14                 & 2166.94                                                 & 1739.29          & 1450.70          \\\midrule
    \multirow{4}{*}{0.8 load factor} & 1                    & 772.08                  & 2313.14                                                 & 1945.56          & 1530.62          \\
                                     & 8                    & 1309.63                 & 3858.59                                                 & 3206.12          & 2695.23          \\
                                     & 16                   & \textbf{1425.00}        & \textbf{4033.52}                                        & \textbf{3546.12} & \textbf{3155.51} \\
                                     & 32                   & 1243.22                 & 2334.66                                                 & 2181.45          & 2050.83          \\\midrule
    \multirow{4}{*}{0.9 load factor} & 1                    & 580.73                  & 1987.13                                                 & 1646.54          & 1323.49          \\
                                     & 8                    & 1193.83                 & 3502.35                                                 & 2643.75          & 2081.43          \\
                                     & 16                   & \textbf{1344.38}        & \textbf{3799.58}                                        & \textbf{2921.74} & \textbf{2359.95} \\
                                     & 32                   & 1206.40                 & 2260.30                                                 & 1839.16          & 1550.71          \\
    \bottomrule
  \end{tabular}
  \caption{Average insertion and query throughput (different positive query percentages) for BCHT when: 1) the number of keys is fixed at 50M keys and 2) the load factor is fixed at 0.8 and 0.9.}
  \label{tab:bcht_results_summary}
\end{table}

\subsubsection{BCHT}  
\begin{figure*}
    \centering
    \begin{subfigure}{0.24\textwidth}
      \includegraphics[width=\textwidth]{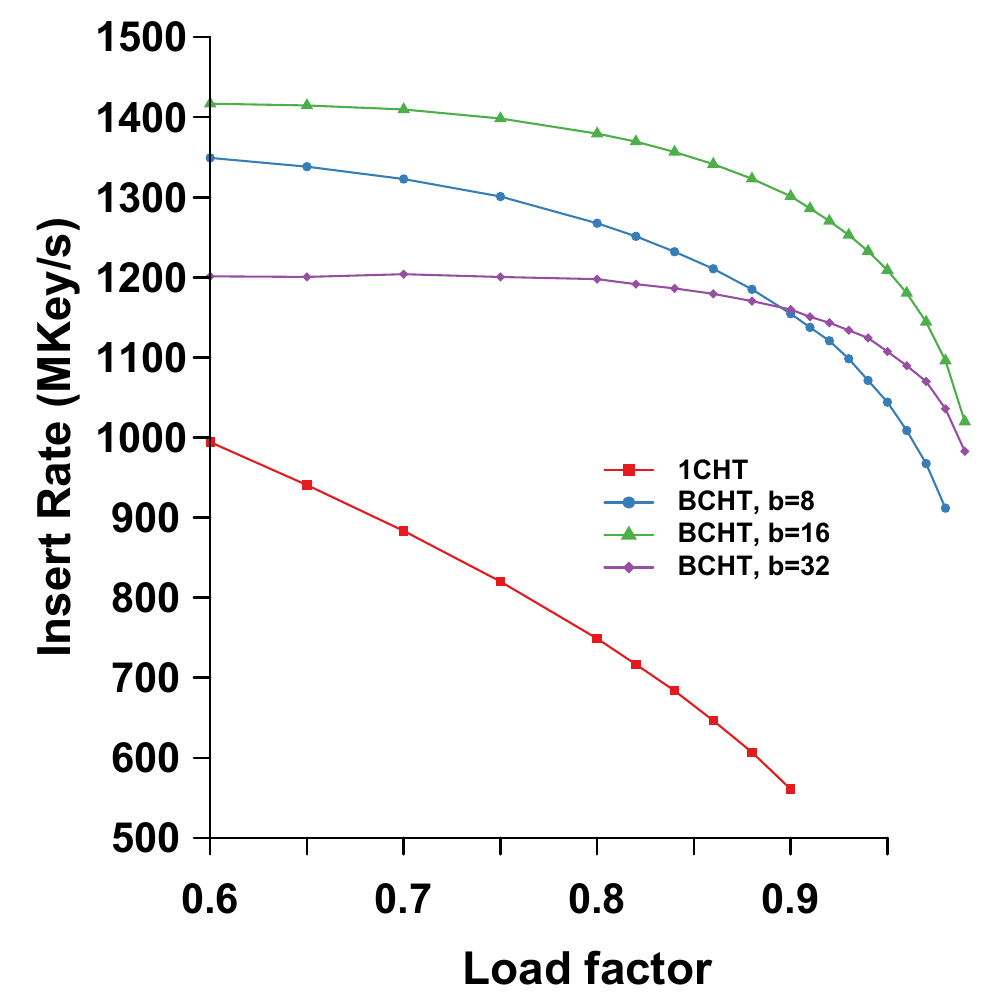}
      \caption{Insertion}
    \end{subfigure}
    \begin{subfigure}{0.24\textwidth}
      \includegraphics[width=\textwidth]{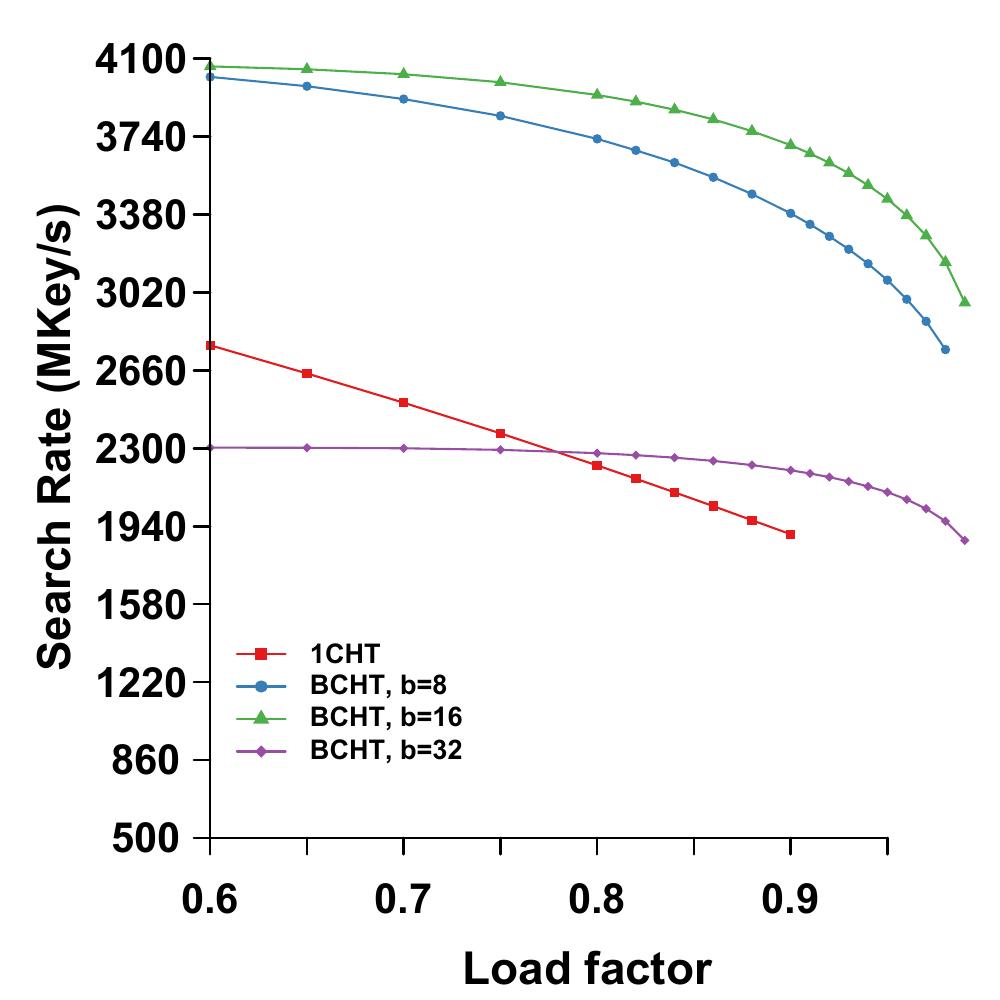}
      \caption{Find (100\% positive queries)}
    \end{subfigure}
    \begin{subfigure}{0.24\textwidth}
      \includegraphics[width=\textwidth]{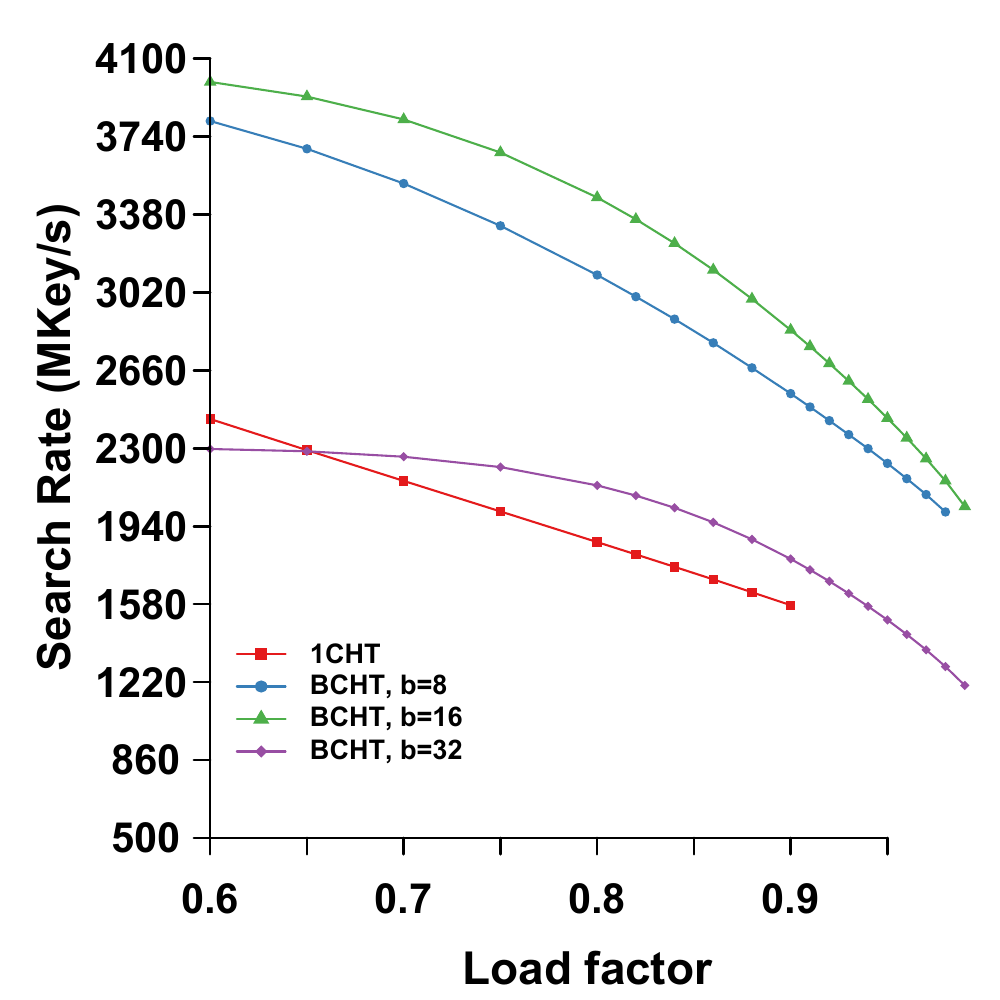}
      \caption{Find (50\% positive queries)}
    \end{subfigure}
    \begin{subfigure}{0.24\textwidth}
      \includegraphics[width=\textwidth]{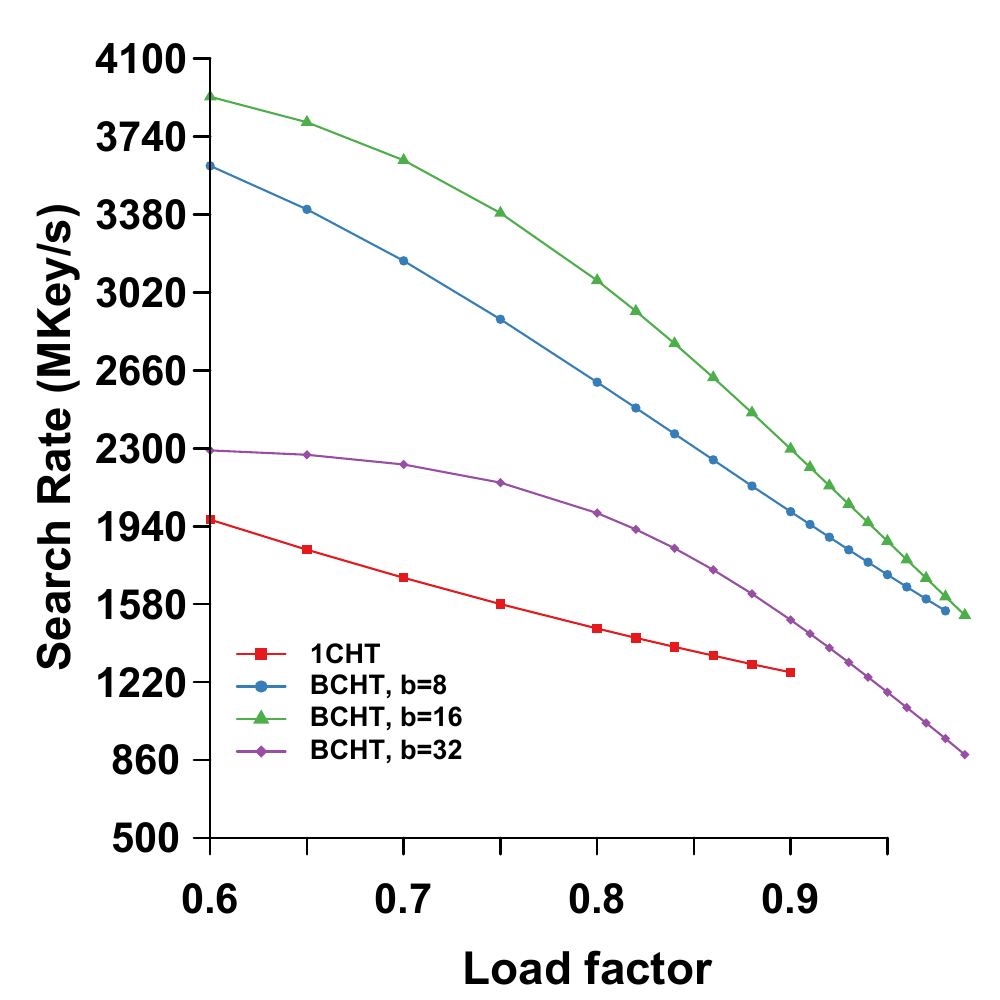}
      \caption{Find (0\% positive queries)}
    \end{subfigure}
    \caption{BCHT insertion and query rates for different positive query ratios and 50M keys.}
    \label{fig:config_bcht_fixed_num_keys}
  \end{figure*}

Figure~\ref{fig:config_bcht_fixed_num_keys} shows the throughput for the fixed number of keys scenario while varying the load factor. It shows that in all of the scenarios, a BCHT with a bucket size of 16 outperforms all the alternative bucket sizes. These results match our expectation since from a hardware perspective, we strongly prefer that the bucket size matches the GPU's cache line size, i.e., 128 bytes or 16 4-byte key-value pairs. For either query or insertion, the main factor affecting the performance is the load factor.  Higher load factors yield lower throughput for BCHT\@. This is due to the increase in the number of cuckoo chains (or evictions) we need to perform per key insertion. An additional factor that decreases the performance of query throughput is the fraction of query keys that exist in the hash table. Negative queries require inspecting more than one bucket, especially at high load factors (i.e., when most buckets are full), leading to a steeper decrease in performance for scenarios where more queries do not find a key.

\begin{figure*}
\setlength\tabcolsep{1pt}
\settowidth\rotheadsize{Radcliffe Cam}
\setkeys{Gin}{width=\hsize}
\begin{tabularx}{\textwidth}{l XXXX }
\rothead{\centering load factor = 0.8}
                        &   \includegraphics[valign=m]{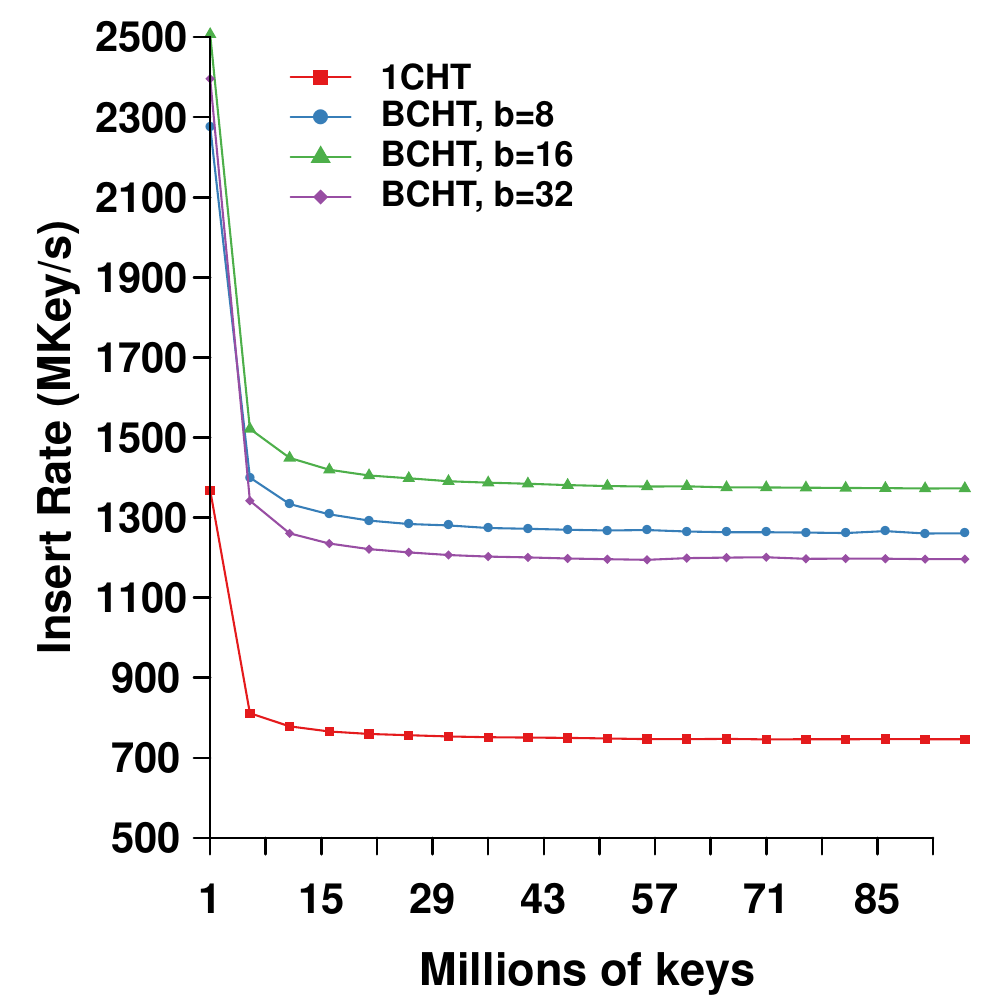}
                        &   \includegraphics[valign=m]{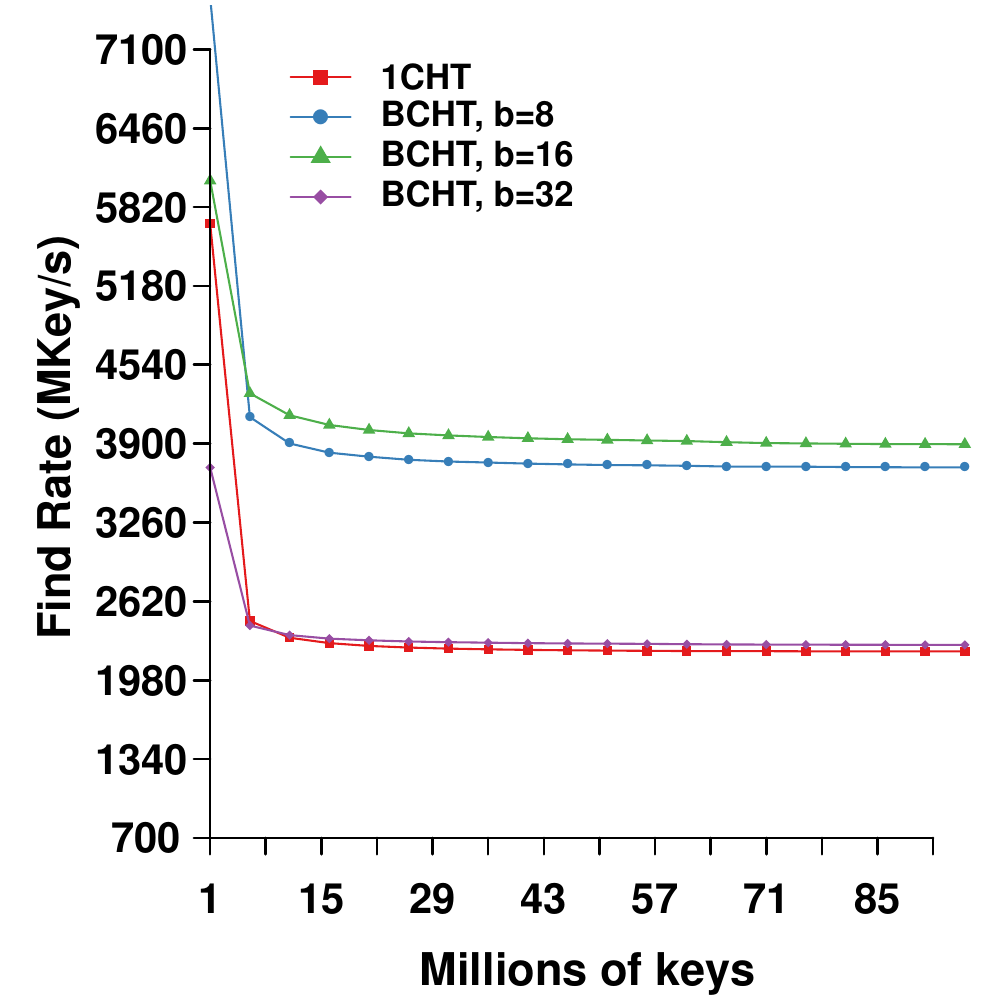}
                        &   \includegraphics[valign=m]{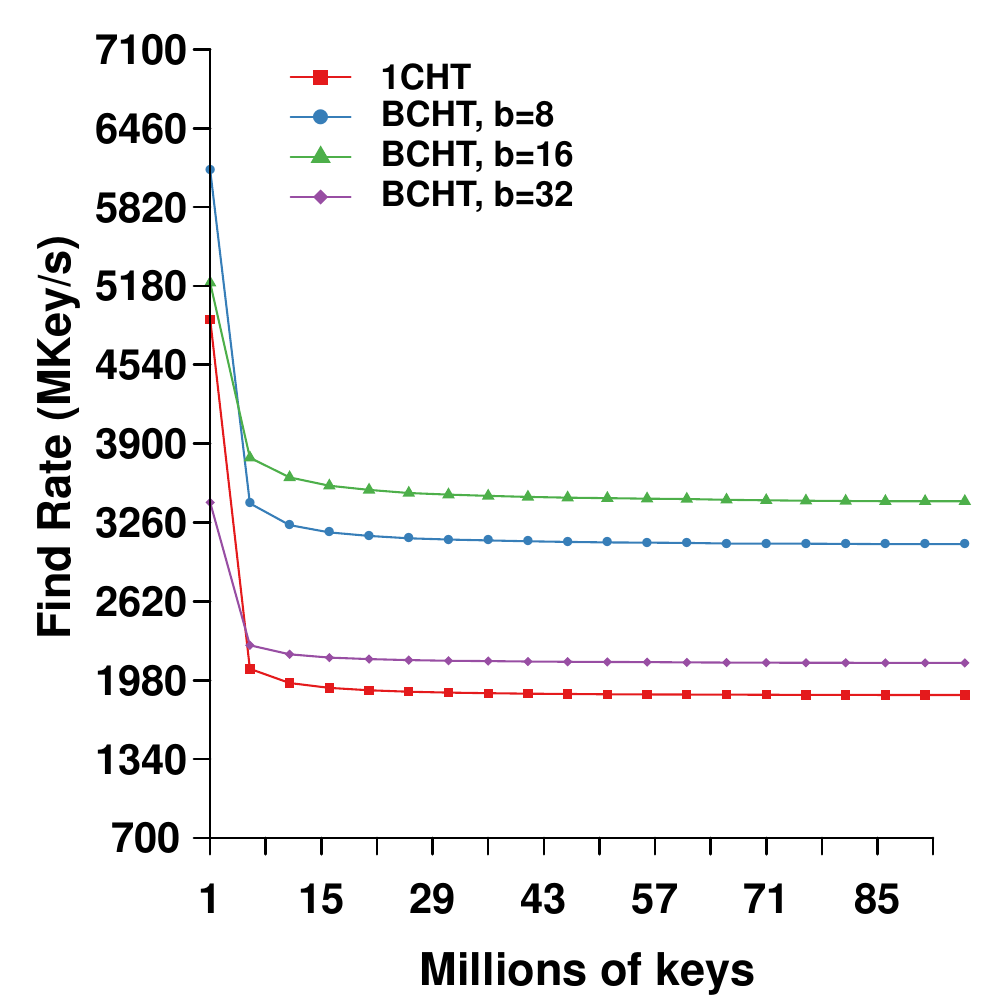}
                        &   \includegraphics[valign=m]{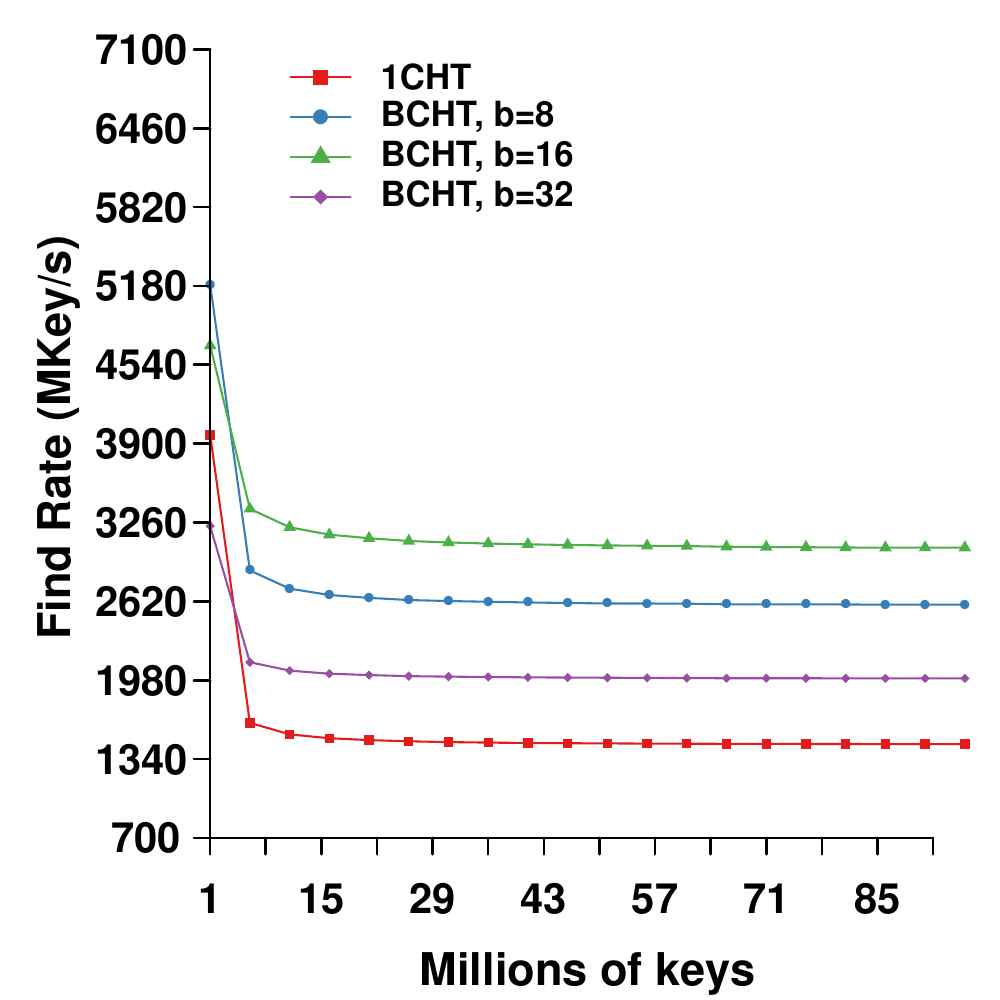}  \\ \addlinespace[2pt]
\rothead{\centering load factor = 0.9}
                        &   \includegraphics[valign=m]{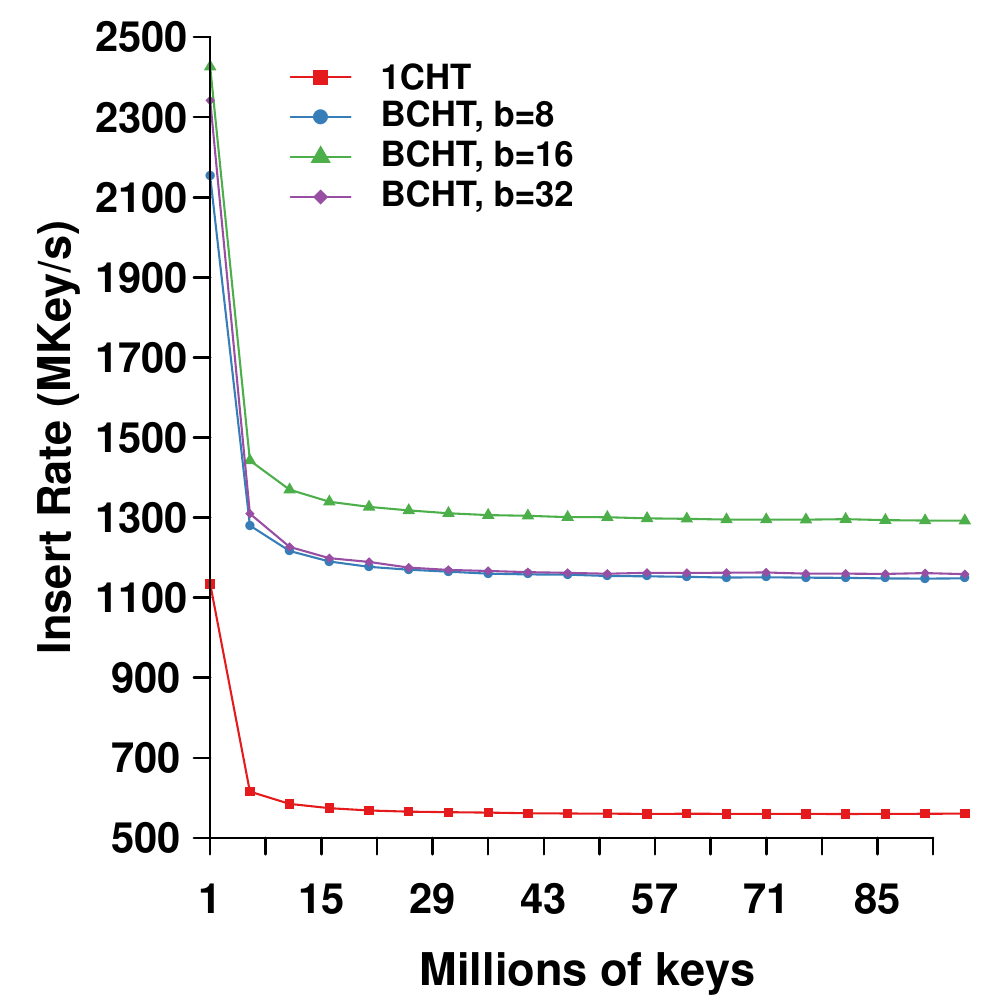}
                        &   \includegraphics[valign=m]{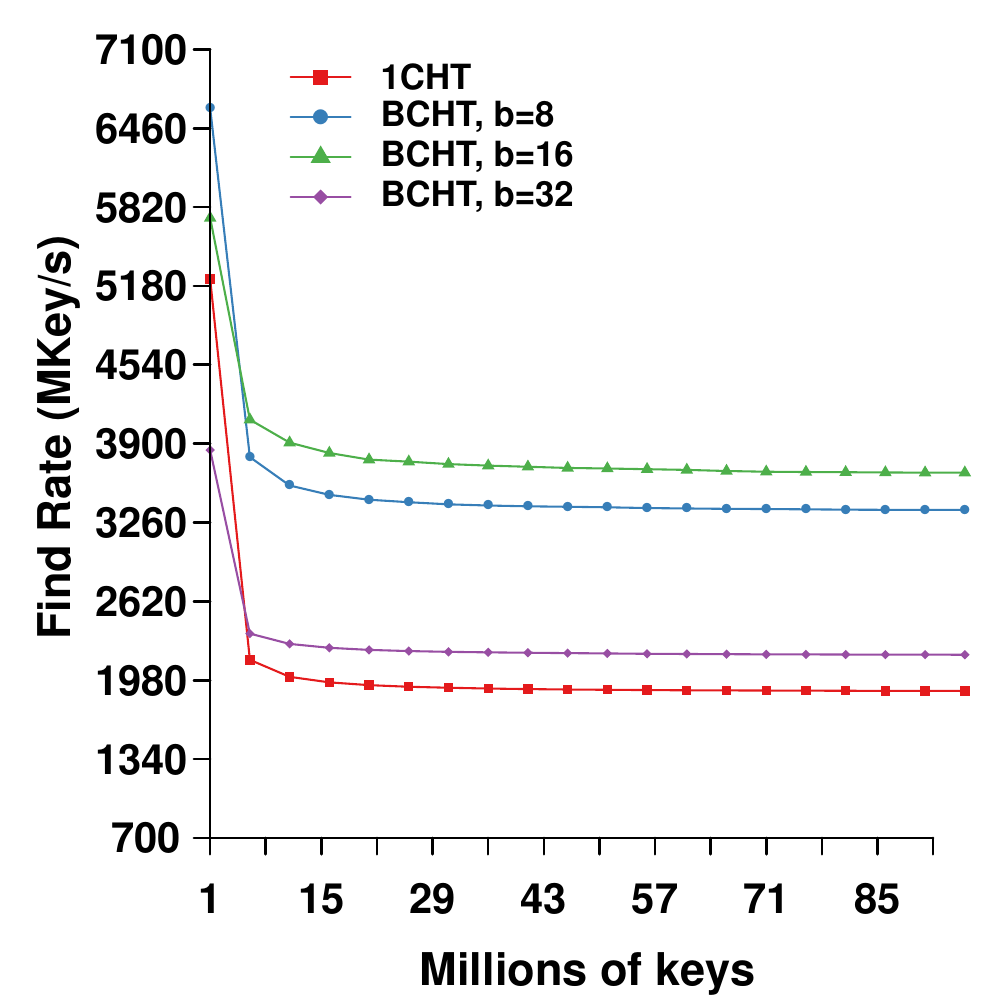}
                        &   \includegraphics[valign=m]{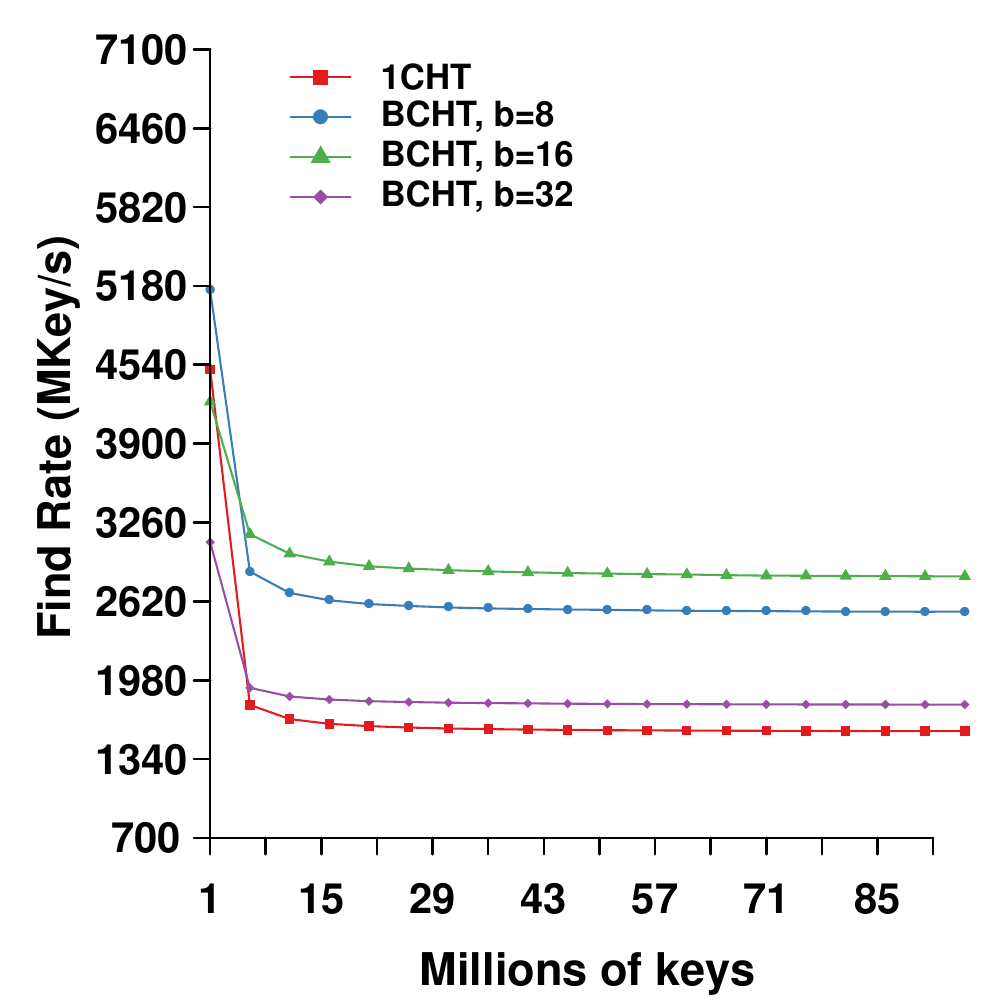}
                        &   \includegraphics[valign=m]{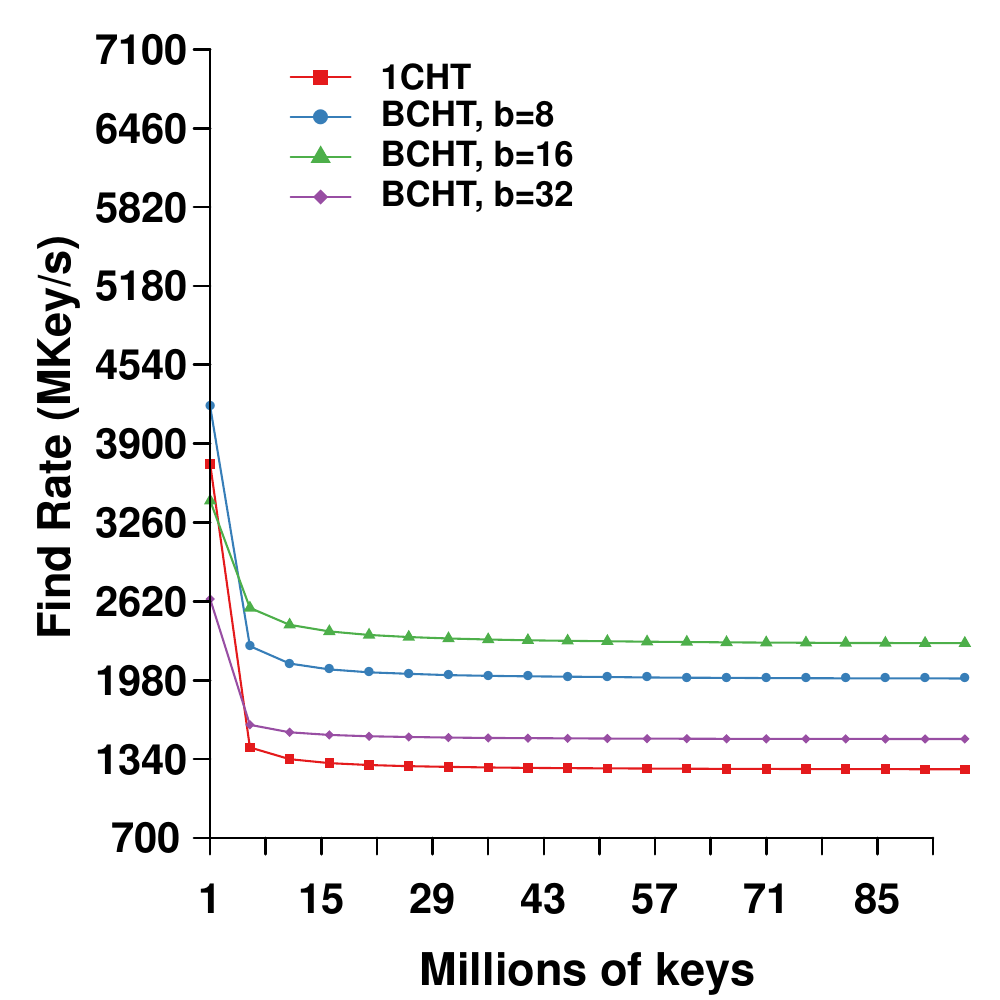}
\end{tabularx}
\caption{BCHT throughput for insertion, 100\%, 50\%, and 0\% positive queries (from left to right).}
\label{fig:config_bcht_fixed_load_factor}
\end{figure*}

Figure~\ref{fig:config_bcht_fixed_load_factor} shows a comparison between the different BCHT bucket sizes for two different load factors (0.8 and 0.9). In it, we see that when the hash table is large enough to exceed the L2 cache size (here, our L2 cache can hold up to 589k pairs) the throughput is almost constant for all the bucket sizes. Table~\ref{tab:bcht_results_summary} provides an averaged summary for the different bucket sizes' results.

\begin{figure*}
  \centering
  \begin{subfigure}{0.24\textwidth}
    \includegraphics[width=\textwidth]{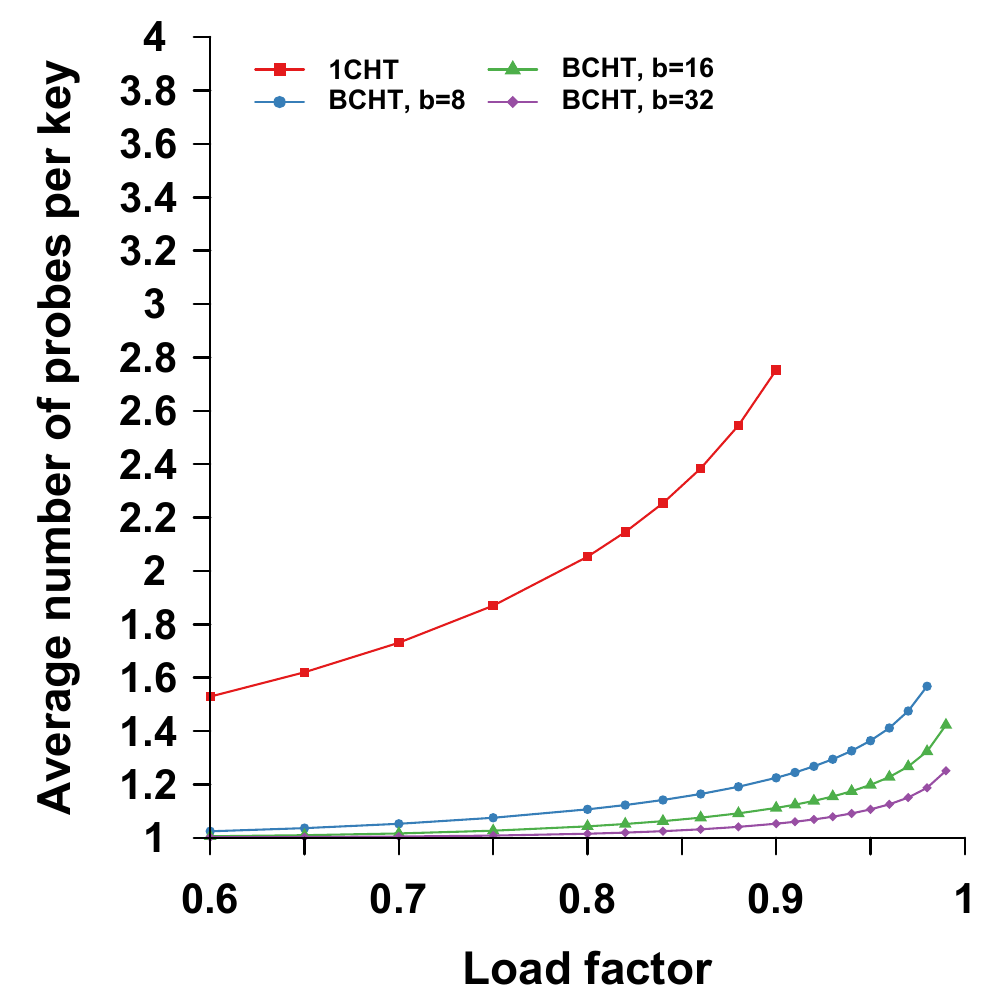}
    \caption{Insertion}
  \end{subfigure}
  \begin{subfigure}{0.24\textwidth}
    \includegraphics[width=\textwidth]{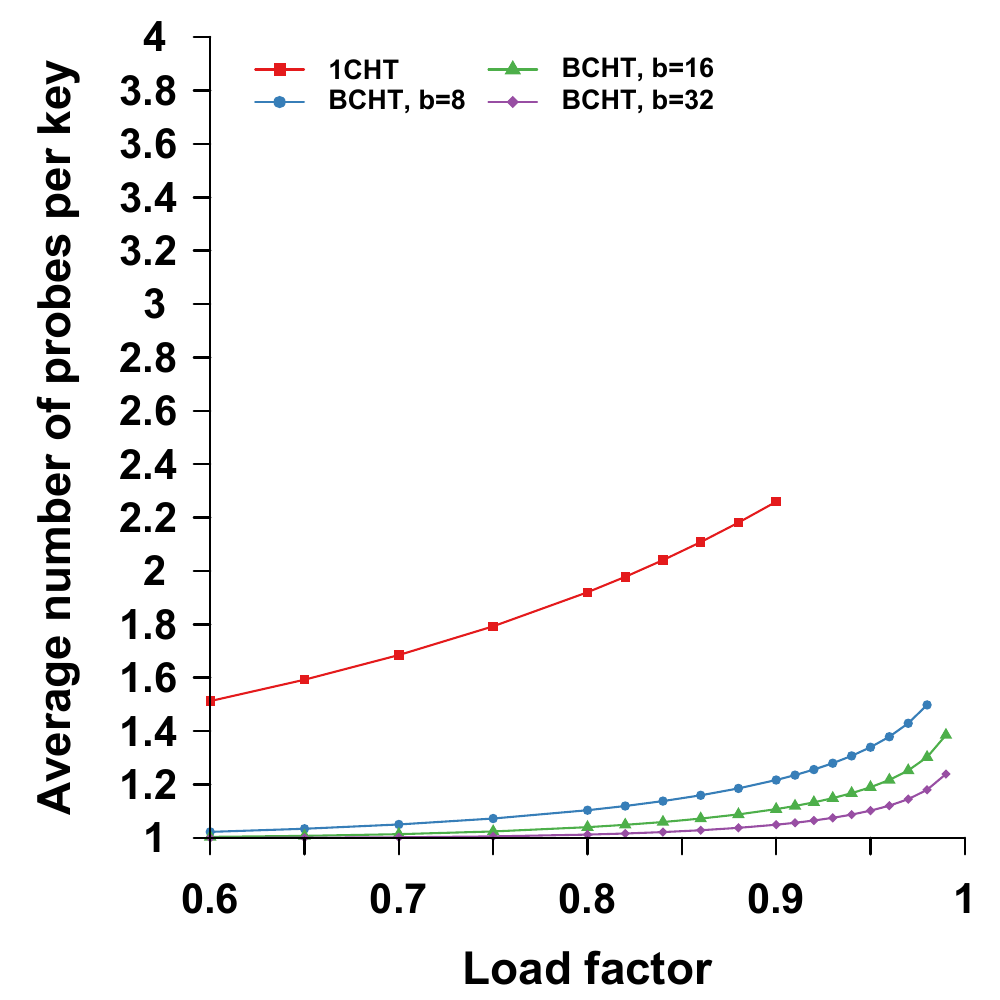}
    \caption{Find (100\% positive queries)}
  \end{subfigure}
  \begin{subfigure}{0.24\textwidth}
    \includegraphics[width=\textwidth]{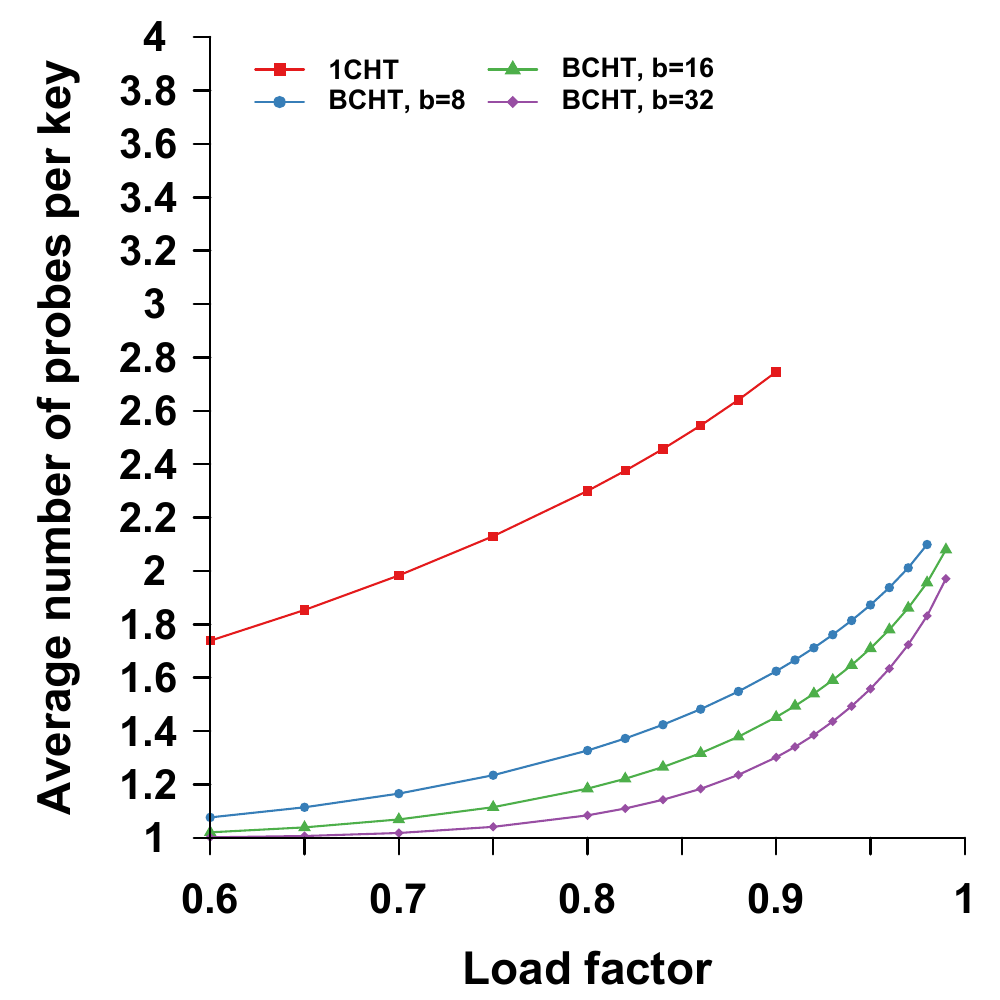}
    \caption{Find (50\% positive queries)}
  \end{subfigure}
  \begin{subfigure}{0.24\textwidth}
    \includegraphics[width=\textwidth]{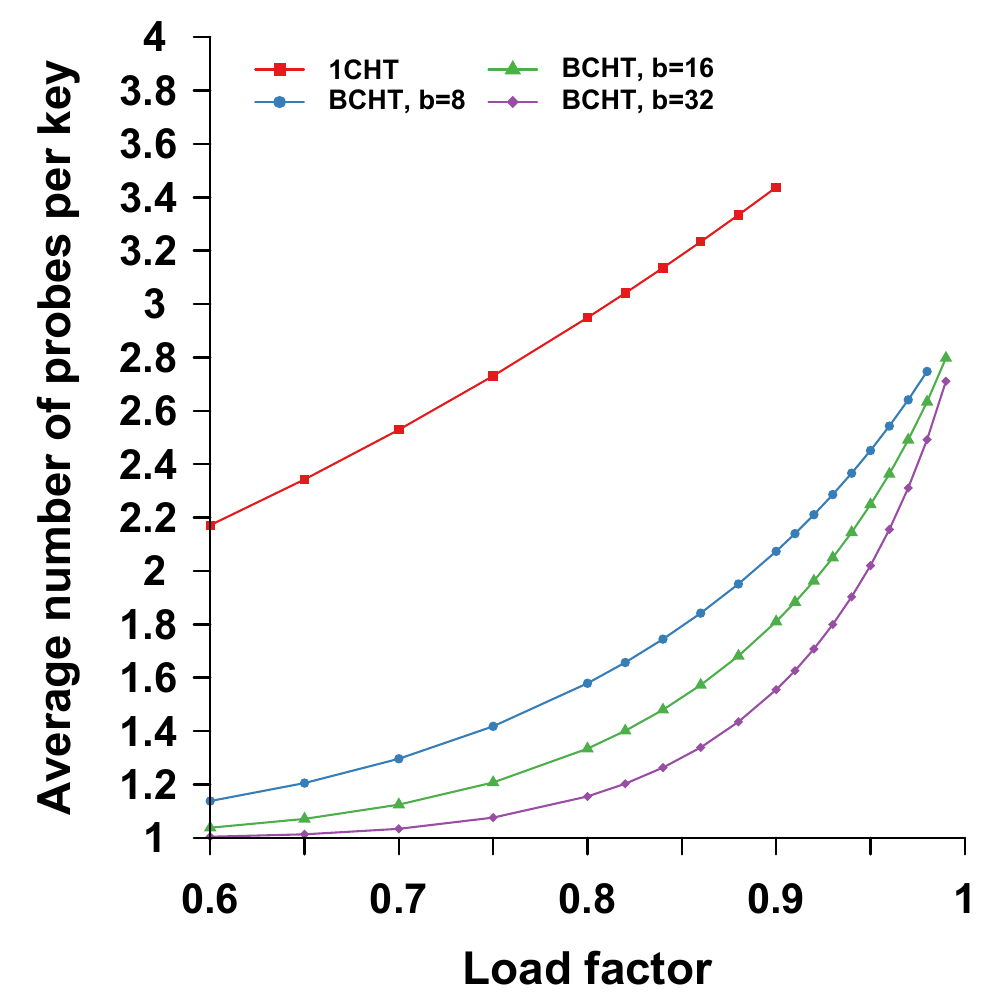}
    \caption{Find (0\% positive queries)}
  \end{subfigure}
  \caption{BCHT insertion and query average probes per key for 50M keys.}
  \label{fig:config_bcht_probes_per_keys}
\end{figure*}

Figure~\ref{fig:config_bcht_probes_per_keys} shows the average number of probes for the different hash table operations. For all operations, we see a clear advantage when using bucketed hash tables over 1CHT\@. For example, the average probe count for insertion drops from 2.75 to 1.23 when the bucket size increases from one to eight (at a load factor of 0.9), increasing the bucket size beyond eight yields minimal improvement resulting in an average probe count of \{1.11, 1.05\} for bucket sizes \{16, 32\}. However, as we discussed earlier, a bucket size of 16 matches the GPU's cache line size, yielding better throughput. The number of probes per query is bounded by the number of hash functions for any bucket size. In the best-case scenario (i.e., when a query finds the key in the first bucket), the number of probes will be close to one. When all queries are positive, we see that bucketed techniques require up to 1.5 probes per key at load factors as high as 0.98. However, in the worst-case scenario (i.e., high load factor and negative query), the query will load all the possible buckets. For example, when the bucket size is 16 and at a load factor of 0.99, the average number of probes per query is 2.8.

\begin{table}
  \centering
  \begin{tabular}{cccccc}
    \toprule
    \multirow{2}{*}{Fixed parameter} & \multirow{2}{*}{$b$} & \multirow{2}{*}{Insert} & \multicolumn{3}{c}{\% of queries present in hash table}                                       \\[0.2em]\cmidrule(l){4-6}
                                     &                      &                         & 100\%                                                   & 50\%             & 0\%              \\ \midrule
    \multirow{2}{*}{50M keys}        & 16                   & \textbf{1058.26}        & \textbf{3091.04}                                        & \textbf{2507.89} & \textbf{2123.46} \\
                                     & 32                   & 849.54                  & 1738.48                                                 & 1410.54          & 1190.29          \\\midrule
    \multirow{2}{*}{0.8 load factor} & 16                   & \textbf{1100.27}        & \textbf{3169.81}                                        & \textbf{2583.22} & \textbf{2194.49} \\
                                     & 32                   & 878.6                   & 1778.99                                                 & 1446.50          & 1222.44          \\\midrule
    \multirow{1}{*}{0.9 load factor} & 32                   & \textbf{886.15}         & \textbf{1785.97}                                        & \textbf{1452.68} & \textbf{1227}    \\
    \bottomrule
  \end{tabular}
  \caption{Average insertion and query throughput (different positive query percentages) for BP2HT when: 1) the number of keys is fixed at 50M keys and 2) the load factor is fixed at 0.8 and 0.9.}
  \label{tab:p2cht_results_summary}
\end{table}

\subsubsection{BP2HT}
\begin{figure*}
  \centering
  \begin{subfigure}{0.24\textwidth}
    \includegraphics[width=\textwidth]{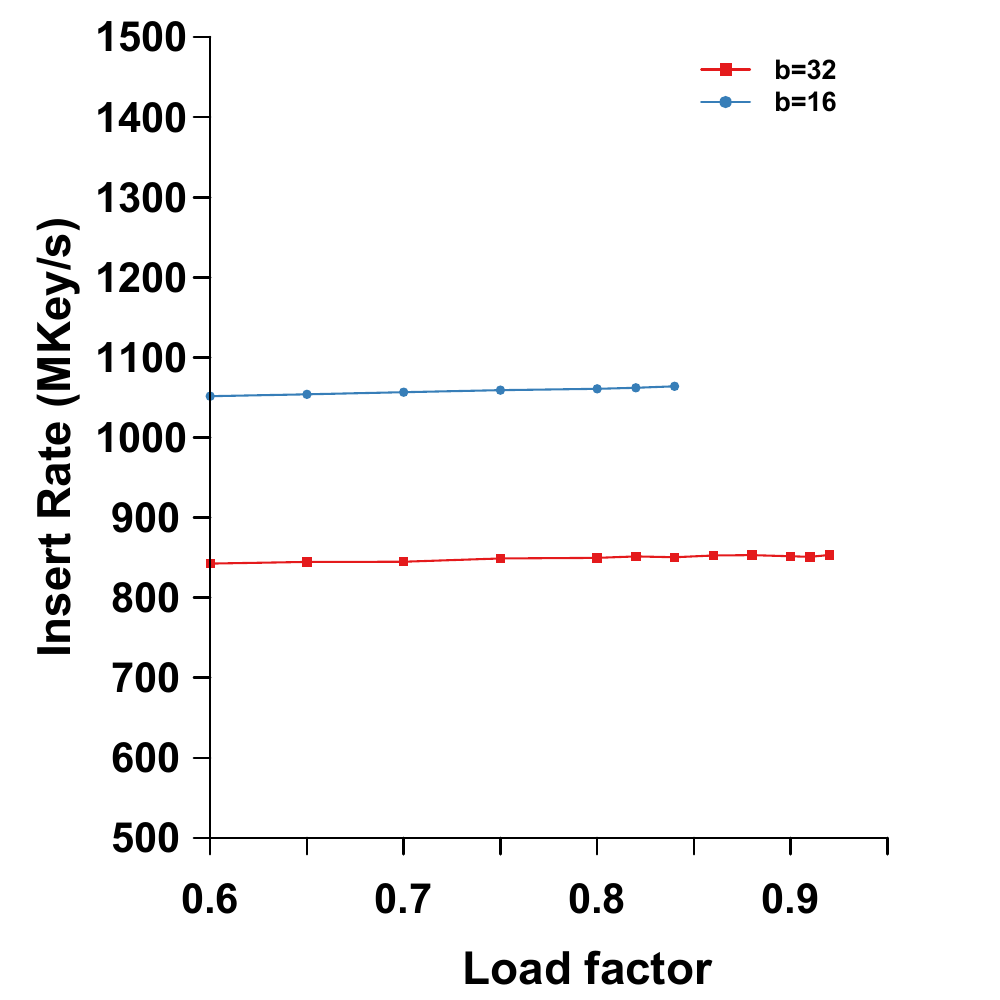}
    \caption{Insertion}
  \end{subfigure}
  \begin{subfigure}{0.24\textwidth}
    \includegraphics[width=\textwidth]{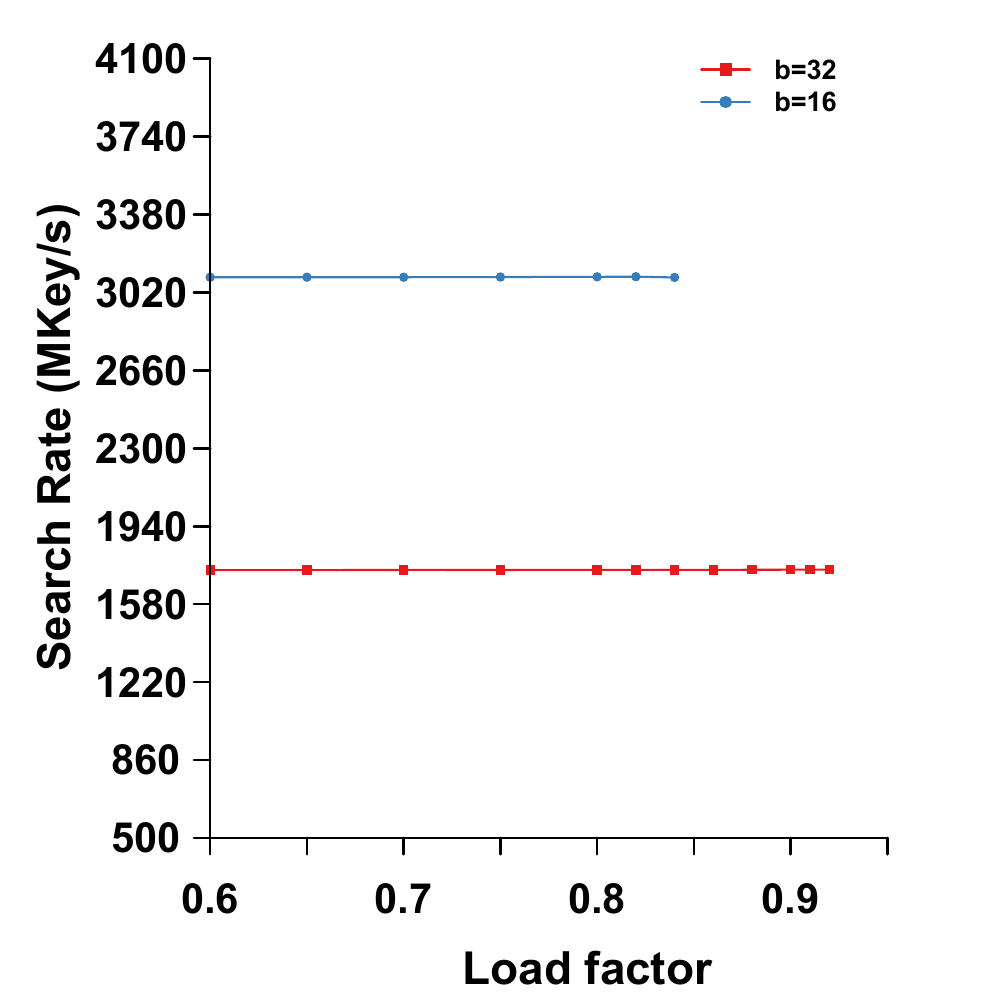}
    \caption{Find (100\% positive queries)}
  \end{subfigure}
  \begin{subfigure}{0.24\textwidth}
    \includegraphics[width=\textwidth]{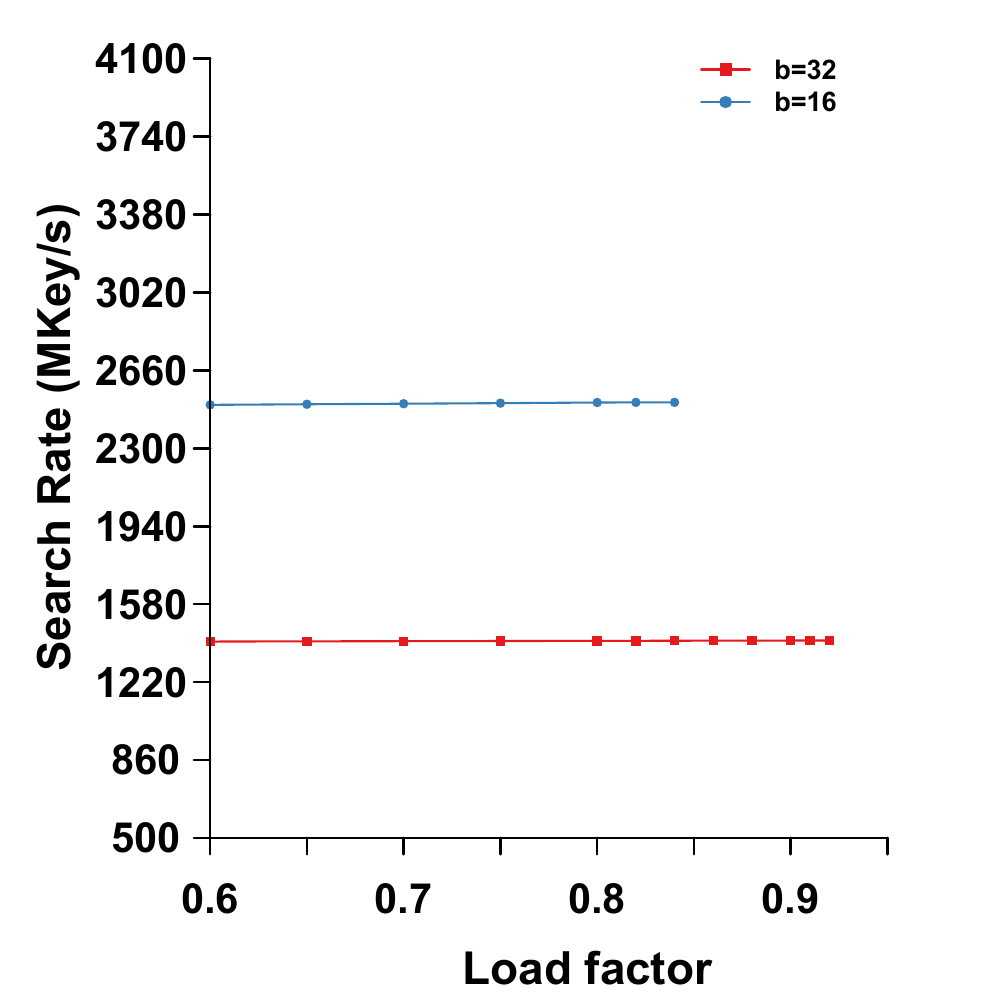}
    \caption{Find (50\% positive queries)}
  \end{subfigure}
  \begin{subfigure}{0.24\textwidth}
    \includegraphics[width=\textwidth]{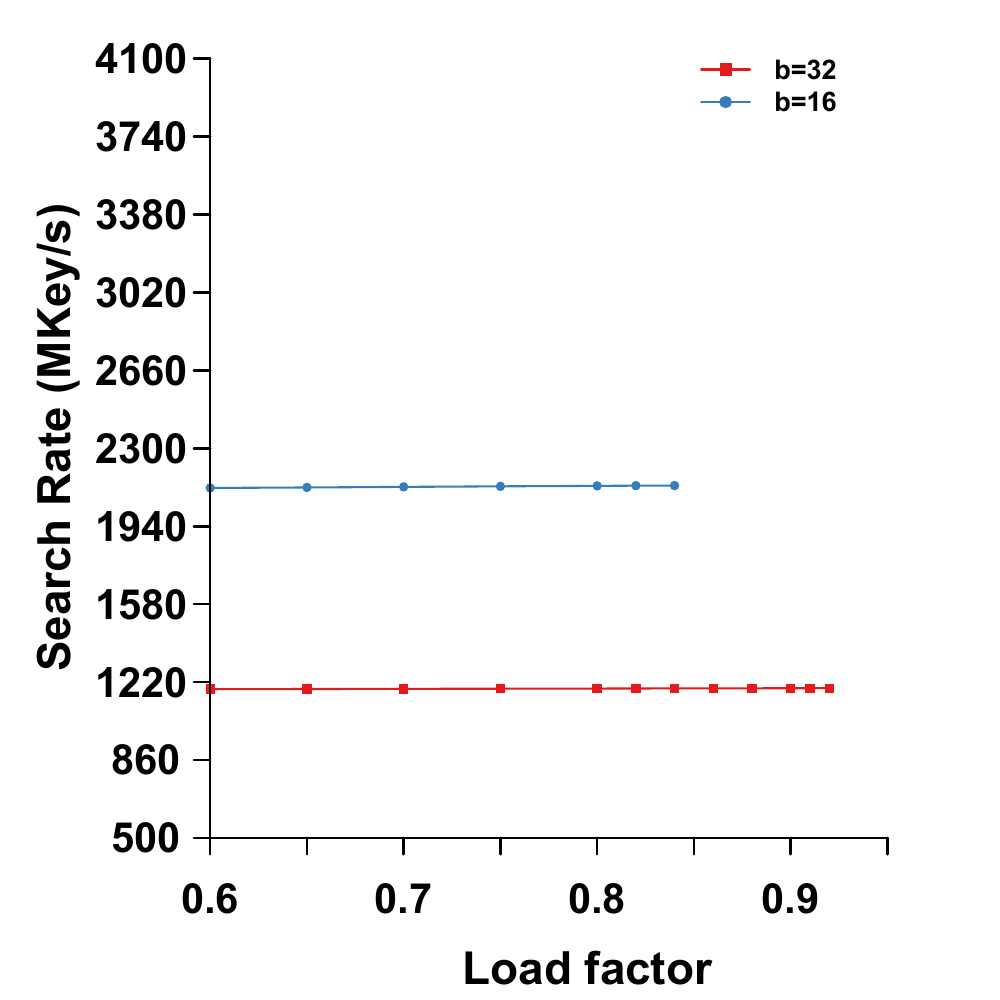}
    \caption{Find (0\% positive queries)}
  \end{subfigure}
  \caption{BP2HT insertion and query rates for different positive query ratios and 50M keys.}
  \label{fig:config_p2cht_fixed_num_keys}
\end{figure*}

\begin{figure*}
  \setlength\tabcolsep{1pt}
  \settowidth\rotheadsize{Radcliffe Cam}
  \setkeys{Gin}{width=\hsize}
  \begin{tabularx}{\textwidth}{l XXXX }
    \rothead{\centering load factor = 0.8}
     & \includegraphics[valign=m]{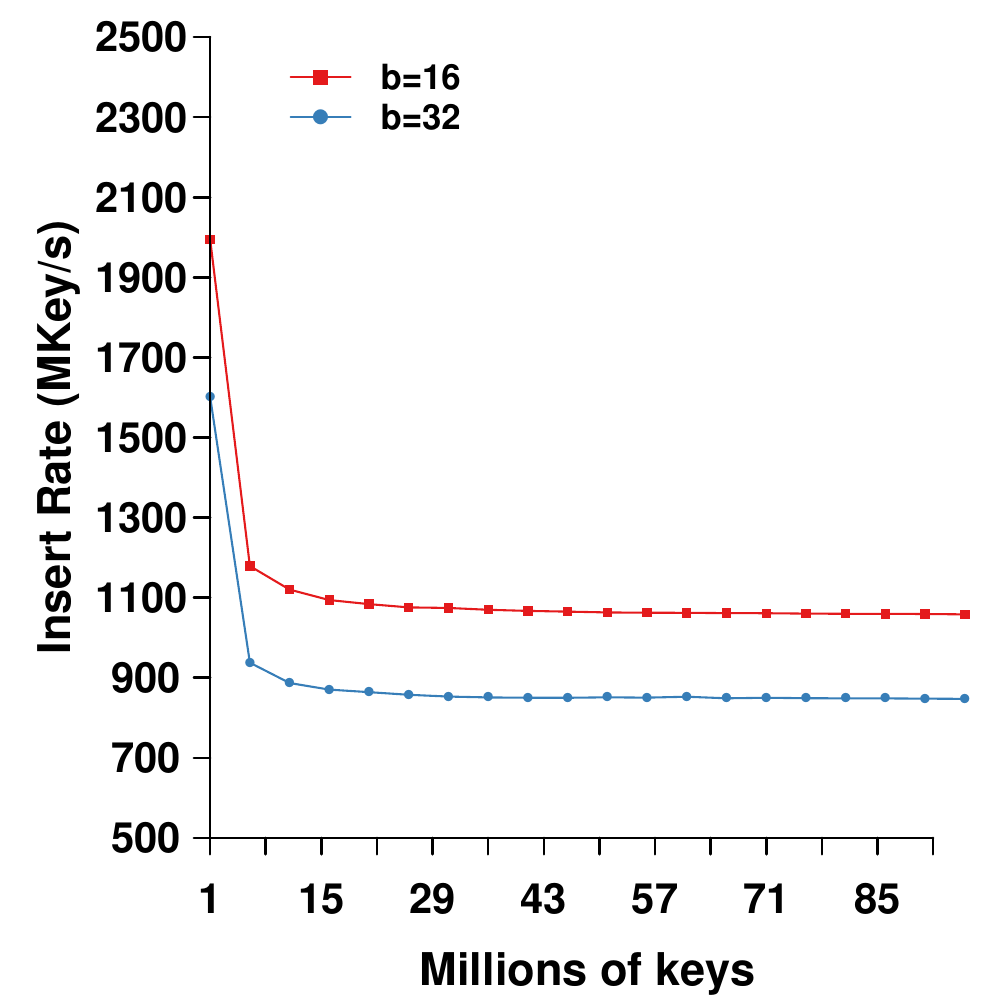}
     & \includegraphics[valign=m]{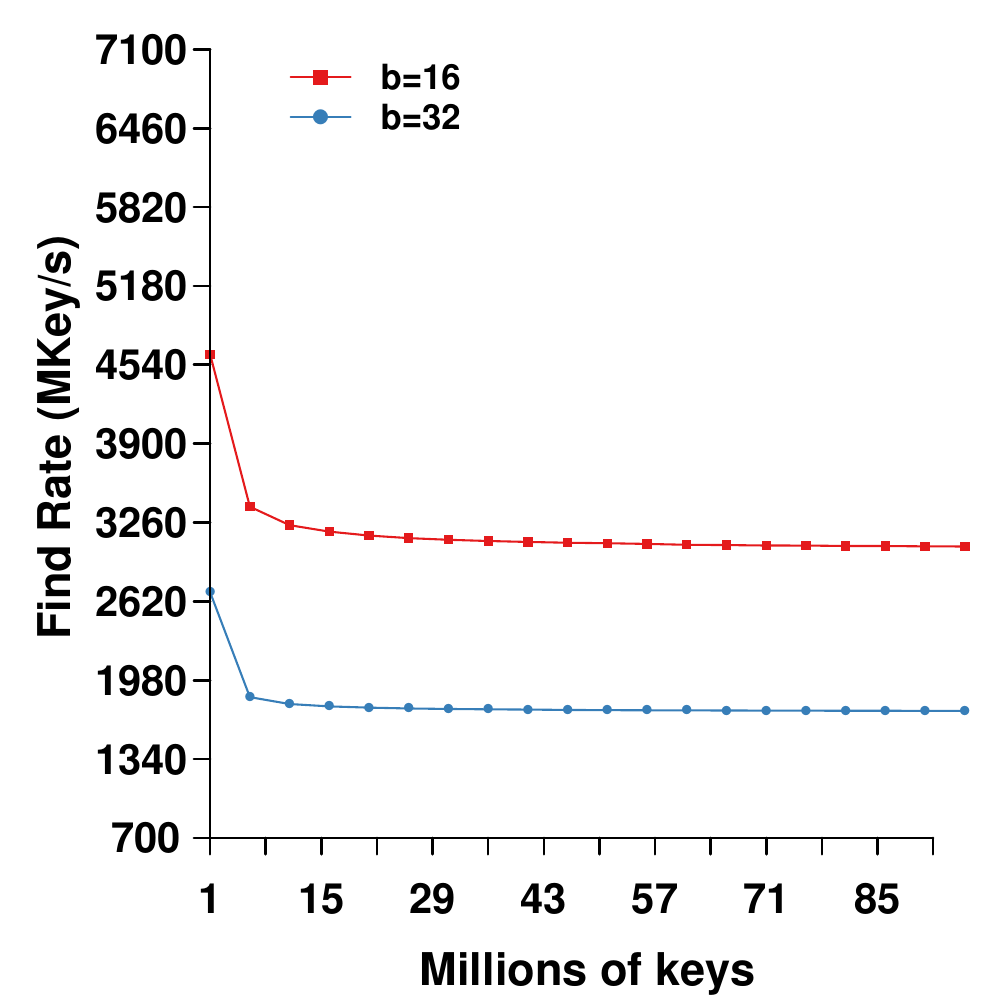}
     & \includegraphics[valign=m]{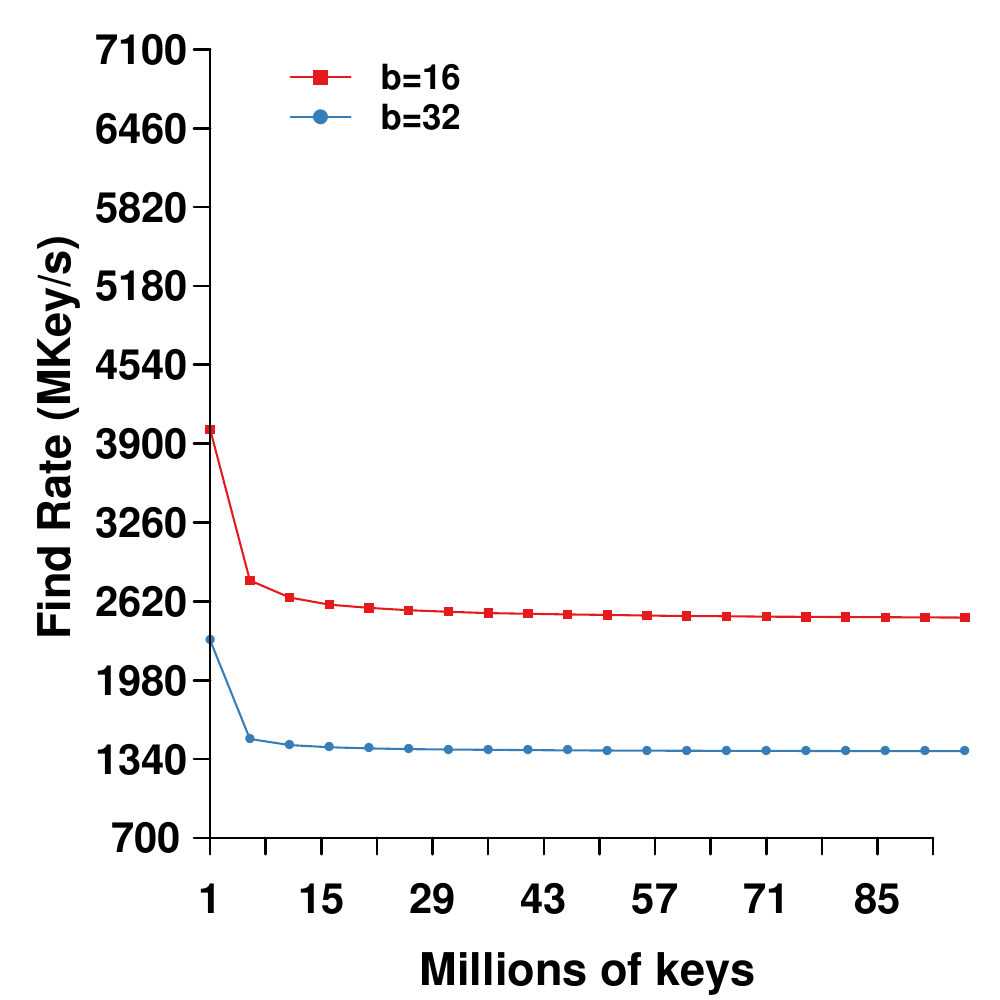}
     & \includegraphics[valign=m]{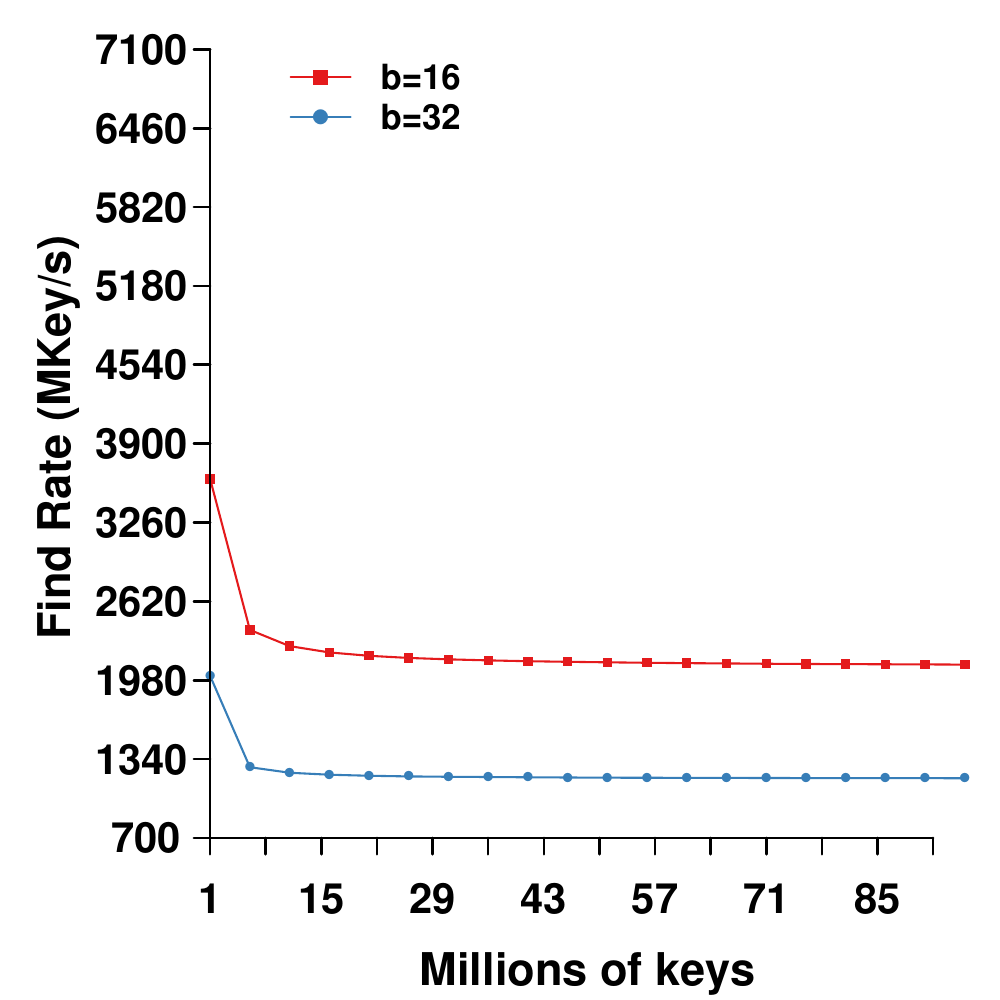}   \\ \addlinespace[2pt]
    \rothead{\centering load factor = 0.9}
     & \includegraphics[valign=m]{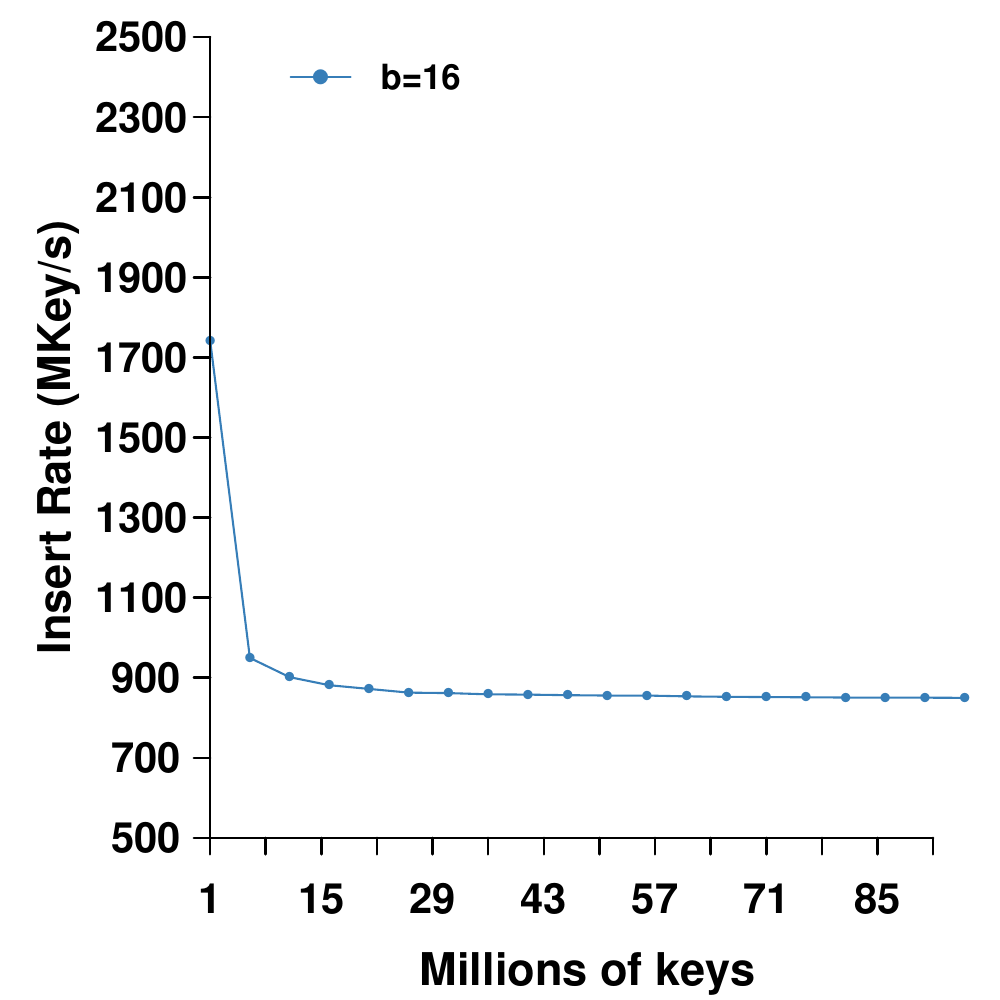}
     & \includegraphics[valign=m]{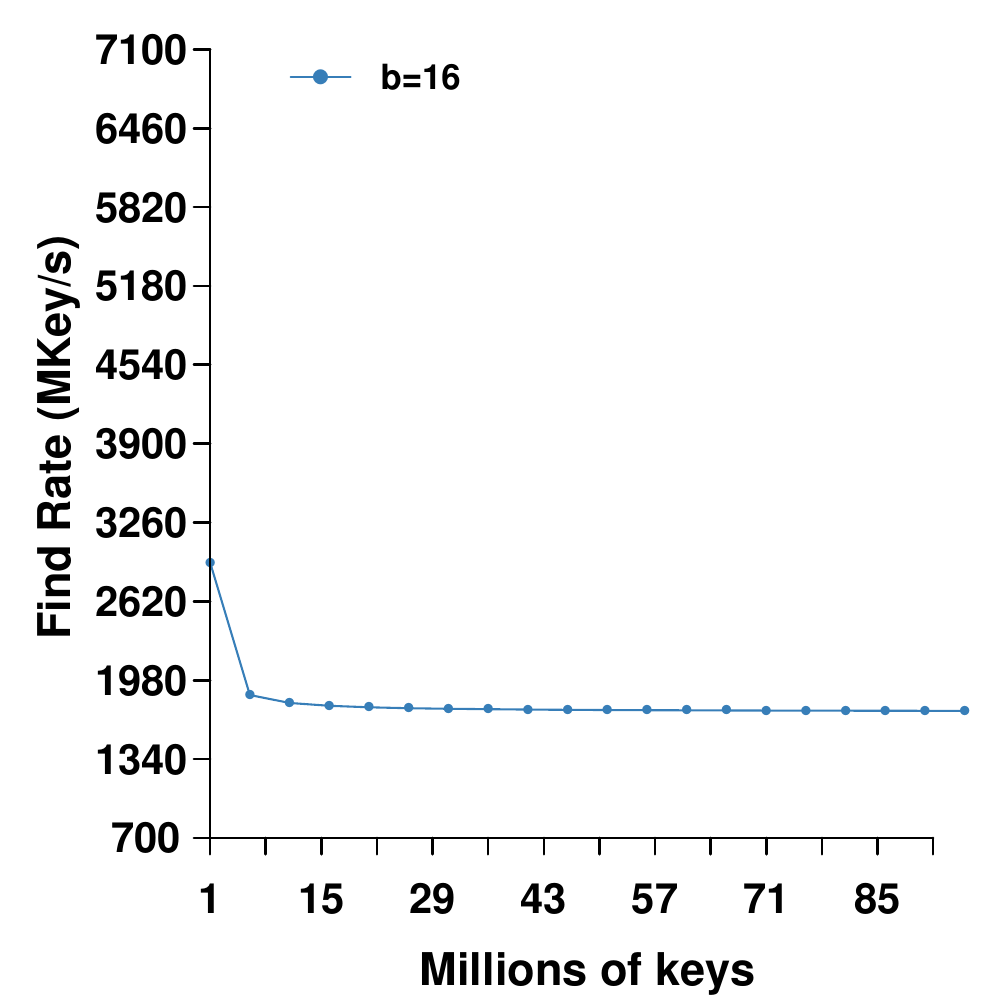}
     & \includegraphics[valign=m]{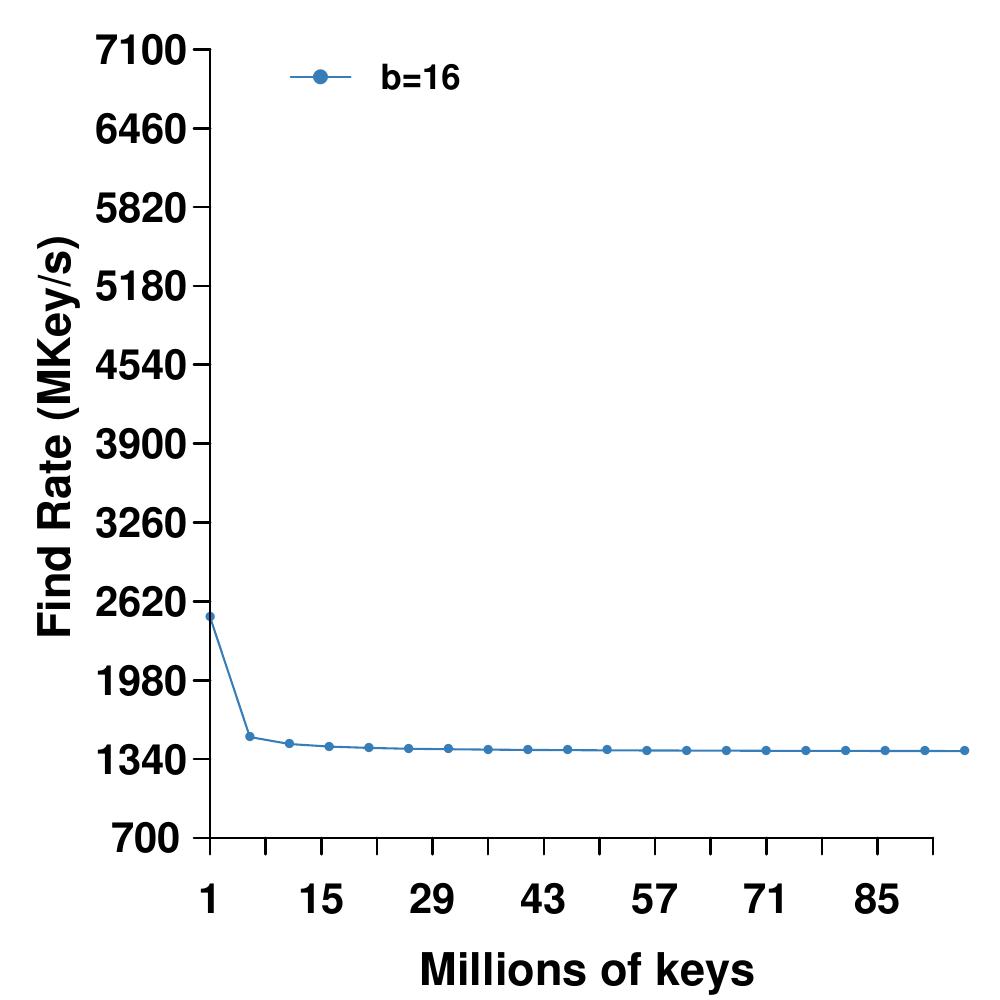}
     & \includegraphics[valign=m]{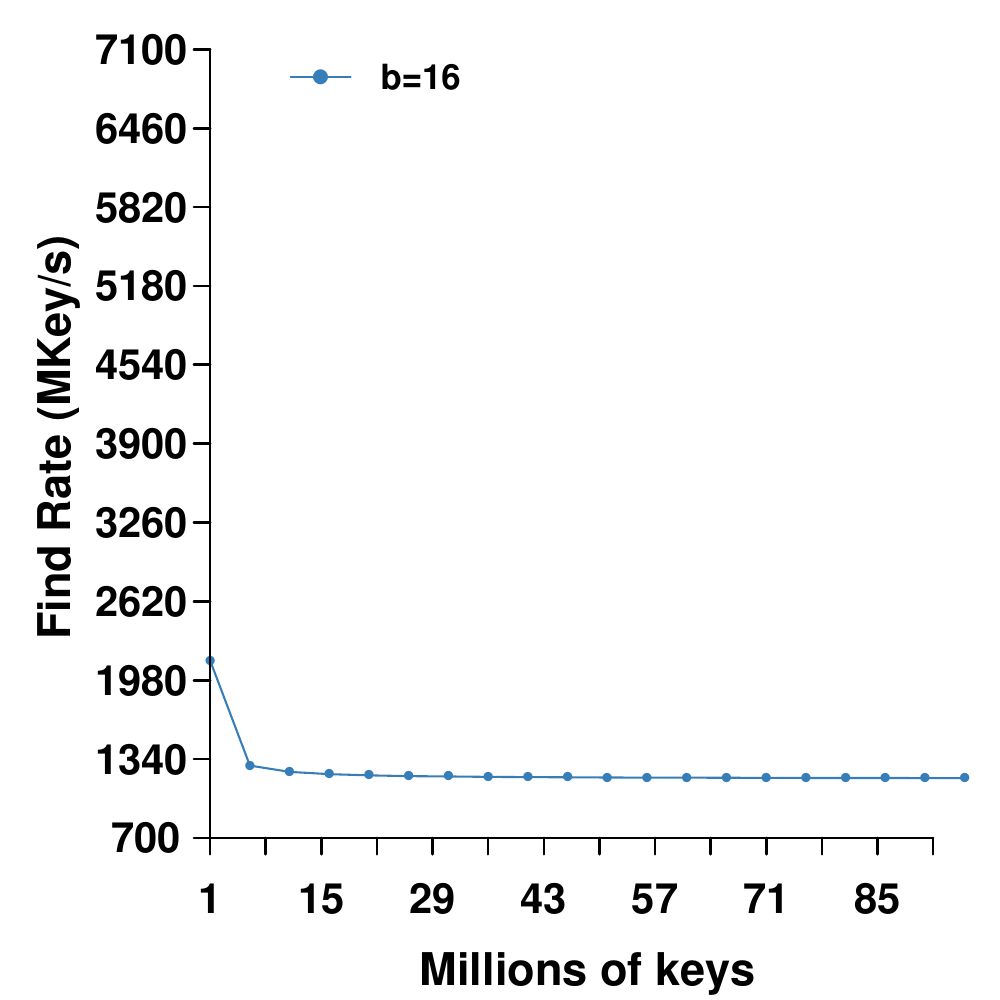}
  \end{tabularx}
  \caption{BP2HT throughput for insertion, 100\%, 50\%, and 0\% positive queries (from left to right).}
  \label{fig:config_p2cht_fixed_load_factor}
\end{figure*}

We experimented with different bucket sizes for BP2HT\@. The achieved load factors for bucket sizes of \{8, 16, 32\} are \{0.65, 0.84, 0.92\}. We pick the two bucket sizes that achieve high load factors and use them throughout our following experiments. Figure~\ref{fig:config_p2cht_fixed_num_keys} shows the performance of BP2HT when the number of keys is constant; clearly $b=16$ shows better performance on all tests than $b=32$, but cannot reach the high peak load factor of $b=32$.  Since insertion in BP2HT requires loading only two buckets then performing an atomic compare-and-swap operation in the least loaded bucket, the performance is constant for any load factor. Similarly, when the load factor is constant, the throughput is constant for all large input sizes. Figure~\ref{fig:config_p2cht_fixed_load_factor} shows the results for two load factors (0.8, and 0.9). Again, $b=16$ has better performance than $b=32$ but cannot reach a load factor of 0.9. We summarize the results of the BP2HT results in Table~\ref{tab:p2cht_results_summary}.

\begin{figure*}
  \centering
  \begin{subfigure}{0.24\textwidth}
    \includegraphics[width=\textwidth]{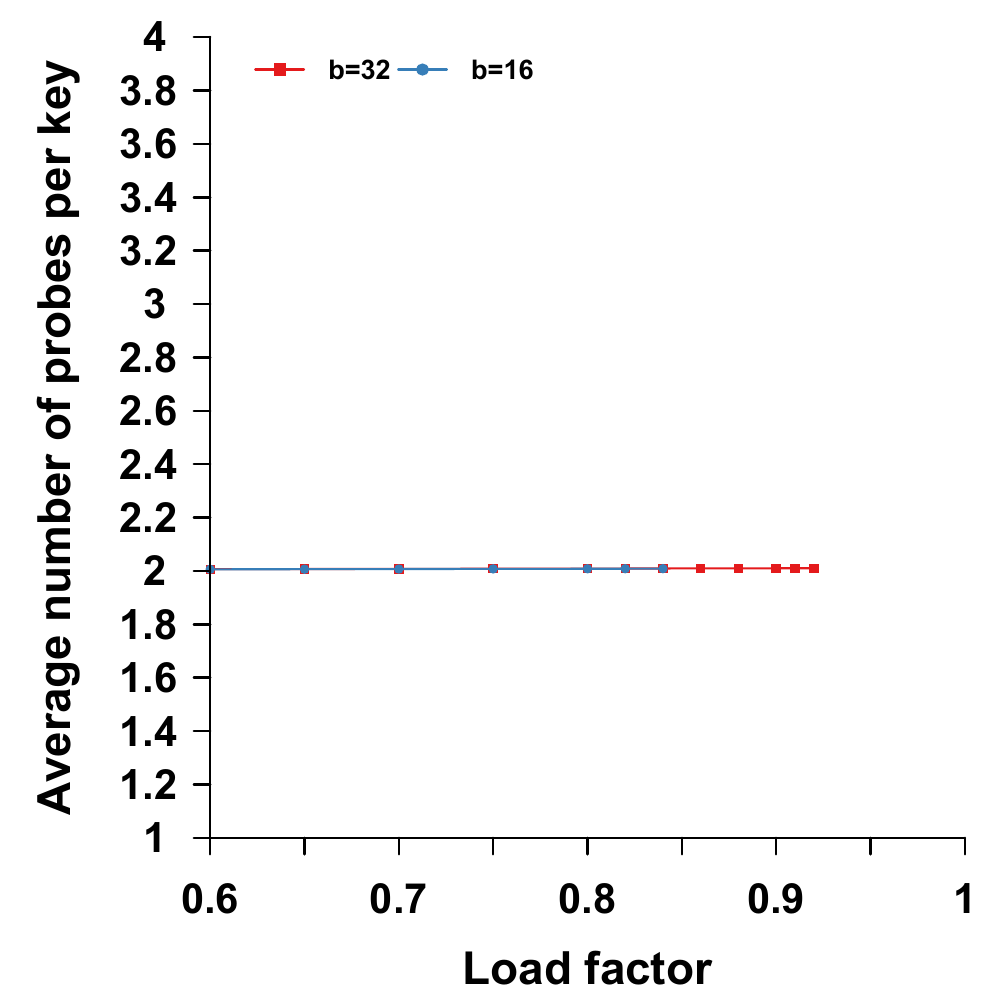}
    \caption{Insertion}
  \end{subfigure}
  \begin{subfigure}{0.24\textwidth}
    \includegraphics[width=\textwidth]{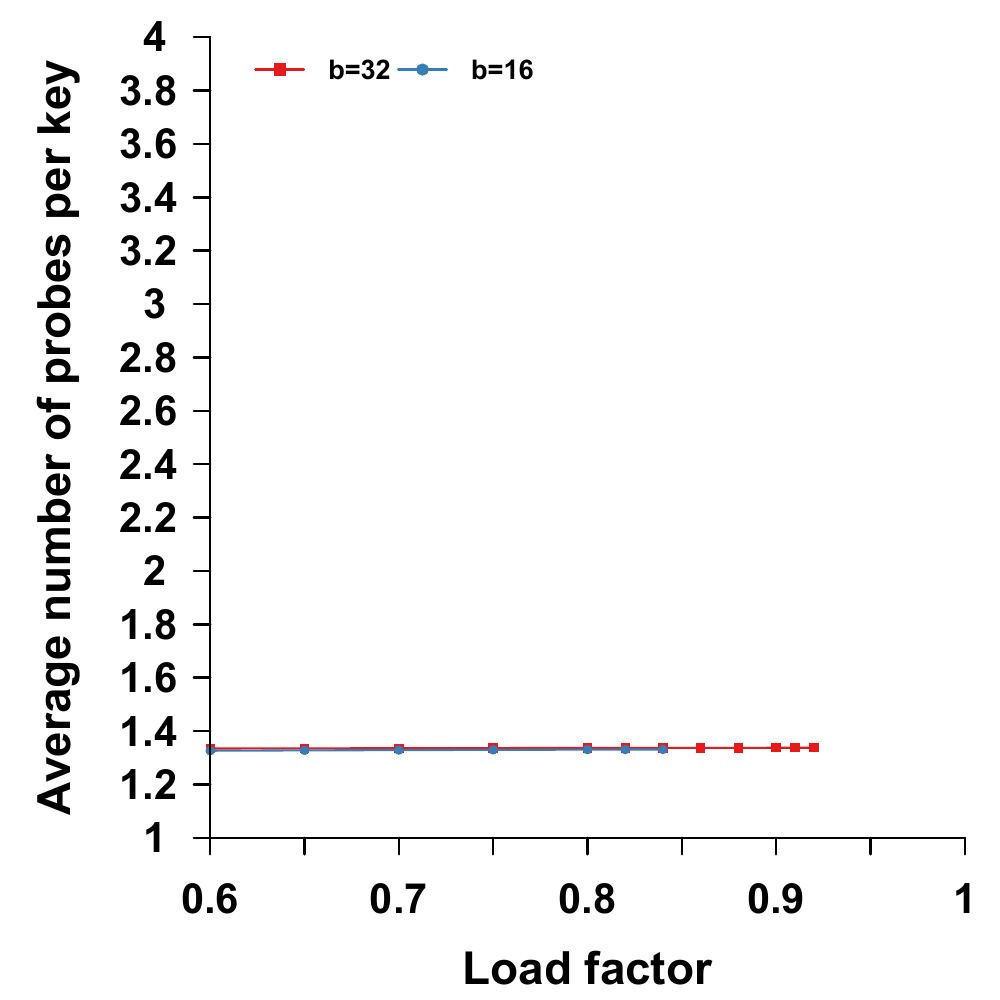}
    \caption{Find (100\% positive queries)}
  \end{subfigure}
  \begin{subfigure}{0.24\textwidth}
    \includegraphics[width=\textwidth]{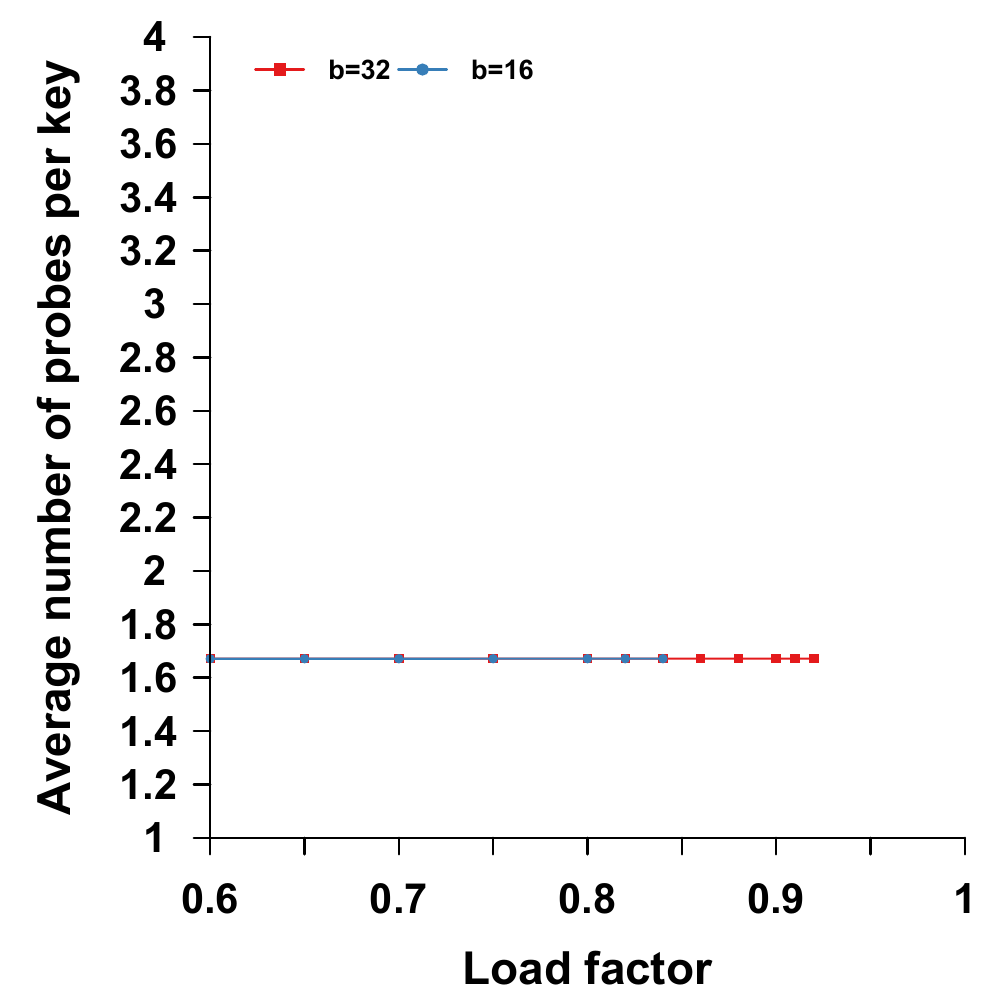}
    \caption{Find (50\% positive queries)}
  \end{subfigure}
  \begin{subfigure}{0.24\textwidth}
    \includegraphics[width=\textwidth]{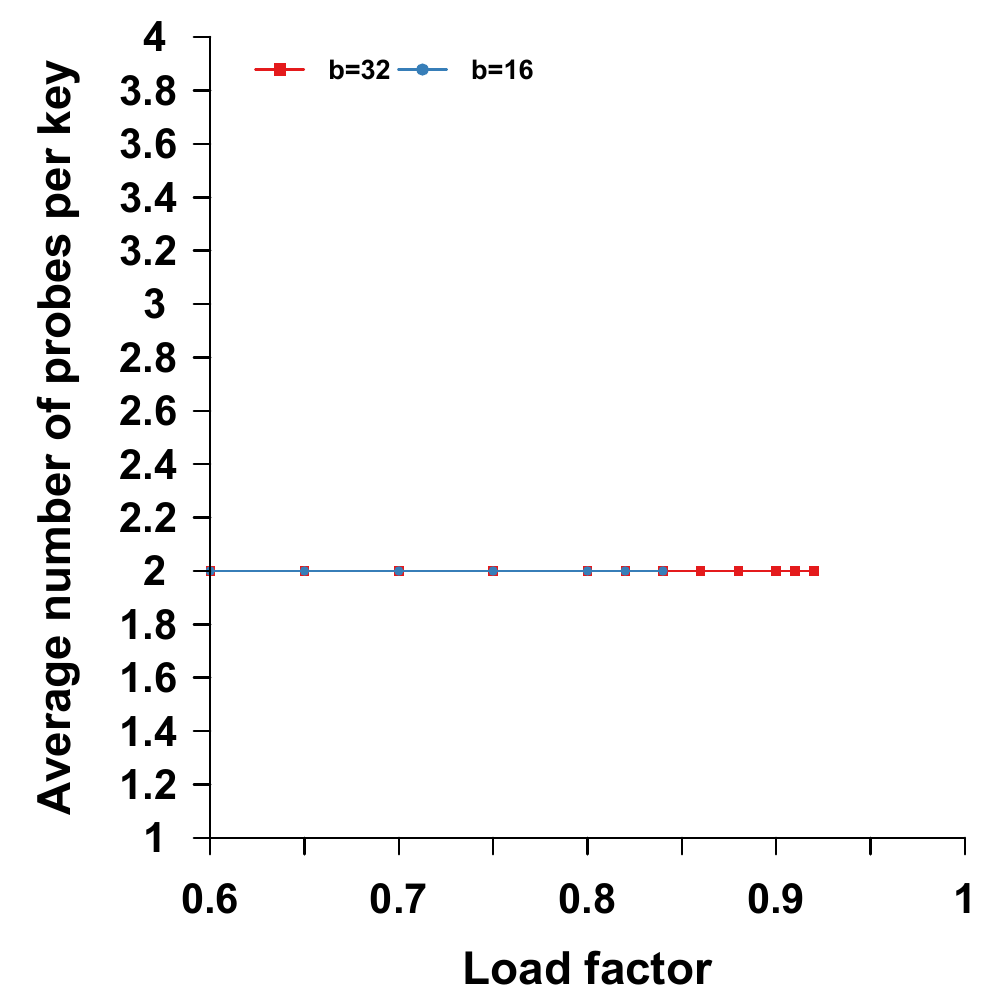}
    \caption{Find (0\% positive queries)}
  \end{subfigure}
  \caption{BP2HT insertion and query average probes per key for 50M keys.}
  \label{fig:config_p2cht_probes_per_keys}
\end{figure*}

Inserting in a BP2HT requires loading two buckets; thus, the average number of probes is two regardless of the bucket size or the load factor. Queries, on the other hand, only require 1.33 probes per key when all queries are positive. In other words, only 33\% of the time, the query reads the second bucket. When 50\% of the queries are positive, the number of probes per key increases to 1.67. In the worst-case scenario (i.e., all queries are negative), BP2HT loads both buckets before evaluating the query. Figure~\ref{fig:config_p2cht_probes_per_keys} shows the average probe count for insertion and query operations.

\begin{table}
  \centering
  \begin{tabular}{cccccc}
    \toprule
    \multirow{2}{*}{Fixed parameter} & \multirow{2}{*}{$b$} & \multirow{2}{*}{Insert} & \multicolumn{3}{c}{\% of queries present in hash table}                                       \\[0.2em]\cmidrule(l){4-6}
                                     &                      &                         & 100\%                                                   & 50\%             & 0\%              \\ \midrule
    \multirow{2}{*}{50M keys}        & 16                   & \textbf{1297.18}        & \textbf{3392.59}                                        & \textbf{1922.64} & \textbf{1436.26} \\
                                     & 32                   & 1087.01                 & 1946.14                                                 & 1088.09          & 803.85           \\\midrule
    \multirow{2}{*}{0.8 load factor} & 16                   & \textbf{1319.87}        & \textbf{3350.79}                                        & \textbf{1941.20} & \textbf{1485.75} \\
                                     & 32                   & 1145.62                 & 2042.45                                                 & 1131.60          & 825.79           \\\midrule
    \multirow{1}{*}{0.9 load factor} & 32                   & \textbf{1073.92}        & \textbf{1860.21}                                        & \textbf{1075.96} & \textbf{829.85}  \\
    \bottomrule
  \end{tabular}
  \caption{Average insertion and query throughput (different positive query percentages) for IHT when: 1) the number of keys is fixed at 50M keys and 2) the load factor is fixed at 0.8 and 0.9.}
  \label{tab:iht_results_summary}
\end{table}

\subsubsection{IHT}
\begin{figure*}
    \centering
    \begin{subfigure}{0.24\textwidth}
      \includegraphics[width=\textwidth]{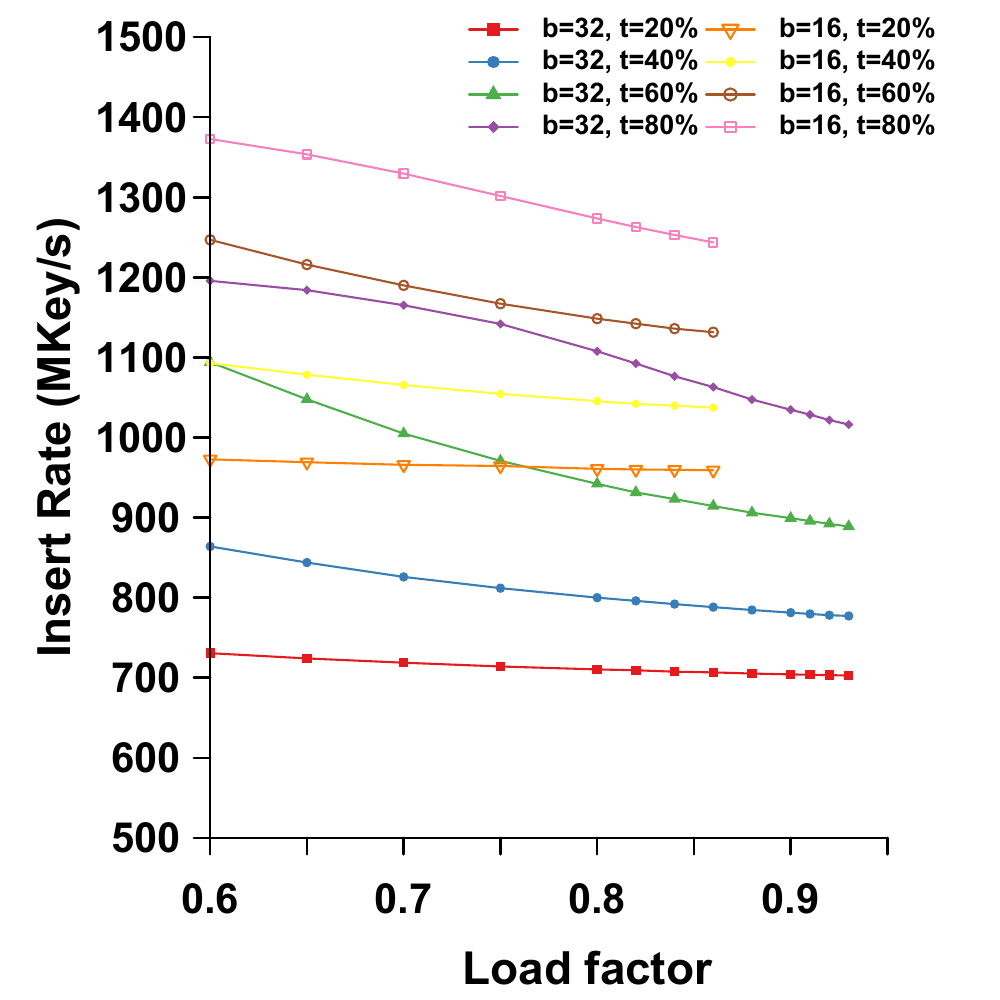}
      \caption{Insertion}
    \end{subfigure}
    \begin{subfigure}{0.24\textwidth}
      \includegraphics[width=\textwidth]{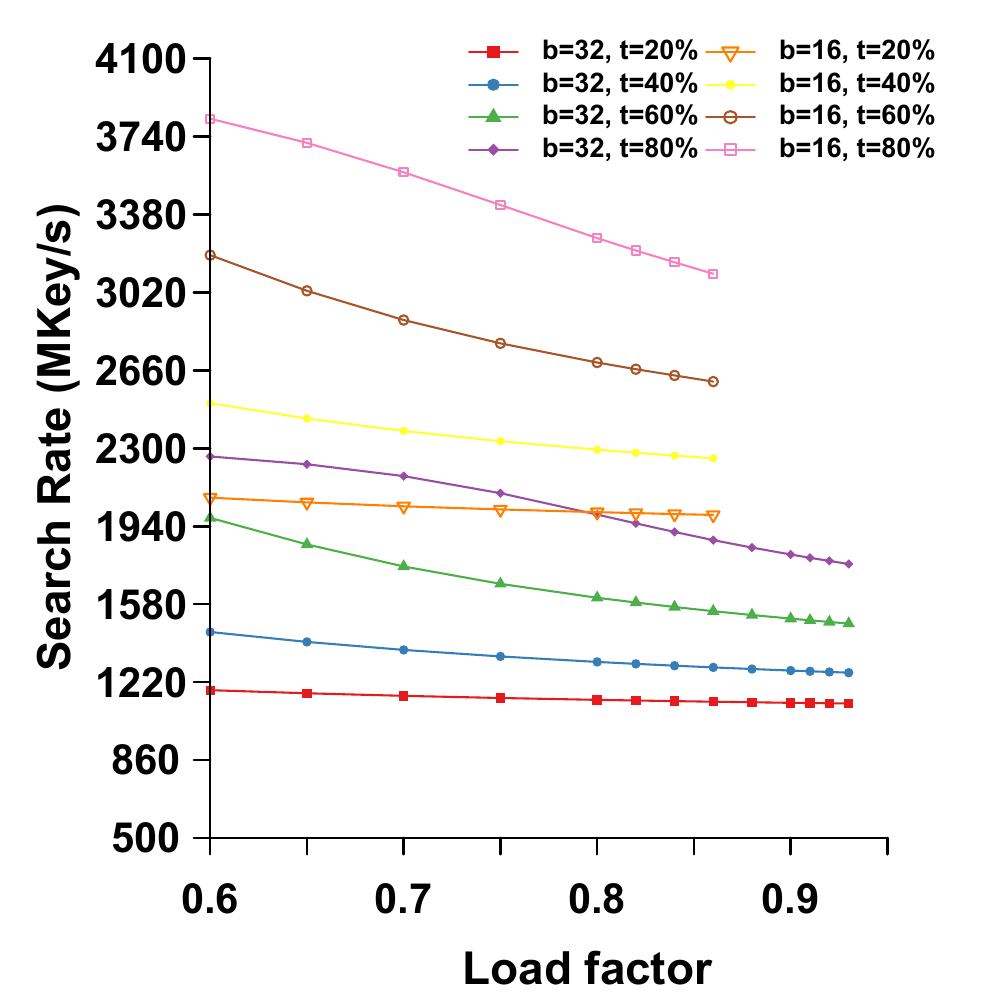}
      \caption{Find (100\% positive queries)}
    \end{subfigure}
    \begin{subfigure}{0.24\textwidth}
      \includegraphics[width=\textwidth]{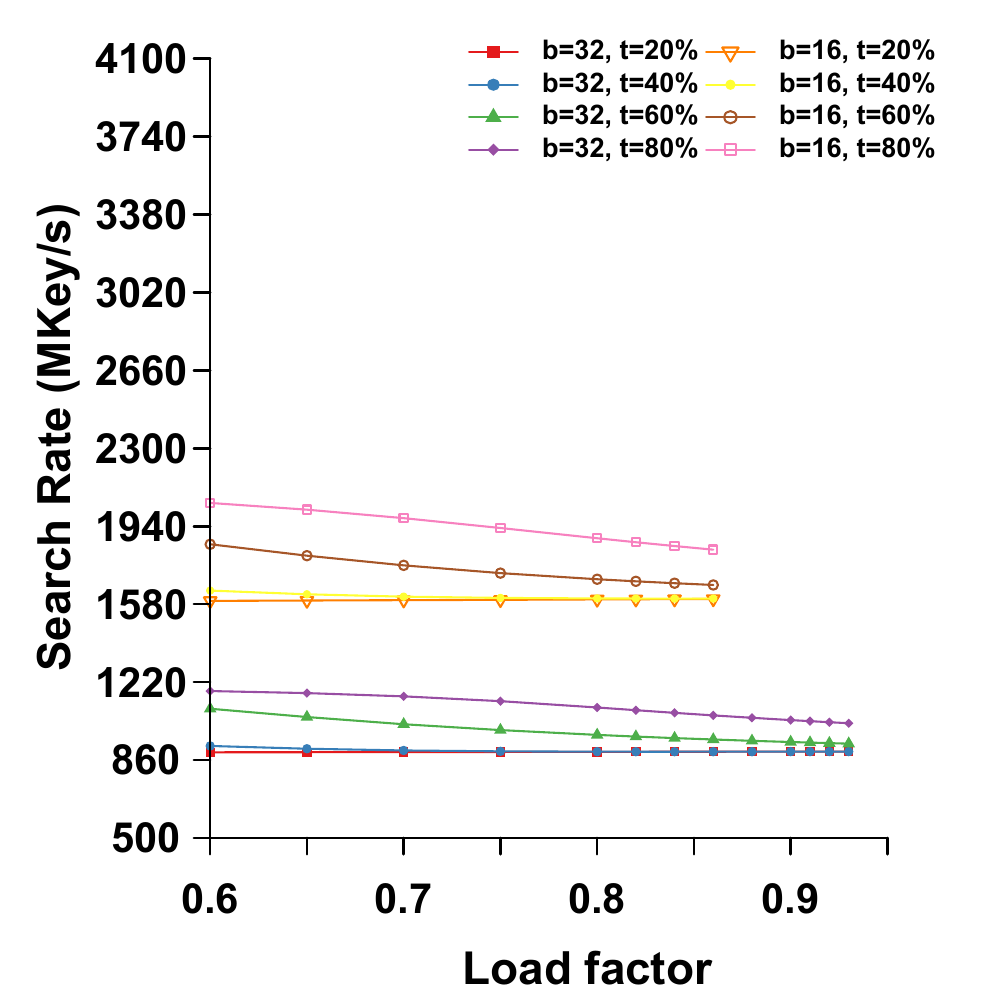}
      \caption{Find (50\% positive queries)}
    \end{subfigure}
    \begin{subfigure}{0.24\textwidth}
      \includegraphics[width=\textwidth]{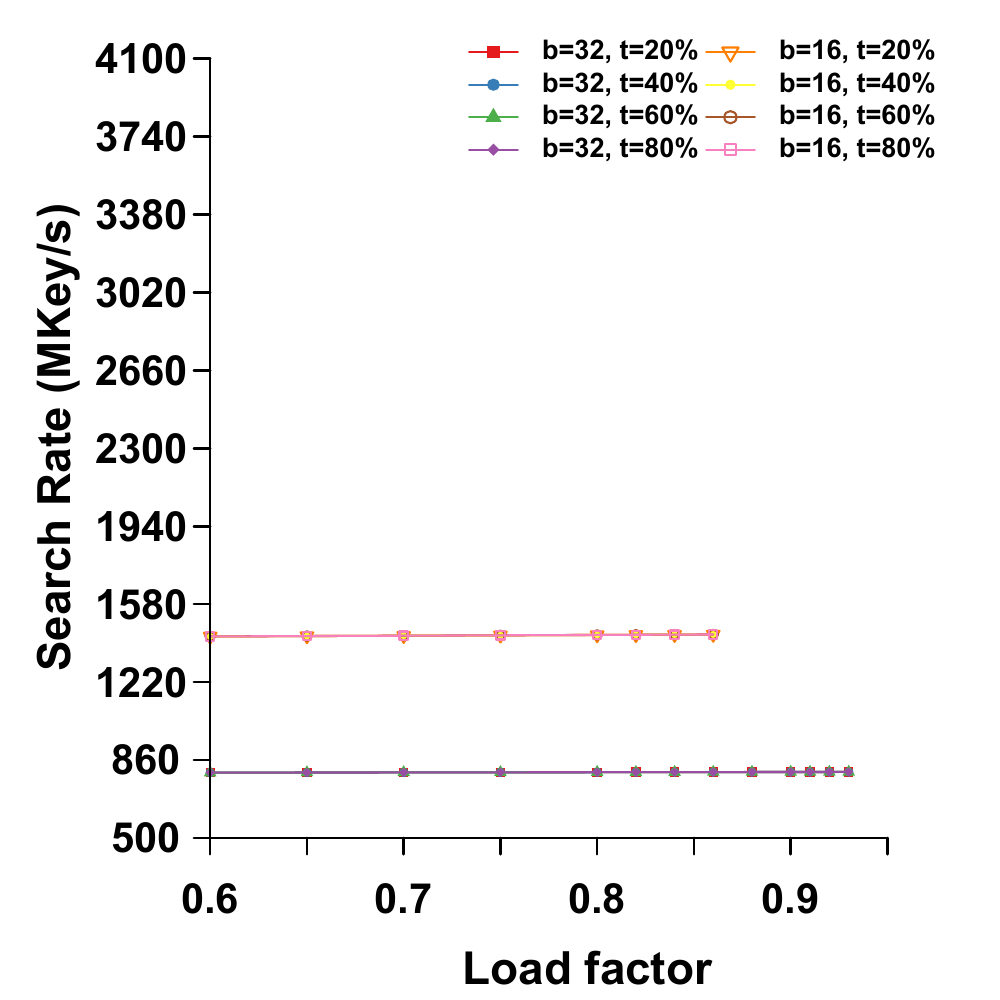}
      \caption{Find (0\% positive queries)}
    \end{subfigure}
    \caption{IHT insertion and query rates for different positive query ratios and 50M keys.}
    \label{fig:config_iht_fixed_num_keys}
  \end{figure*}

We experiment with the parameter space of IHT by varying both the threshold $t$ and block size $b$. We perform the sweep of parameters for the first experiment where the number of keys is fixed in Figure~\ref{fig:config_iht_fixed_num_keys}. We find that IHT requires larger bucket sizes to achieve high load factors; specifically, bucket sizes of \{32, 16, 8\} achieve load factors up to \{0.93, 0.86, 0.7\}, respectively. In general, as the load factor increases, more insertions (or queries) require accessing the two secondary buckets, which leads to a decrease in performance. During queries, IHT requires inspecting all three possible buckets unless the lookup operation finds the key in the first or second bucket (note that when all the queries are negative, the query needs to inspect all three buckets).

We use bucket sizes of 16 and 32 throughout our experiments as they achieve comparable load factors to the alternative hash tables. For both bucket sizes we find that as the threshold increases, the achieved throughput, for either insertion or queries, also increases. As the threshold increases, more insertions occur into the primary bucket, thus avoiding loading the secondary buckets and saving bandwidth. The best performing threshold for both bucket sizes is around 80\% of the bucket size. Table~\ref{tab:iht_results_summary} provides a summary of our results for IHT\@.

\begin{figure*}
  \centering
  \begin{subfigure}{0.24\textwidth}
    \includegraphics[width=\textwidth]{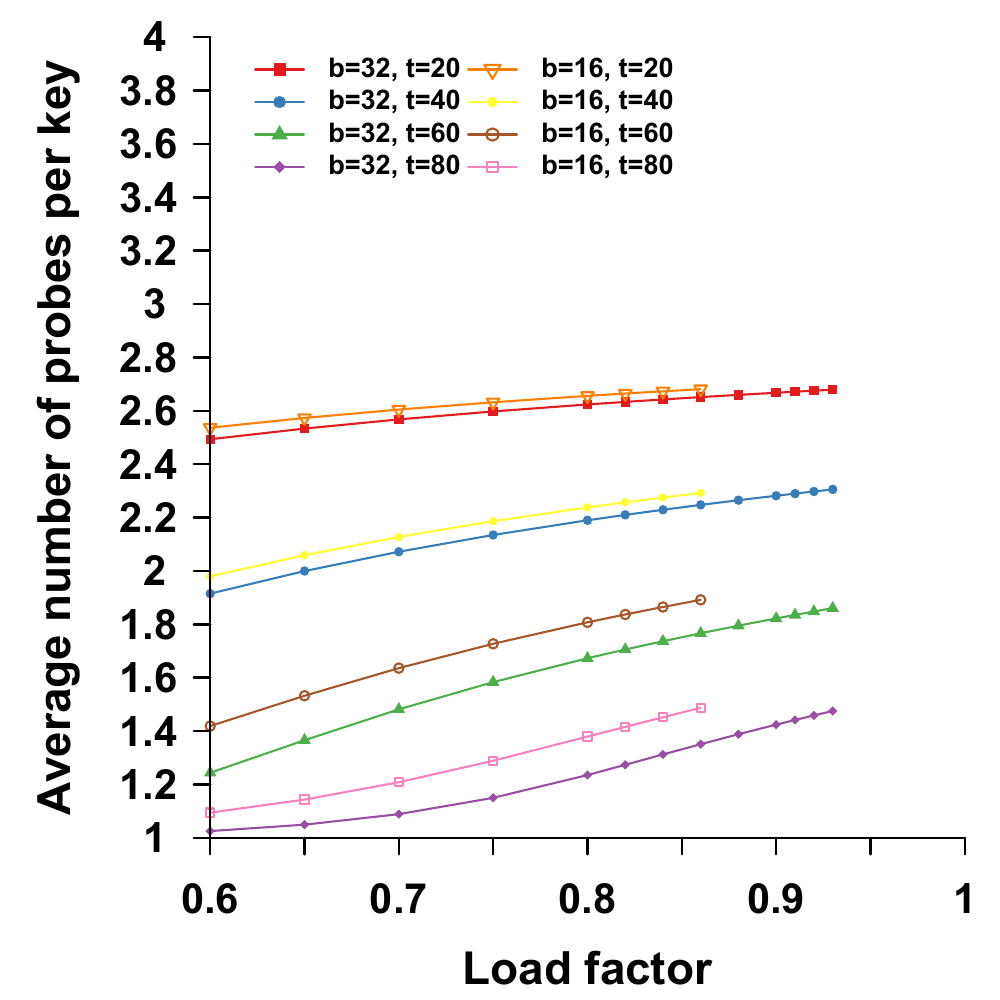}
    \caption{Insertion}
  \end{subfigure}
  \begin{subfigure}{0.24\textwidth}
    \includegraphics[width=\textwidth]{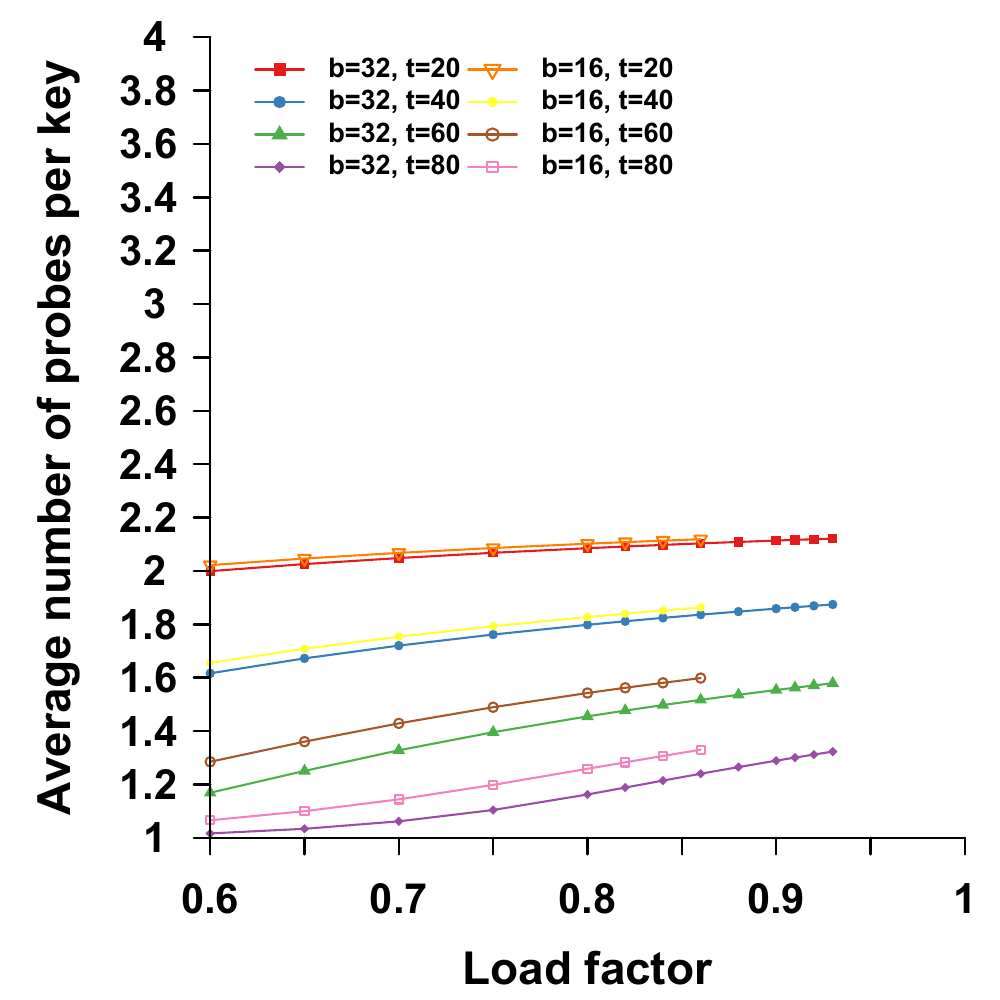}
    \caption{Find (100\% positive queries)}
  \end{subfigure}
  \begin{subfigure}{0.24\textwidth}
    \includegraphics[width=\textwidth]{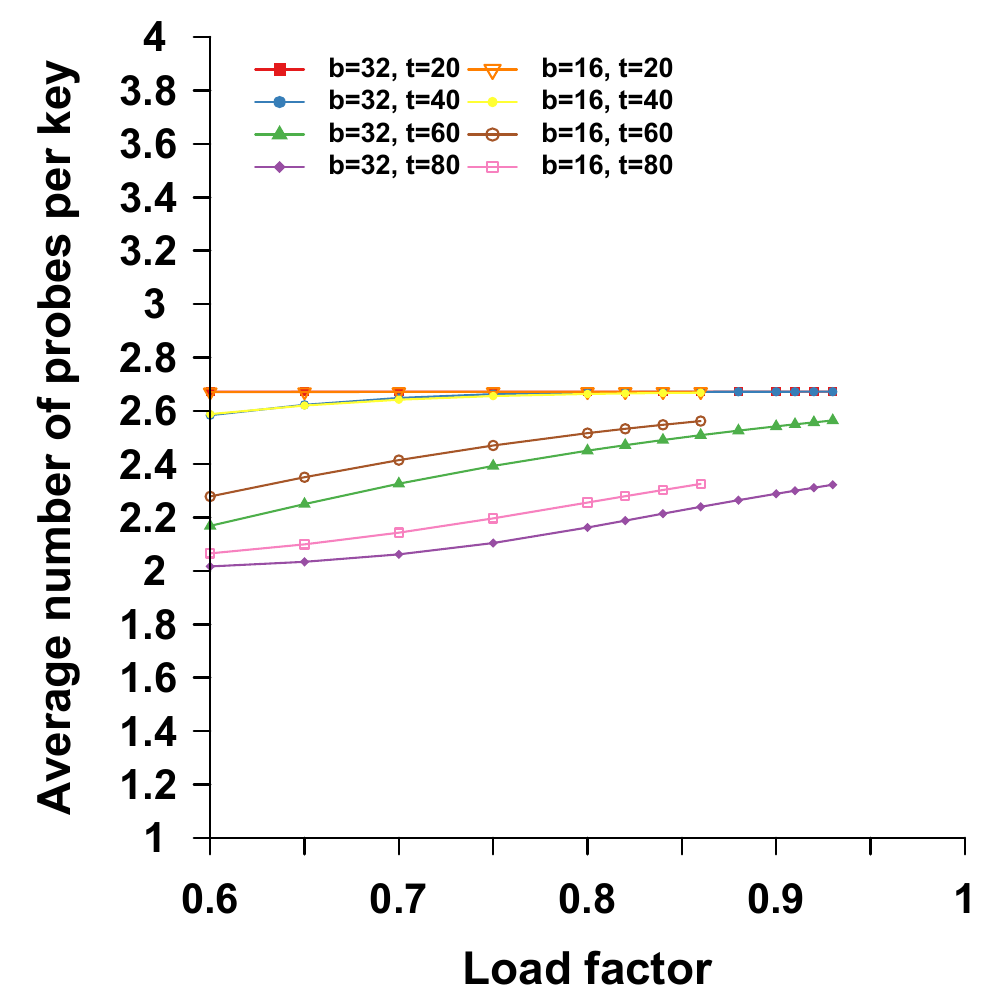}
    \caption{Find (50\% positive queries)}
  \end{subfigure}
  \begin{subfigure}{0.24\textwidth}
    \includegraphics[width=\textwidth]{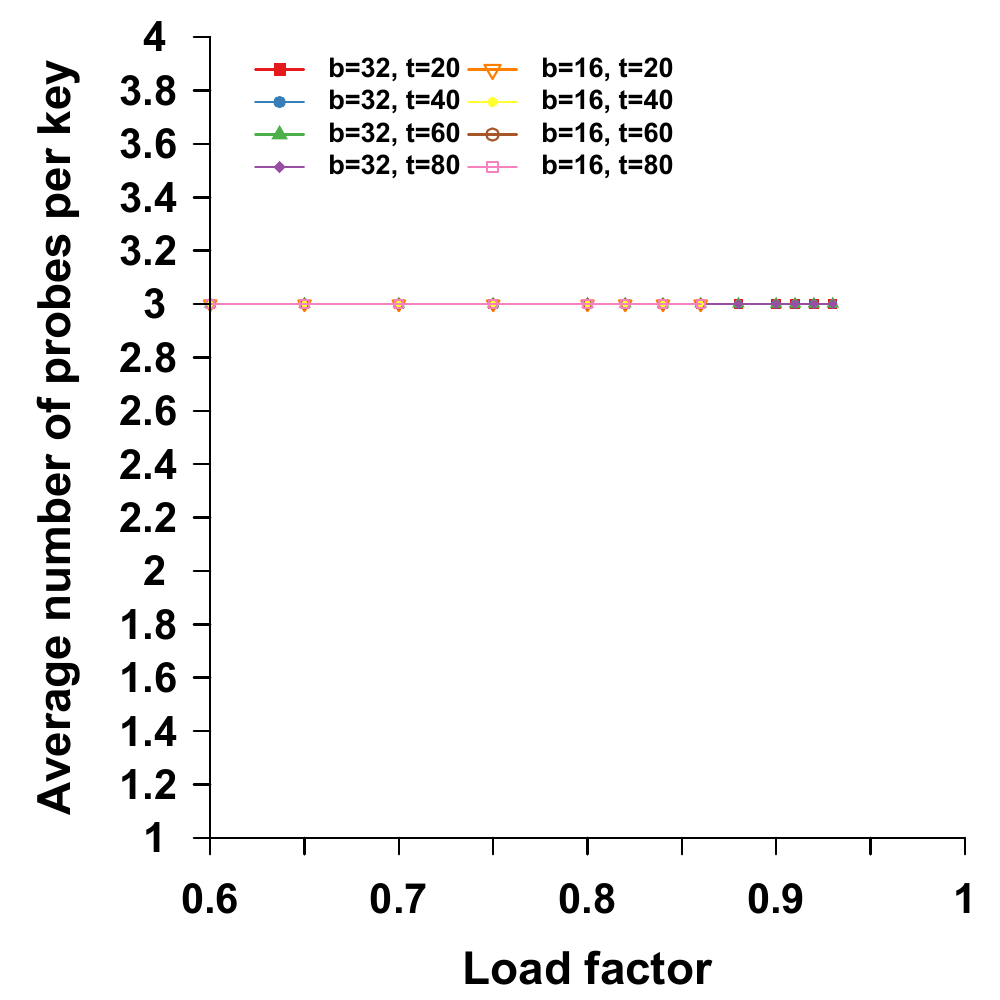}
    \caption{Find (0\% positive queries)}
  \end{subfigure}
  \caption{IHT insertion and query average probes per key for 50M keys.}
  \label{fig:config_iht_probes_per_keys}
\end{figure*}

Figure~\ref{fig:config_iht_probes_per_keys} shows the average number of probes for the different IHT bucket sizes and thresholds. We find that as the threshold increases, the number of probes drops significantly. For example, when inserting into an IHT at a load factor of 0.86 and a bucket size of 16, the average probe counts are \{2.68, 2.29, 1.89, 1.49\} at thresholds \{20\%, 40\%, 60\%, 80\%\}. Increasing the threshold means performing more insertions in the primary bucket without reading the two secondary buckets, which explains the significant drop in the probes count. Similar to our earlier observation in BCHT average probes count, increasing the bucket size from 16 to 32 yields minimal improvement to the probe count. When all queries are positive, and at a load factor of 0.86, IHT with a bucket size of 16 and threshold of \{20\%, 40\%, 60\%, 80\%\} performs \{2.11, 	1.86, 1.6, 1.33\} probes per key. As the percentage of positive queries drops, queries inspect more secondary buckets, inspecting all three possible buckets for the different load factors and bucket sizes.

\subsection{Comparison across implementations}

Given the previous results, we choose the following four configurations of hash tables for comparisons: 1) baseline 1CHT, 2) BCHT with bucket size of 16, 3) BP2HT with bucket size of 32, 4) IHT with bucket size of 32 and threshold 80\%.

\subsubsection{Varying the load factor when the number of keys is constant}

Figure~\ref{fig:recommended_vs_different_load_factors} shows a comparison between the recommended configurations. In it, we see that BCHT offers the best performance for both insertion and queries. BCHT has the advantage that it can use a bucket size of 16 while achieving high load factors (i.e., each bucket is a single cache line), while IHT and BP2HT use a bucket size of 32 to achieve comparable load factors (i.e., each bucket is two cache lines). On the other hand, 1CHT only uses 8 bytes of data from a 128B cache line.

With respect to insertion performance, the required memory traffic for IHT (2 or 6 cache lines in the best and worst scenarios, respectively) and BP2HT (4 cache lines) lower their performance compared to BCHT (1 to 3 cache lines). 1CHT, on the other hand, offers worse performance than both BCHT and IHT but better than BP2HT\@. Only when the load factor reaches 0.72 does BP2HT's performance exceed 1CHT\@. On average the insertion throughput of \{BCHT, IHT, BP2HT, 1CHT\} are \{1273.86, 1087.01, 849.54, 735.84\} Mkey/s.

For all-positive queries (and in the best case scenario), BCHT and 1CHT only require loading a single cache line, while IHT and BP2HT load two cache lines. This result in 1CHT throughput's exceeding IHT's and BP2HT's. BCHT's throughput continues to exceed all other alternatives.

As the ratio of positive queries goes down, each technique must check additional buckets. In the worst-case scenario, BCHT loads 3 cache lines; 1CHT loads 4 cache lines; BP2HT loads 4 cache lines; and IHT loads 6 cache lines. This result in the performance ranking of BCHT (highest), 1CHT, BP2HT, and IHT (lowest). In summary the average query throughput of \{BCHT, IHT, BP2HT, 1CHT\} are \{3642.61, 1946.14, 1738.48, 2235\}, \{2840.57, 1088.09, 1410.54, 1884.83\}, and \{2313.17, 803.85, 1190.29, 1496.21\} Mkey/s for all-positive, 50\% positive, and all-negative queries, respectively.

Note that at low load factors, the performance of IHT is similar to BCHT when the bucket size is the same. However, at high load factors (bucket sizes exceed the threshold), IHT uses the power of two strategy, which requires loading two additional buckets. In contrast, BCHT loads just enough buckets to perform the cuckooing.

\subsubsection{Varying the input size}

\begin{figure*}
    \centering
    \begin{subfigure}{\textwidth}
      \includegraphics[width=0.24\textwidth]{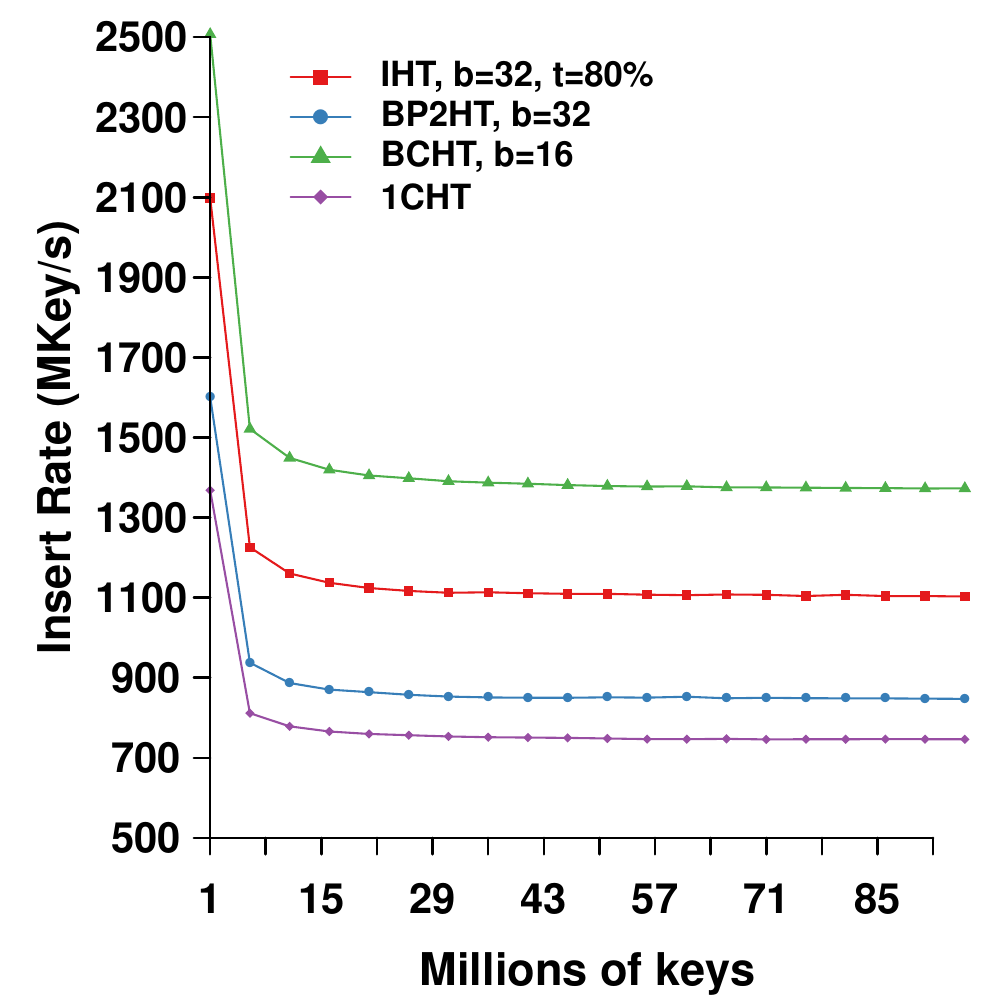}
      \includegraphics[width=0.24\textwidth]{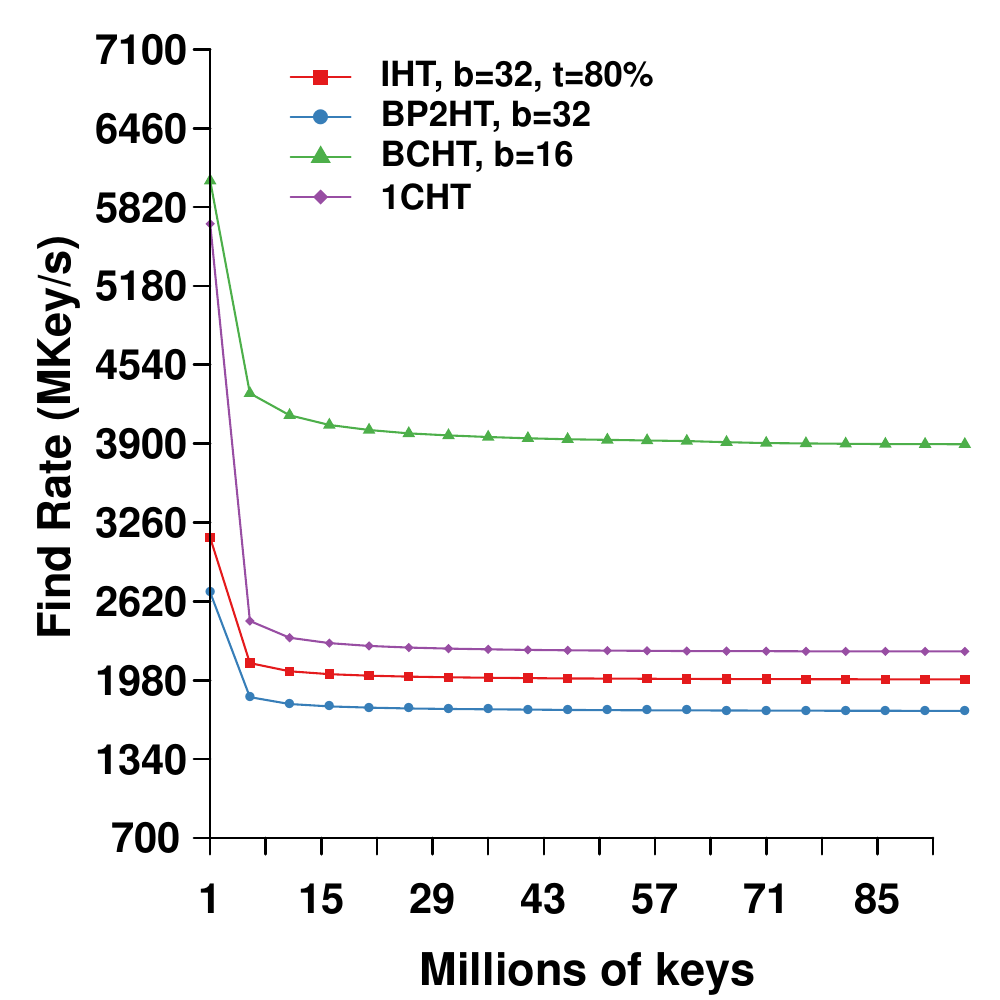}
      \includegraphics[width=0.24\textwidth]{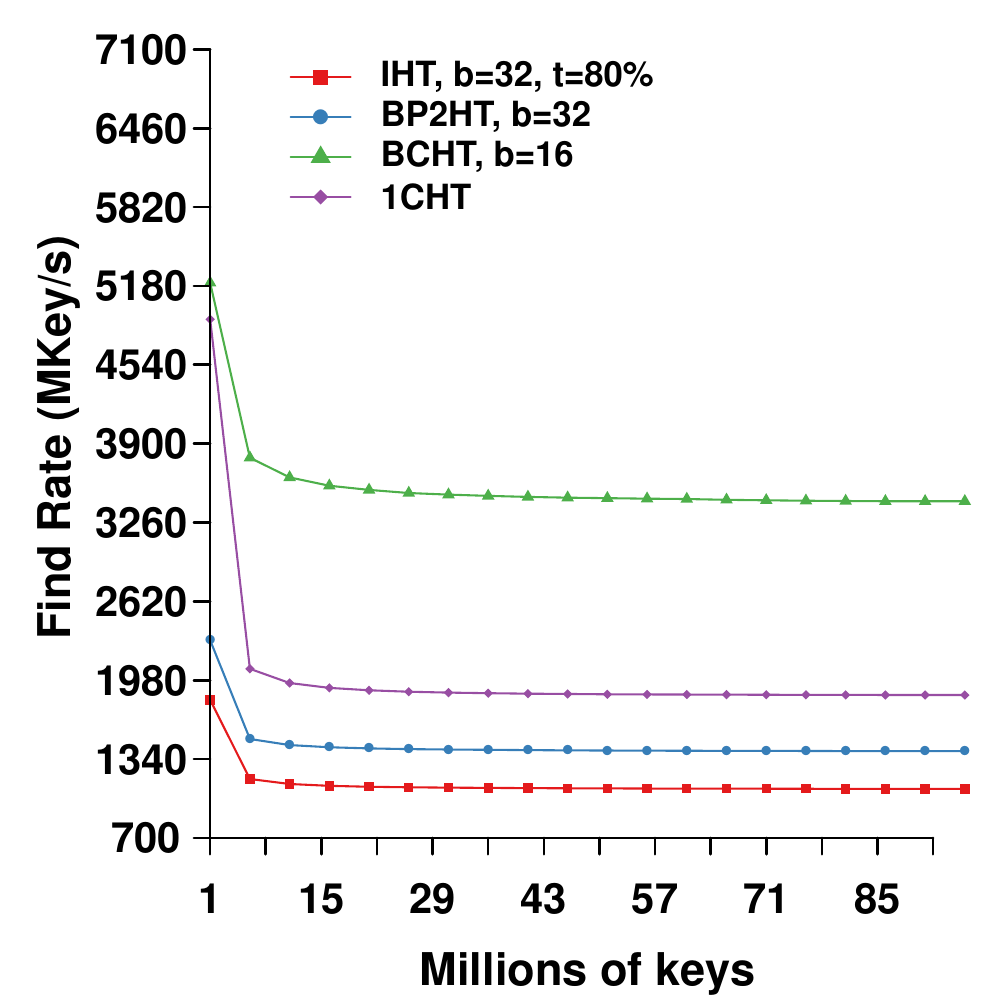}
      \includegraphics[width=0.24\textwidth]{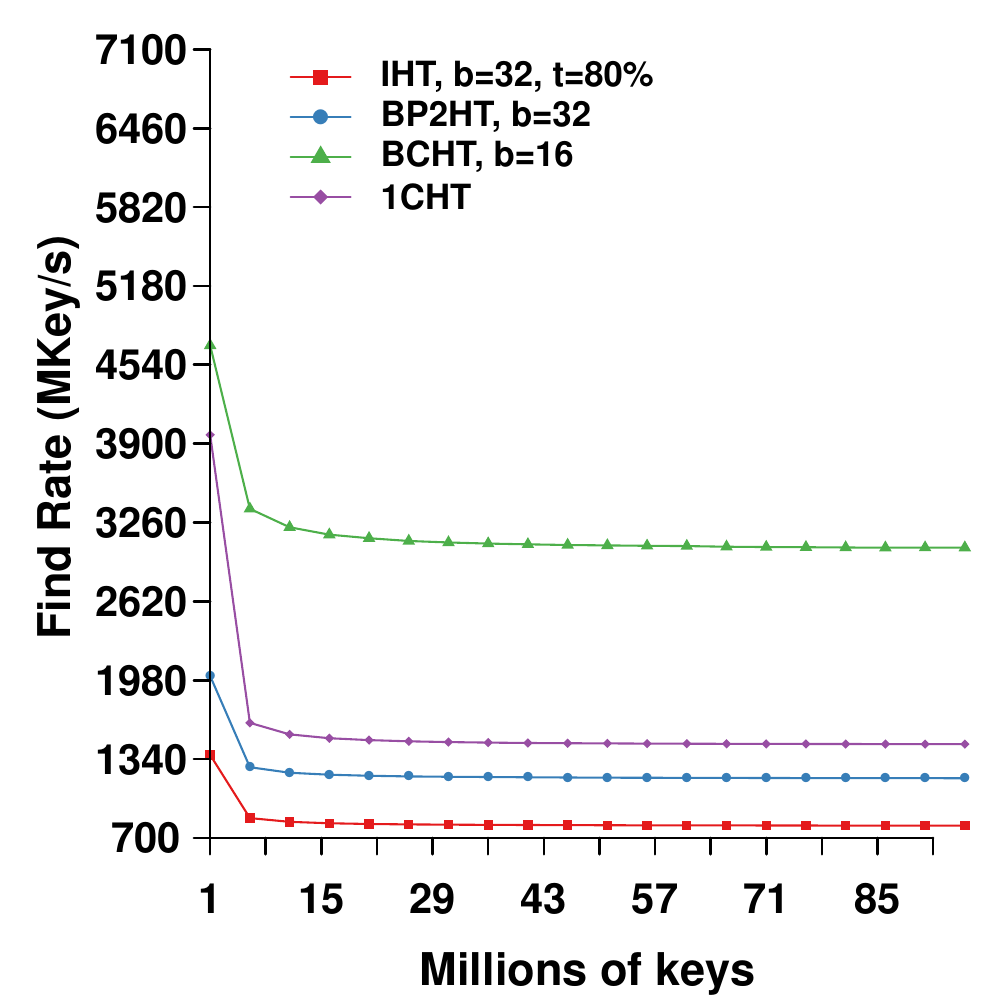}
      \caption{Insertion rate, and query rates 100\%, 50\%, and 0\% positive queries. Load factor is 0.8.}
    \end{subfigure}
    \begin{subfigure}{\textwidth}
      \includegraphics[width=0.24\textwidth]{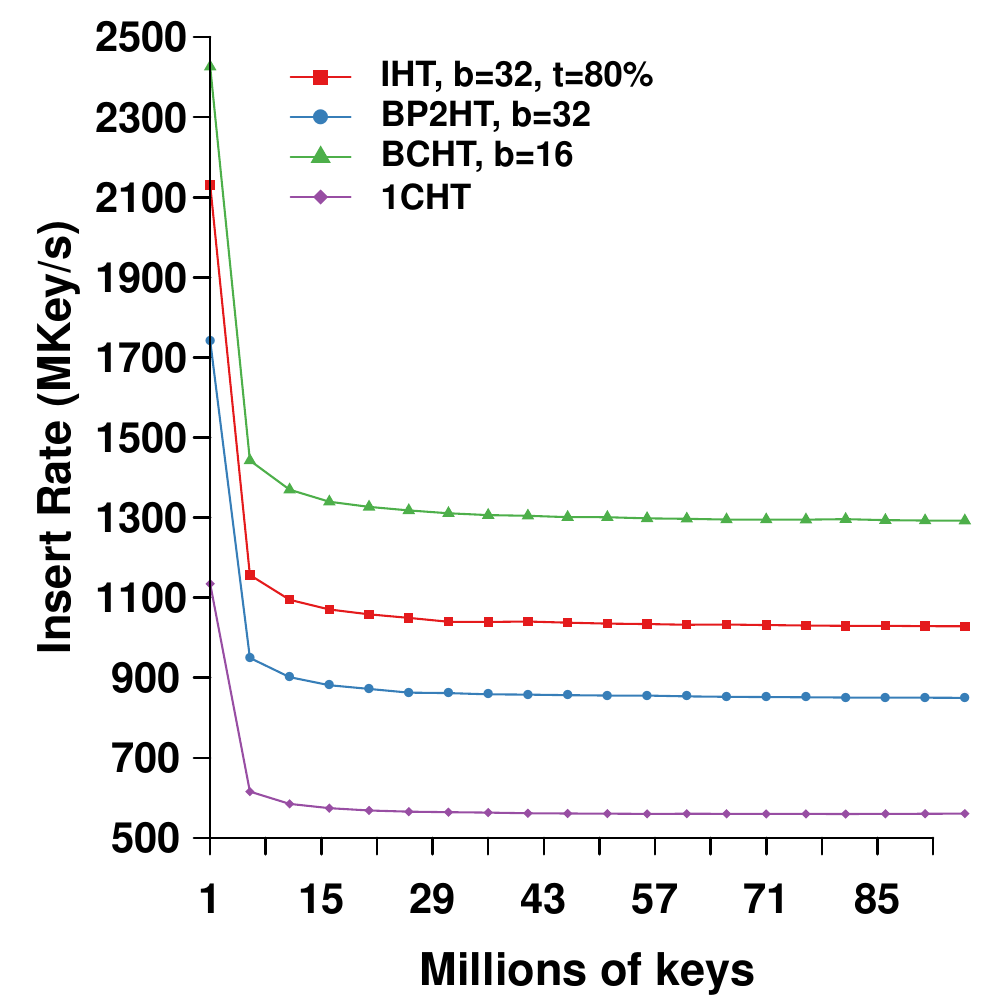}
      \includegraphics[width=0.24\textwidth]{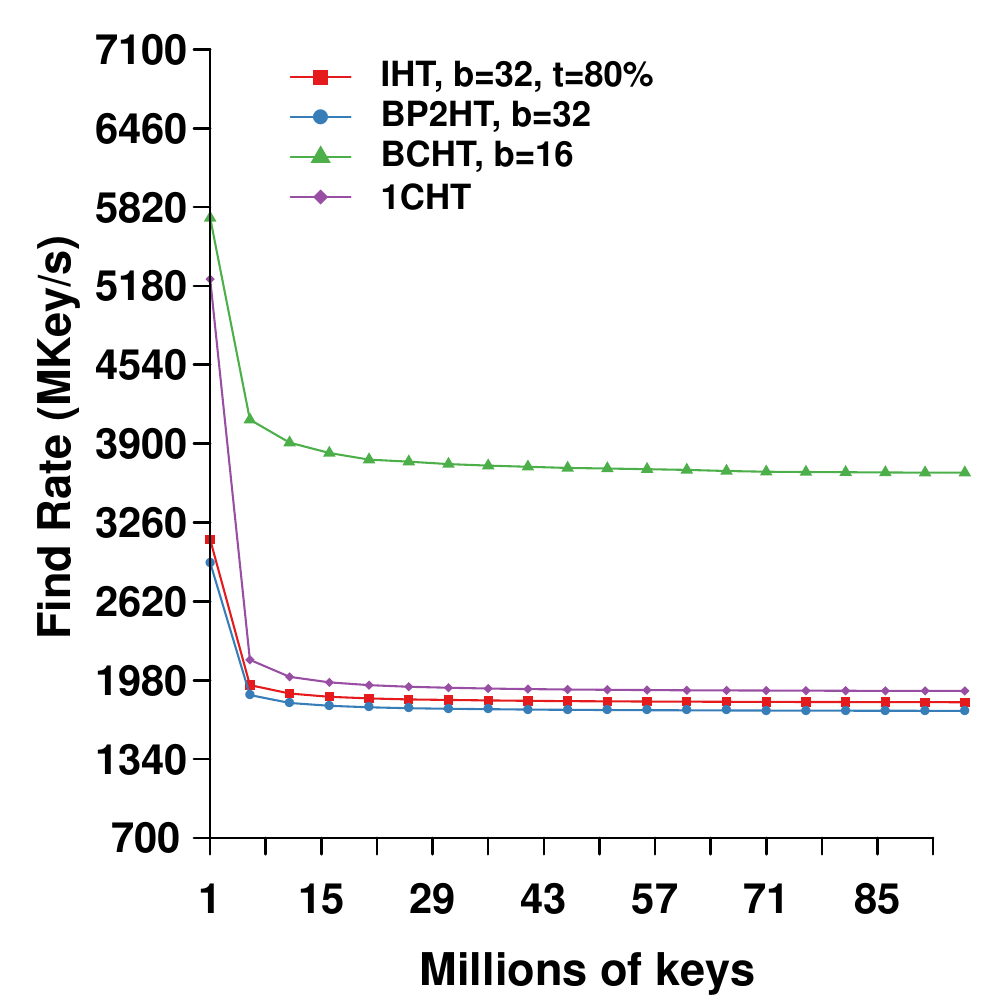}
      \includegraphics[width=0.24\textwidth]{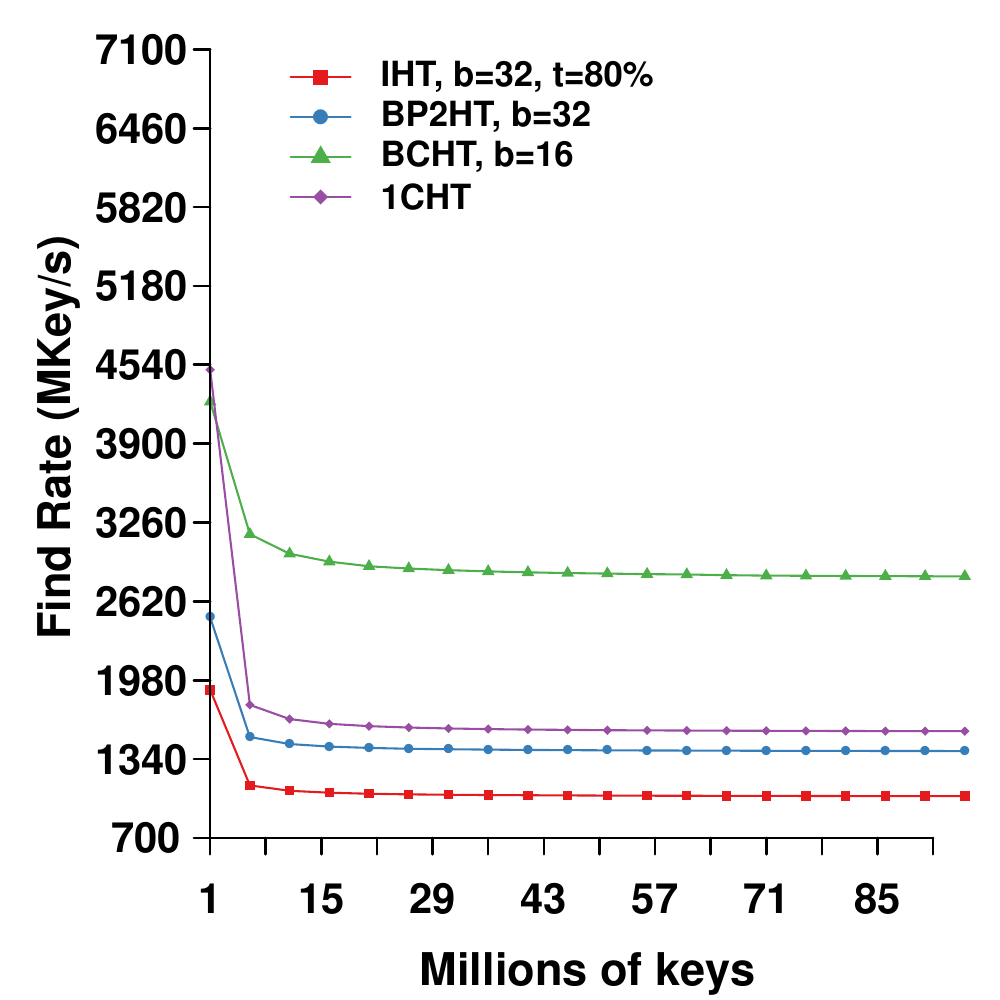}
      \includegraphics[width=0.24\textwidth]{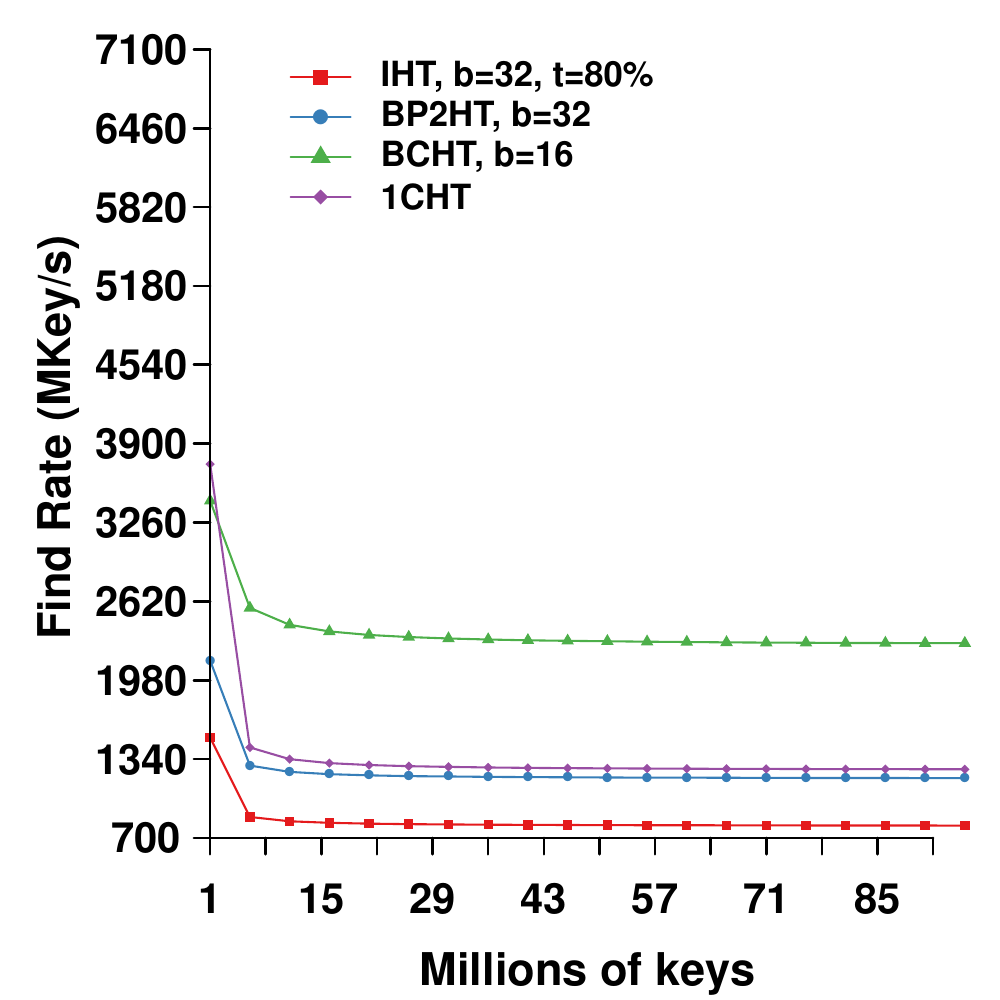}
      \caption{Insertion rate, and query rates 100\%, 50\%, and 0\% positive queries. Load factor is 0.9.}
    \end{subfigure}
    \caption{Average insertion and query rate for different number of keys and the recommended hash table variants. Number of keys is 50M keys.}
    \label{fig:recommended_vs_different_keys}
  \end{figure*}

We now consider the scenario when we vary the input number of keys while the load factor is constant. All methods offer consistent insertion and query throughputs once the key volume exceeds the cache size. Figure~\ref{fig:recommended_vs_different_keys} shows the performance of the 4 hash tables for load factors 0.8 and 0.9. Similar to the previous section, BCHT offers the best performance followed by IHT, BP2HT, then 1CHT\@. In summary, at 0.8 load factor, the insertion throughput of \{BCHT, IHT, BP2HT, 1CHT\} is \{1425, 1145.62, 878.60, 772.08\} Mkey/s.

For queries, the throughput of \{BCHT, IHT, BP2HT, 1CHT\} is \{4033.52, 2042.45, 1778.99, 2313.14\}, \{3546.12, 1131.6, 1446.50, 1945.56\}, and \{3155.51, 825.79, 1222.44, 1530.62\} for all-positive, 50\% positive and all-negative queries, respectively.

\subsection{Success rate}
\begin{figure}
  \centering
  \includegraphics[width=\columnwidth]{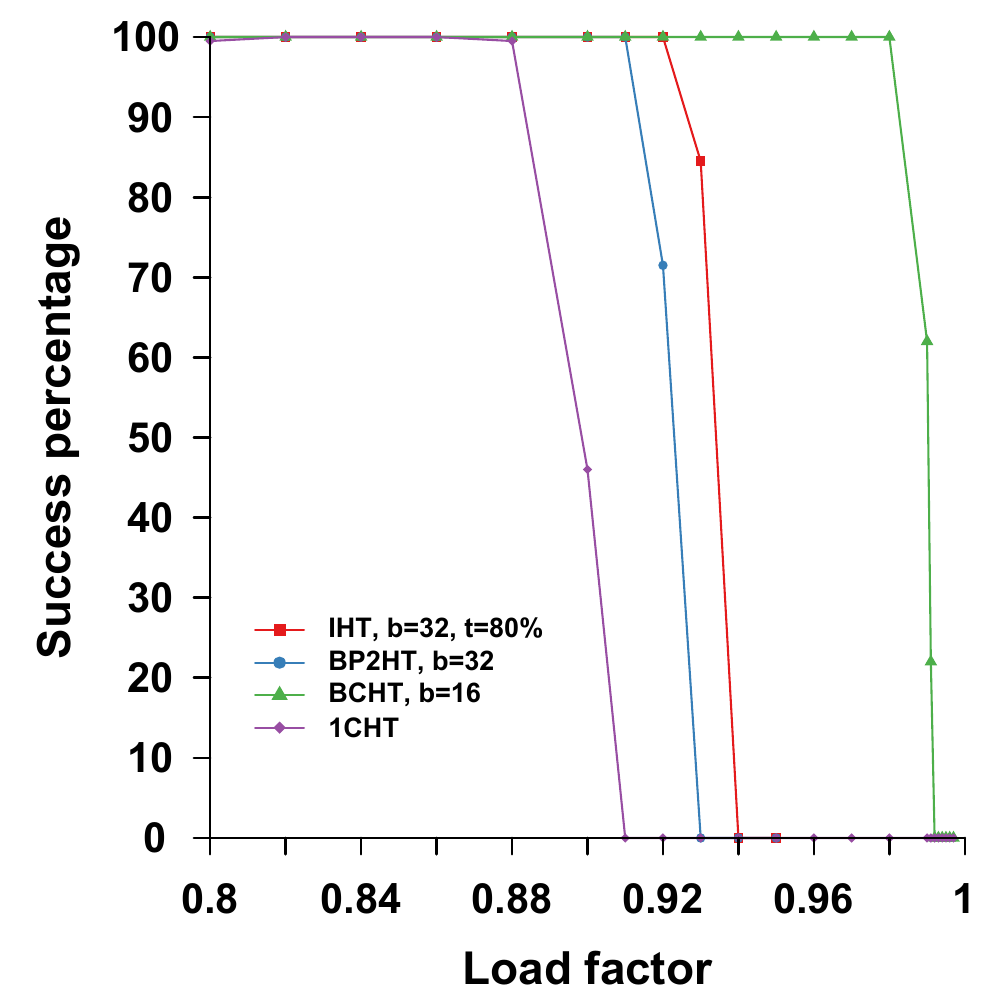}
  \caption{Comparison of success rates across the recommended hash table variants. The input is 50M keys, run over 200 experiments.}
  \label{fig:recommended_success_rate}
\end{figure}

All the hash tables we discussed could fail during building. To evaluate the success rate of each technique, we build each hash table using the same set of input keys in 200 different runs. For each different run, we use different hash functions (with random constants) and we record the success or failure of each run. Figure~\ref{fig:recommended_success_rate} shows the achievable load factors when the success percentage is around 99\%. BCHT achieves the highest load factor of 0.98, followed by IHT and BP2HT with a load factor of 0.91, then 1CHT with a load factor of 0.88.

\subsection{Analysis of the average probes count}

The main factor influencing the performance of insertion (or query) in a hash table is the number of probes each operation performs to complete an insertion (or query). To understand the performance of each operation, we instrument our implementation to count the average number of probes per key each operation needs to perform. Note that buckets have different sizes in each implementation, but here the probe is equivalent to reading a bucket regardless of its size (reading two buckets instead of one does not necessarily result in $2\times$ slower performance). Moreover, insertion in a 1CHT requires a single atomic exchange, whereas, in bucketed techniques, two operations are required: 1) reading a bucket, 2) performing an atomic compare-and-swap operation (and in some cases instead of an atomic exchange). We will investigate these effects in the next section; however, we will only consider the number of probes in this section. By only considering the average number of probes, we can understand the performance of each hash table in a hardware-independent way.

\begin{figure*}
  \centering
  \begin{subfigure}{\textwidth}
    \includegraphics[width=0.24\columnwidth]{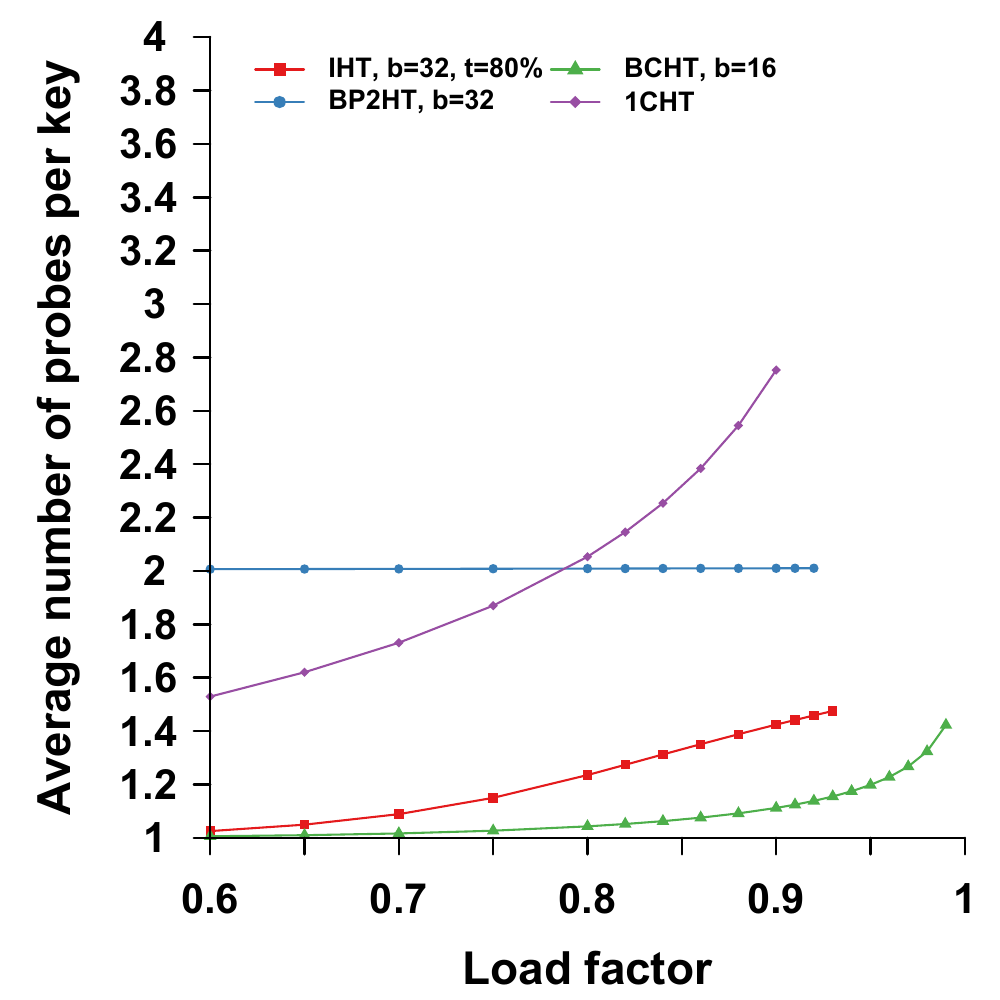}
    \includegraphics[width=0.24\columnwidth]{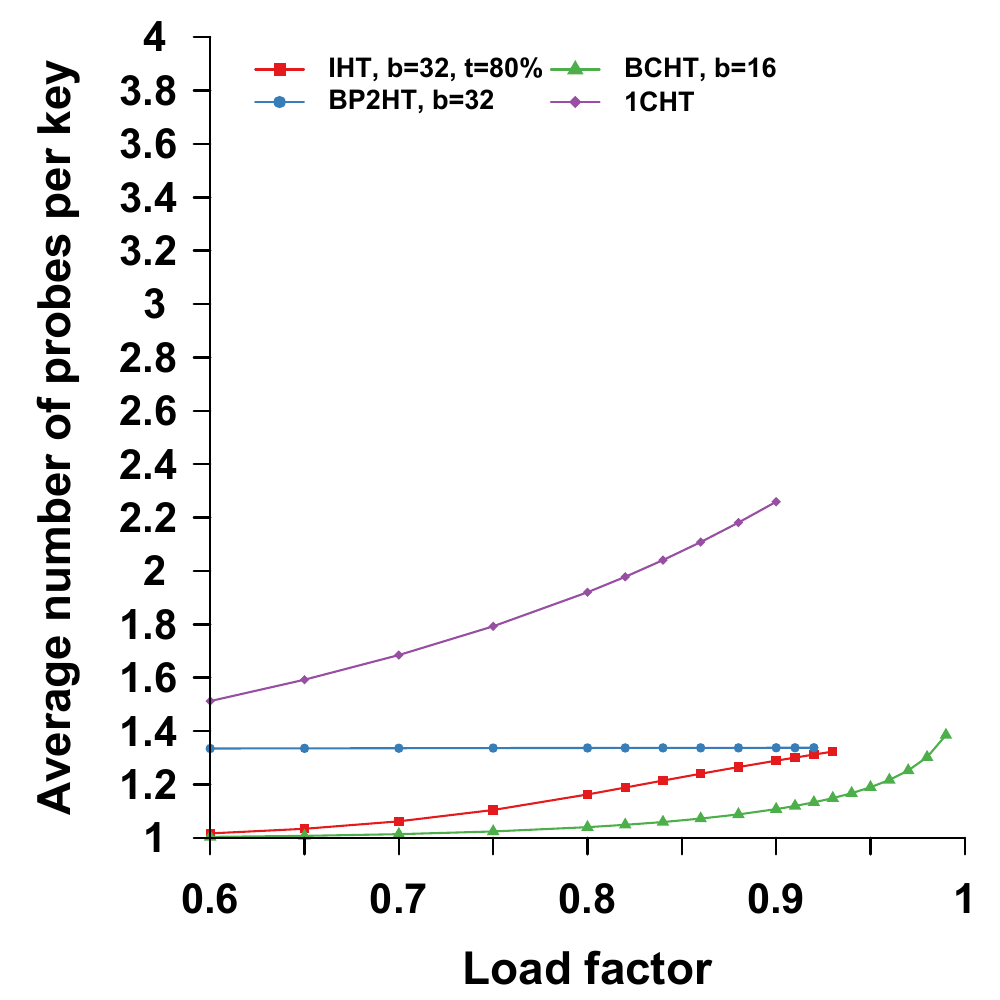}
    \includegraphics[width=0.24\columnwidth]{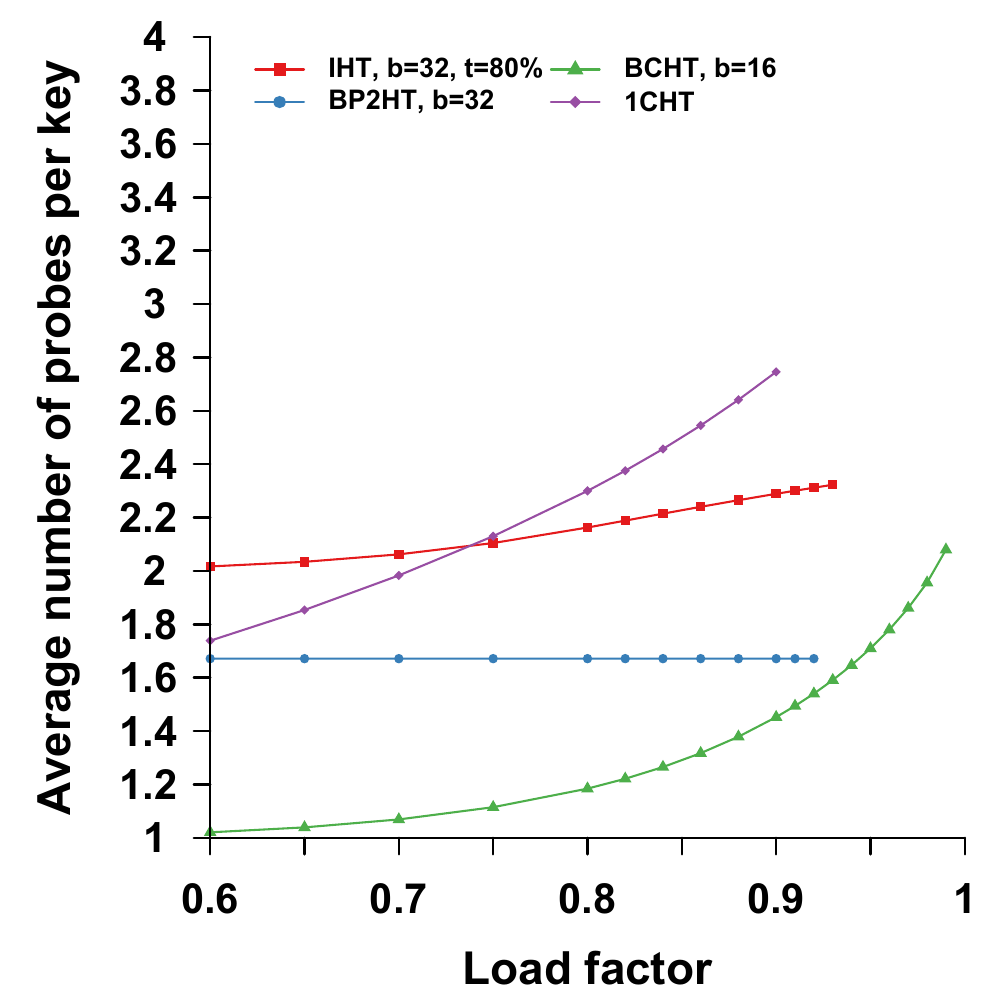}
    \includegraphics[width=0.24\columnwidth]{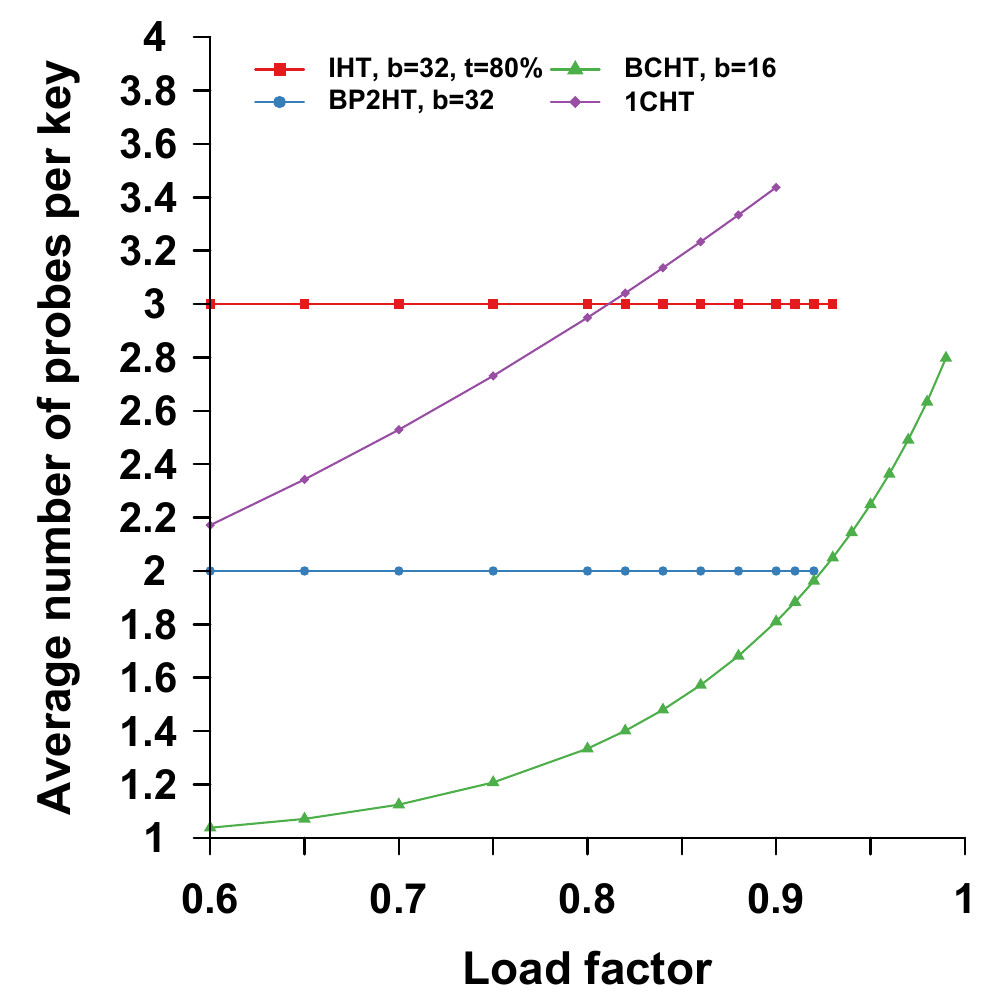}
  \end{subfigure}
  \caption{Average probes per key for insertions and queriers (100\%, 50\%, 0\%  positive queries) and a varying load factor. Number of keys is 50M keys.}
  \label{fig:recommended_avg_probes}
\end{figure*}

Figure~\ref{fig:recommended_avg_probes} shows the average number of probes for insertion and query using different positive query ratios in a table of size 50 million keys. We build each variant for different load factors ranging from 0.6 up to the maximum load factor we can achieve.

\begin{LaTeXdescription}
  \item[Insertion] We find that BCHT enjoys a very low average number of probes per key (up to 1.43 at load factor 0.99), which will be tough to beat for any competing method. IHT has a similarly low number of probes as well (up to 1.46 at load factor 0.92), primarily because it only requires reading the primary bucket unless the primary bucket exceeds the threshold, when it reads 2 additional buckets. In the best-case scenario for IHT, a threshold\% of the total keys will only require reading the primary bucket, whereas ($100 - \text{threshold}$)\% will read the primary and two secondary buckets. BP2HT requires loading two buckets regardless of the load factor. 1CHT with 4 hash functions reads up to 2.75 buckets per insertion at 0.9 load factor.

  \item[Queries, all in table] When all queries exist in the hash table, the number of probes drops for all techniques. Unlike insertions, queries do not have atomic operations that could fail and for all hash tables, queries do not require accessing all the possible buckets. For instance, BP2HT's average probe count drops from 2 in insertion to 1.34 for queries at the highest load factor (0.92). Similarly, BCHT, IHT, 1CHT's average probe count drops to 1.38, 1.32, and 2.26 at the highest achievable load factors respectively.

  \item[Queries, not all in table] However, as the ratio of positive queries drops (i.e., more keys do not exist in the table), the average number of probes increases and it is possible that they can reach up to the number of hash functions in 1CHT and BCHT or 2 and 3 in BP2HT and IHT respectively. Note that unlike 1CHT and BCHT, IHT and BP2HT cannot perform an early exit if one of the possible buckets has an empty spot in them. At a 0\% positive query ratio, \{BCHT, IHT, BP2HT, 1CHT\} reach up to \{2.80, 3, 2, 3.44\} probes per query.
\end{LaTeXdescription}

The results that we showed in the previous sections validate and complement our high-level hash table recommendations in Section~\ref{sec:recommendations}.

\subsection{Analysis of the achieved throughput and memory transfers}

To evaluate our implementations and compare them to the ideal speed-of-light for the GPU, we measure two different metrics: 1) achieved DRAM throughput and 2) the average number of DRAM sectors (defined below) per key (insertion or lookup). Figure~\ref{fig:throughput_and_sectors} shows the profiler results for these analyses. For this experiment, we use a large number of keys (200M keys) to avoid any caching effects, and for fairness, we use the same bucket size across the different hash tables.

\paragraph*{Achieved throughput} For all bucketed hash tables, lookups achieve more than 500~GiB/s (over 75.5\% of the achievable bandwidth). 1CHT only achieves around 267~GiB/s. On the other hand, insertion achieves lower throughput (between 220 and 300~GiB/s) in bucketed hash tables and 141~GiB/s in 1CHT\@.

\paragraph*{Memory transfers}
To understand the slowdown for insertions compared to lookups, we measure the average number of DRAM sectors per key (the profiler unit of measurement for load and store operations is a sector), where a \textit{sector} is defined as 1/4 of a cache line (i.e., 32 bytes).

For a query operation, the dominant factor in the number of sectors per key is the number of buckets a query needs to read to find a key. For example, when BCHT finds a key in the first possible bucket (i.e., one bucket is required to evaluate the query), we expect the number of sectors per key to be four (12 in the worst-case scenario of three buckets). Similarly, BP2HT and IHT will require four sectors in the best-case scenario and 8 and 12 in the worst-case scenario, respectively. 1CHT only requires several sectors between 2 and 8 (note that the access granularity between the DRAM and the L2 cache is 64 bytes, i.e., when an uncoalesced access reads a single sector, it will transfer two sectors).

On the other hand, insertion in bucketed tables requires writing back the pair into the bucket using at least one atomic compare-and-swap operation (and, in some cases, an atomic exchange operation if the bucket is full or in 1CHT)\@. Since we perform these atomic operations on only 8 bytes key-value pair, the atomic operation will add at least one additional sector. We see this effect for both IHT and BCHT in Figure~\ref{fig:throughput_and_sectors}. Note that these atomic operations are divided into two operations: first, moving the required sector to the L2 cache, followed by the update and writeback operations. However, since we just fetched the bucket before the atomic operation, more than 90\% of the time, the bucket is already in the L2 cache.

BP2HT, on the other hand, shows a higher-than-one increment in the number of sectors between lookups and insertion, and the reason behind that is that BP2HT requires loading two buckets before deciding on the bucket in which we will insert the pair, and this increases the number of sectors by an additional four sectors, resulting in a total of five additional sectors for lookups.

We note that although we are using sectors here as our unit, loading a sector from the DRAM has the same cost as loading a full cache line, thus an additional sector (specifically for writing back a pair into the bucket) costs the same as reading an entire cache line.

\begin{figure*}
    \centering
    \begin{subfigure}{\textwidth}
      \includegraphics[width=0.24\columnwidth]{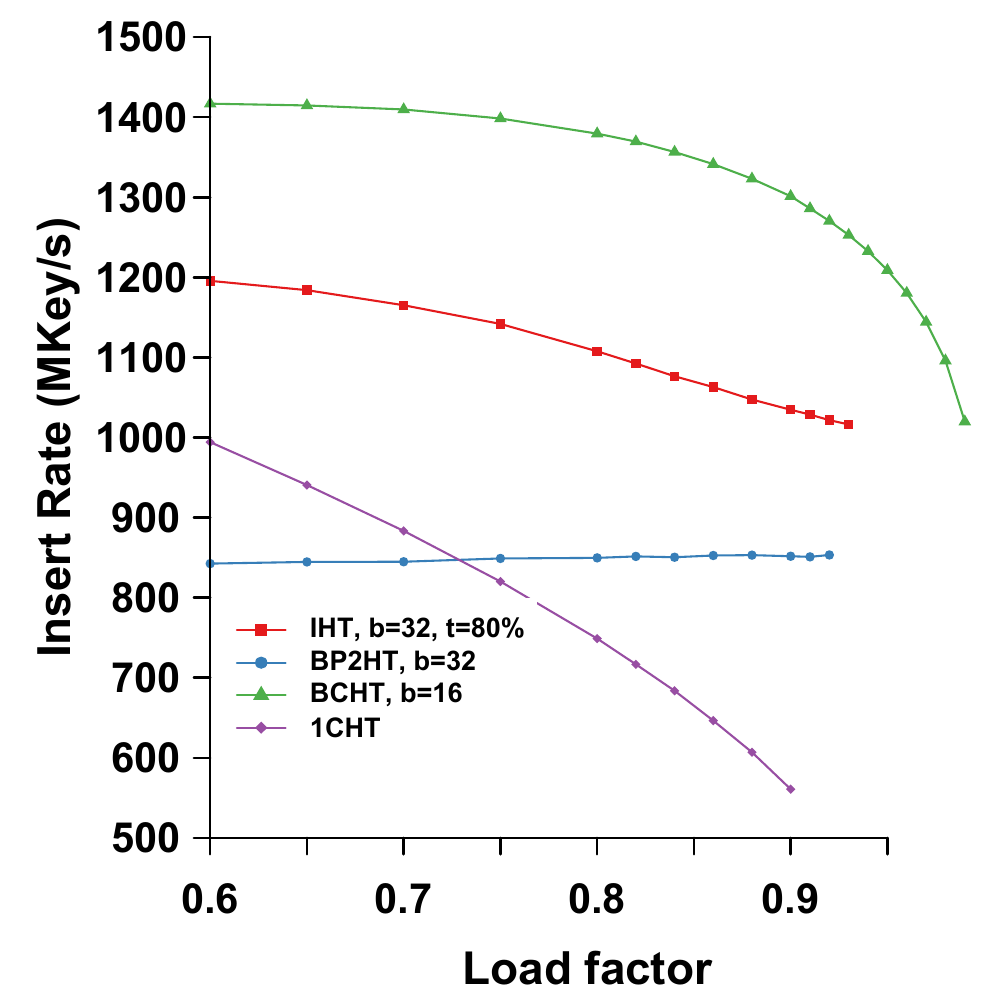}
      \includegraphics[width=0.24\columnwidth]{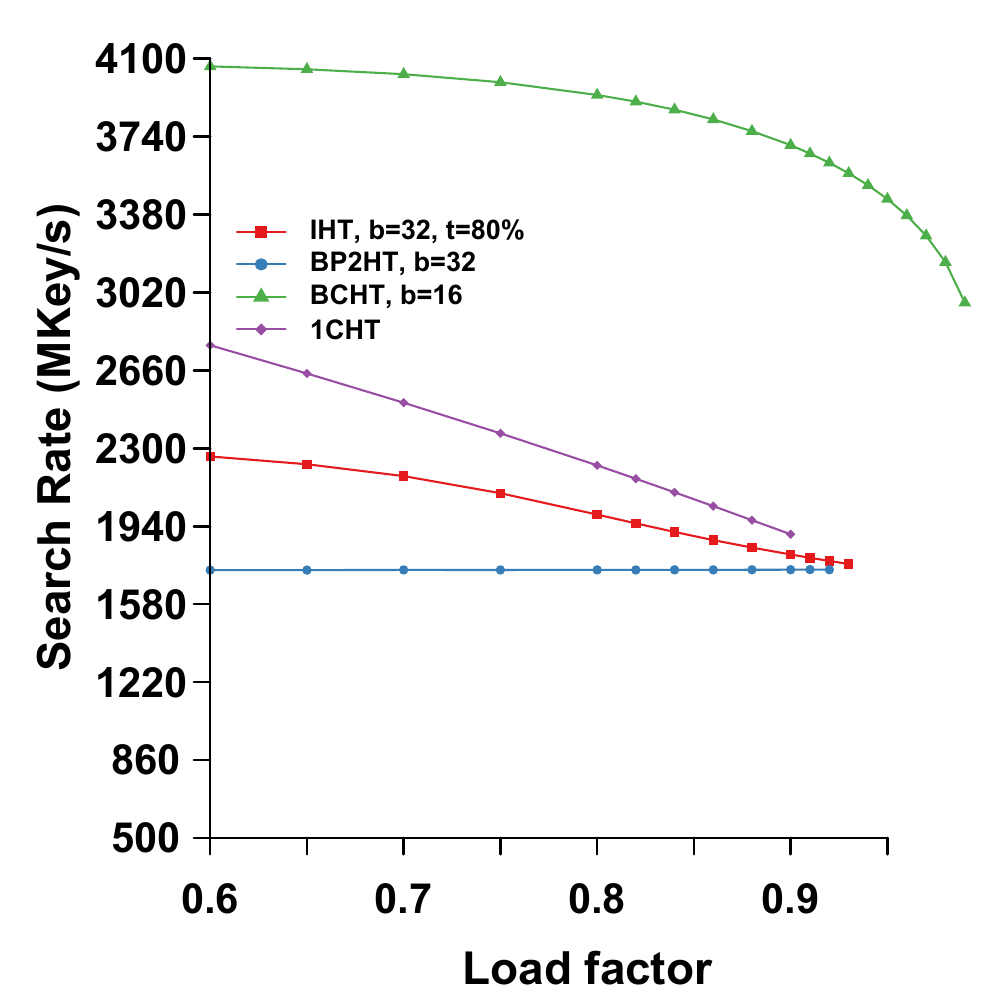}
      \includegraphics[width=0.24\columnwidth]{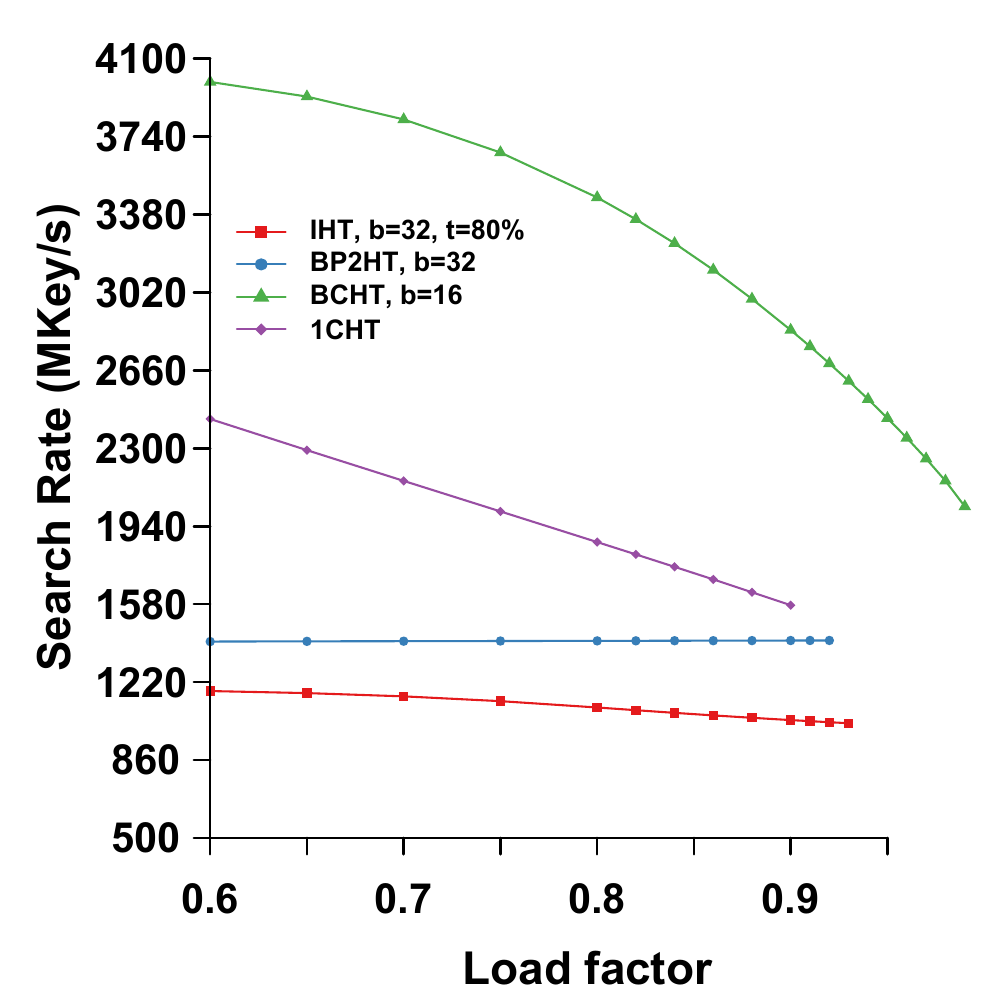}
      \includegraphics[width=0.24\columnwidth]{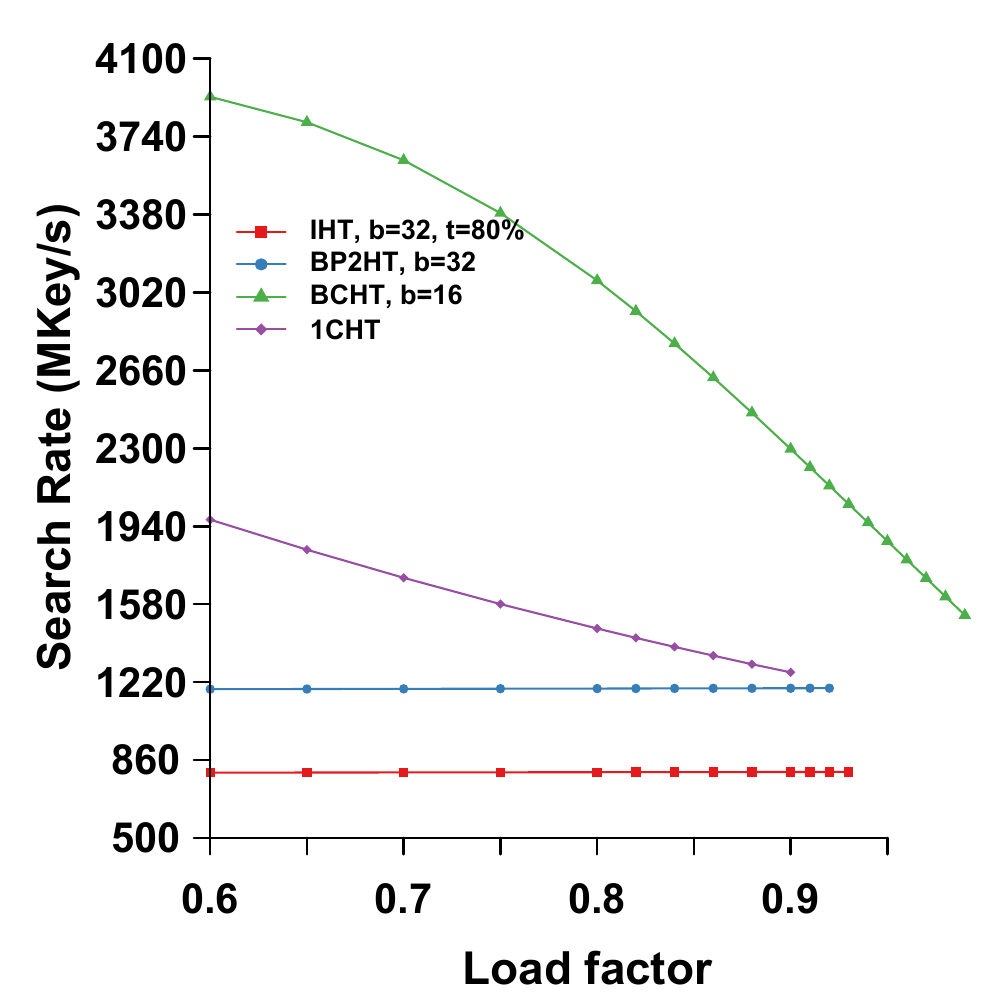}
    \end{subfigure}
    \caption{Average insertion and query rate (100\%, 50\%, 0\%  positive queries) for different load factors and the recommended hash table variants. Number of keys is 50M keys.}
    \label{fig:recommended_vs_different_load_factors}
  \end{figure*}

\begin{figure}
    \centering
    \begin{subfigure}{0.24\textwidth}
      \includegraphics[width=\textwidth]{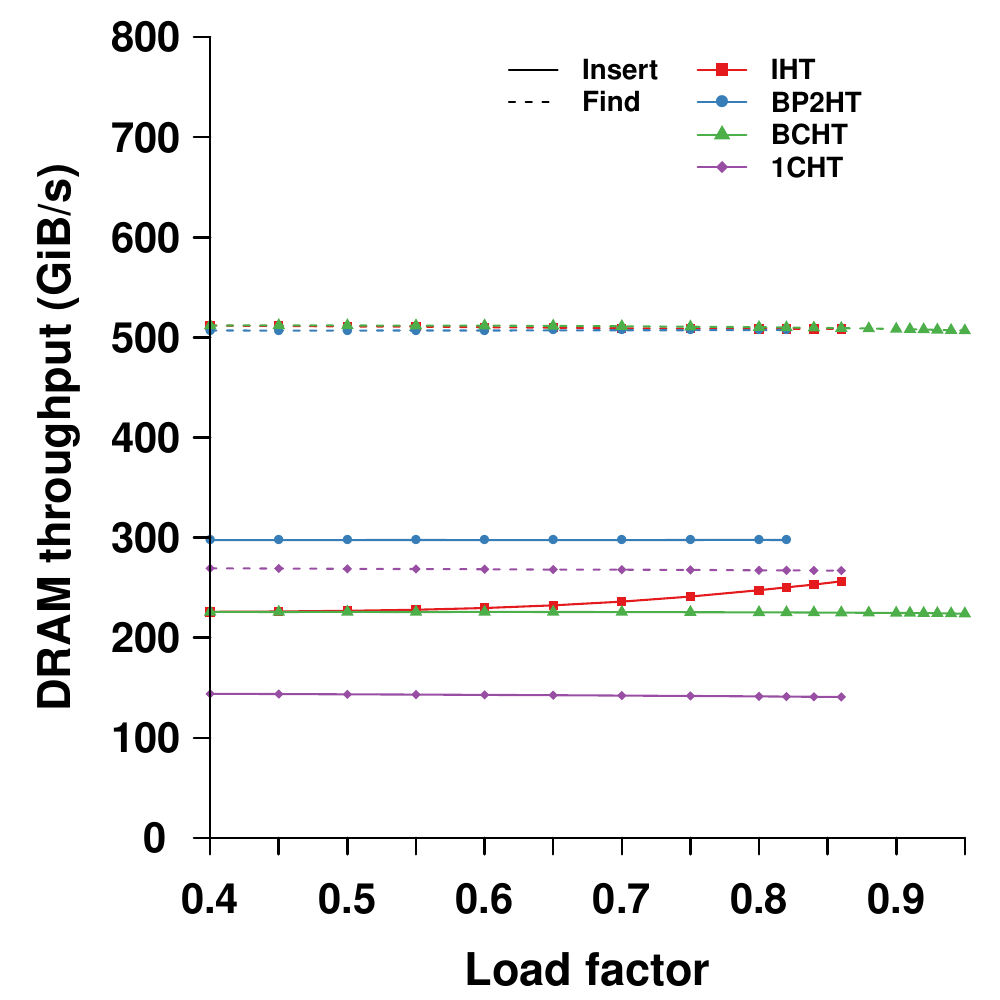}
    \end{subfigure}
    \begin{subfigure}{0.24\textwidth}
      \includegraphics[width=\textwidth]{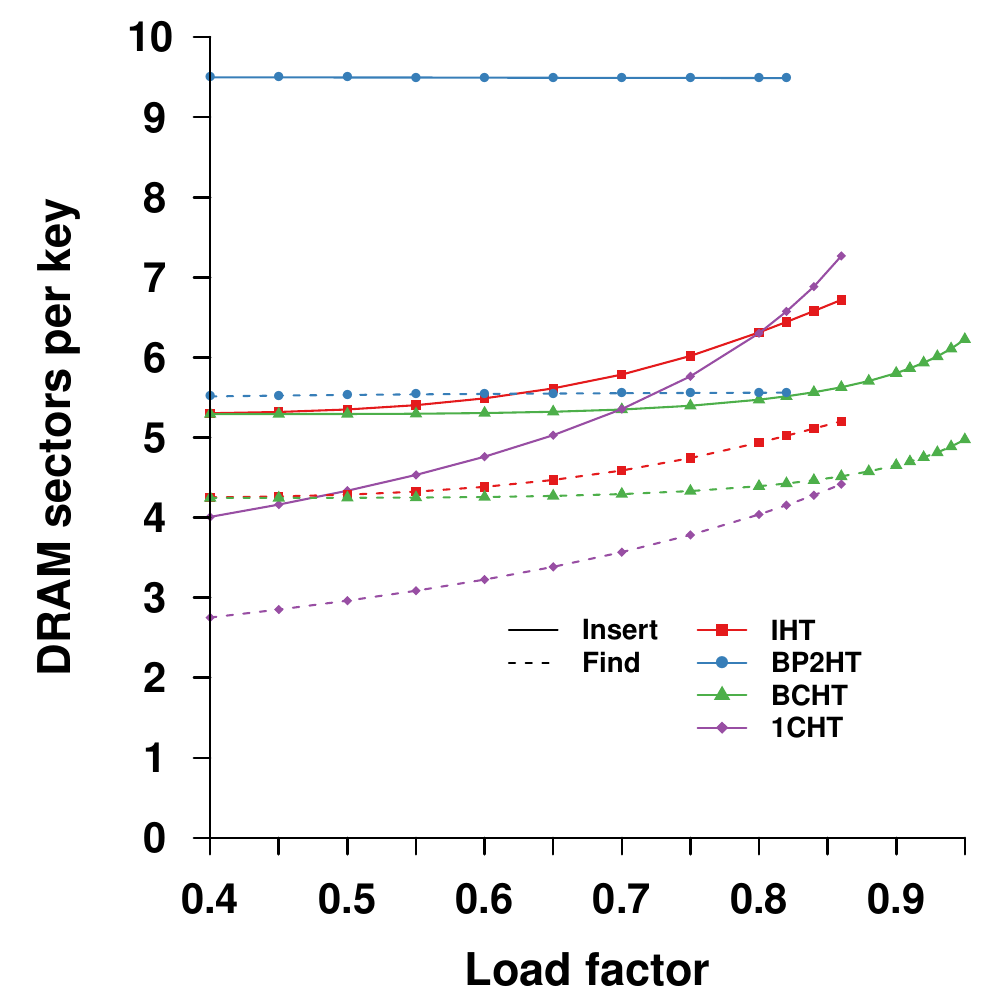}
    \end{subfigure}
    \caption{Achieved throughput and the average number of sectors per key (a sector is 32 bytes) for IHT, BCHT, and BP2HT (each with a bucket of 16 and IHT uses a threshold of 13 keys per primary bucket). The number of keys is 200M keys, and all lookups exist in the hash table. Notice the relationship between the average number of probes per key (Figure~\ref{fig:recommended_avg_probes}) and the average number of sectors per key. A single probe into a 128-bytes bucket corresponds to loading four sectors. Also, notice the inverse relationship between the average number of sectors per key and the average insertion or query throughput in Figure~\ref{fig:recommended_vs_different_load_factors}.}
    \label{fig:throughput_and_sectors}
  \end{figure}

Our conclusion from this analysis is that we effectively utilize the DRAM bandwidth, and the main performance limiter is the number of sectors per key that a hash table operation must load (or store). Our analysis here closely matches the insertion (or query) throughput results in Figure~\ref{fig:recommended_vs_different_load_factors}.



\begin{figure*}
\setlength\tabcolsep{1pt}
\settowidth\rotheadsize{Radcliffe Cam}
\setkeys{Gin}{width=\hsize}
\begin{tabularx}{\textwidth}{l XXXX }
\rothead{\centering load factor = 0.8}
                        &   \includegraphics[valign=m]{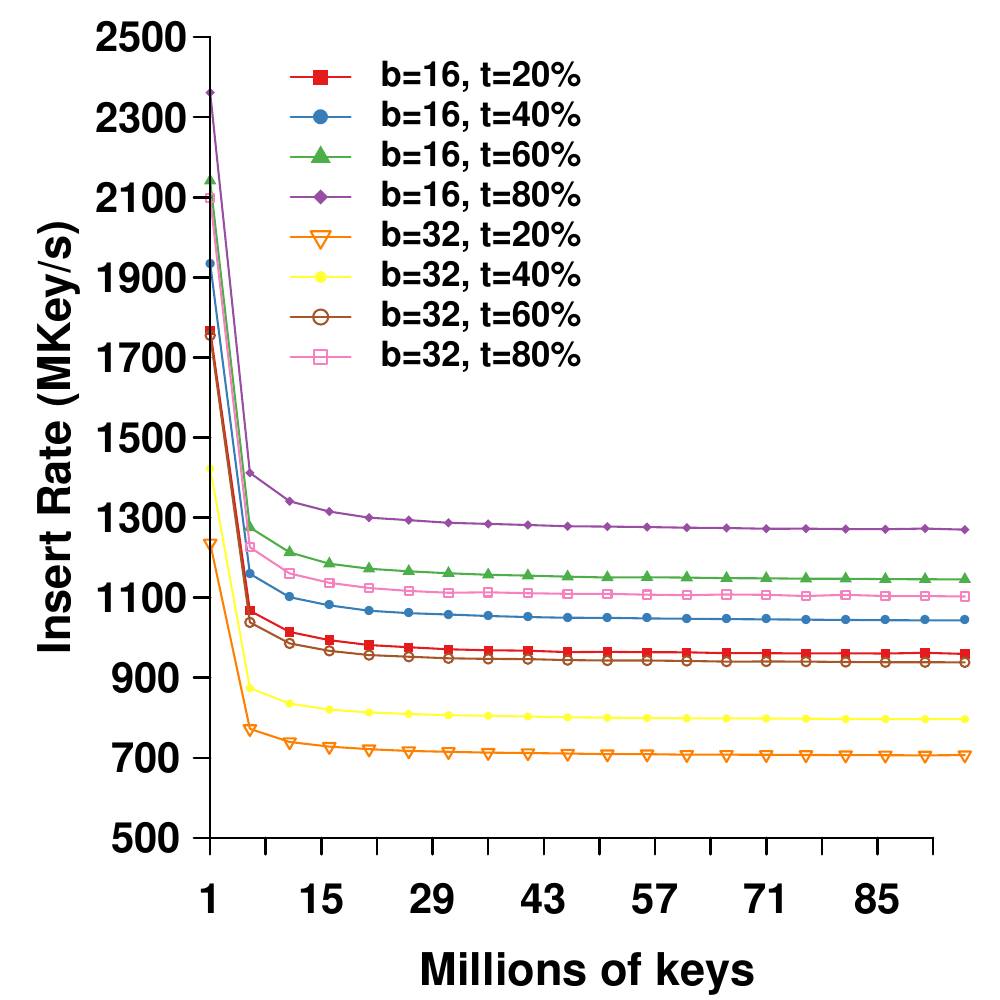}
                        &   \includegraphics[valign=m]{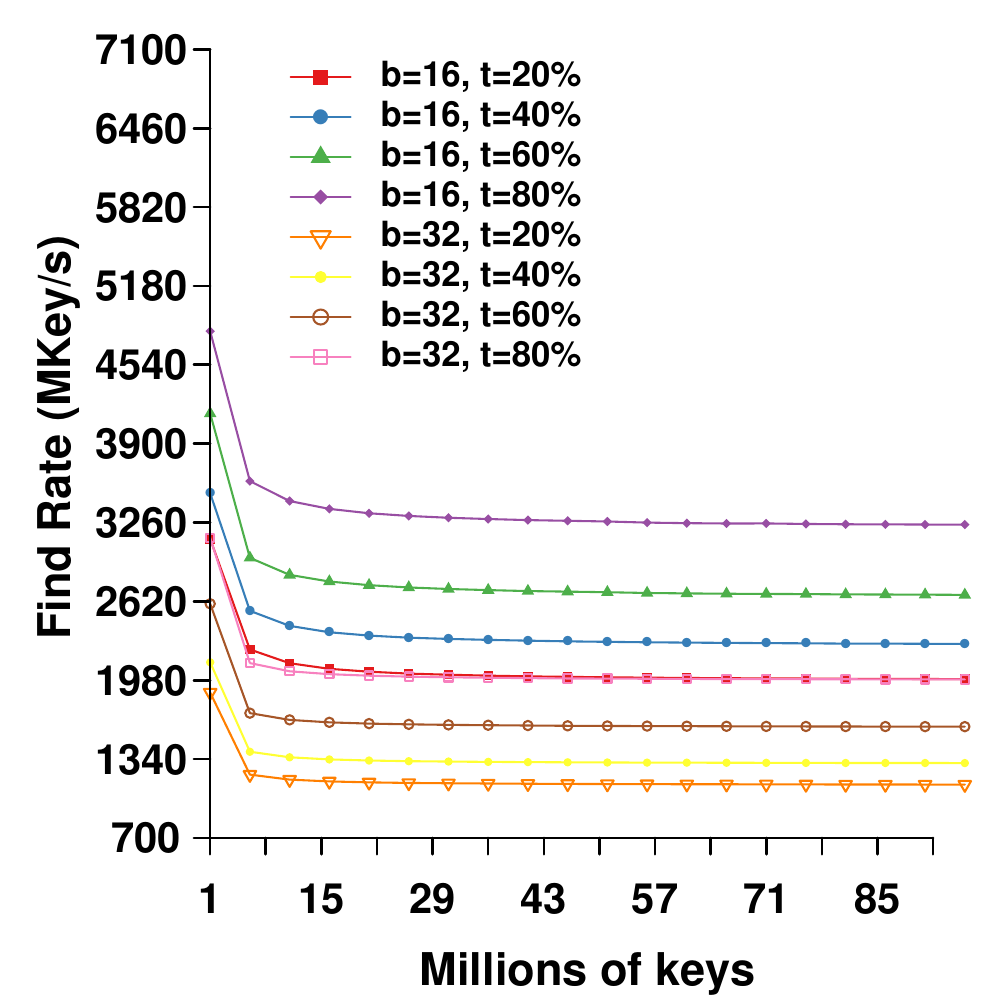}
                        &   \includegraphics[valign=m]{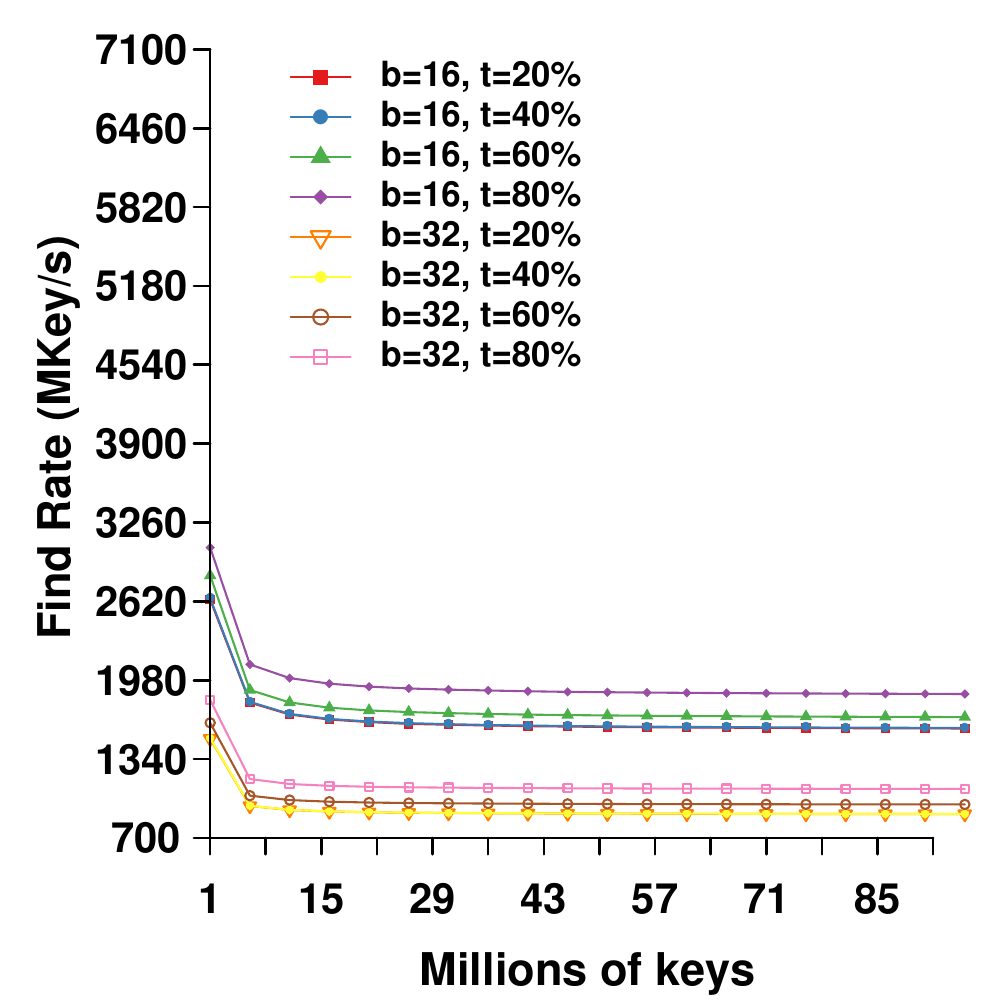}
                        &   \includegraphics[valign=m]{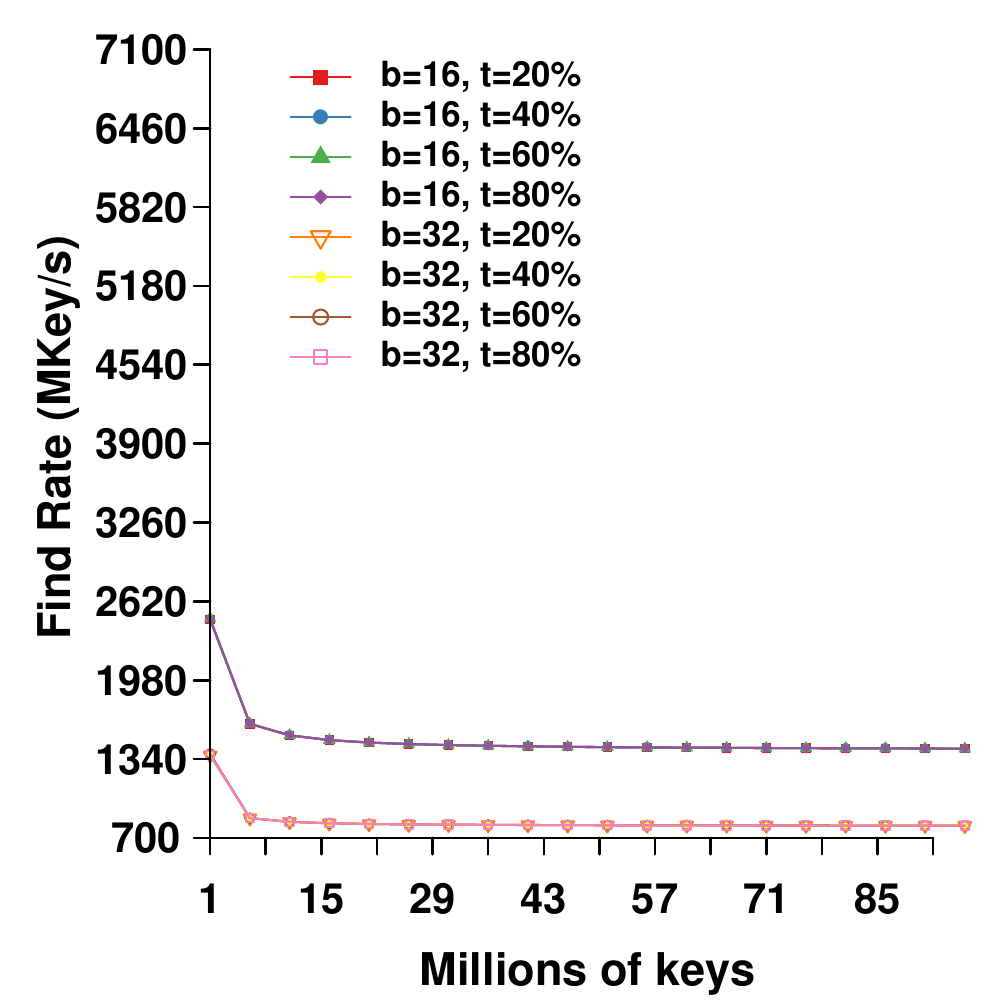}  \\ \addlinespace[2pt]
\rothead{\centering load factor = 0.9}
                        &   \includegraphics[valign=m]{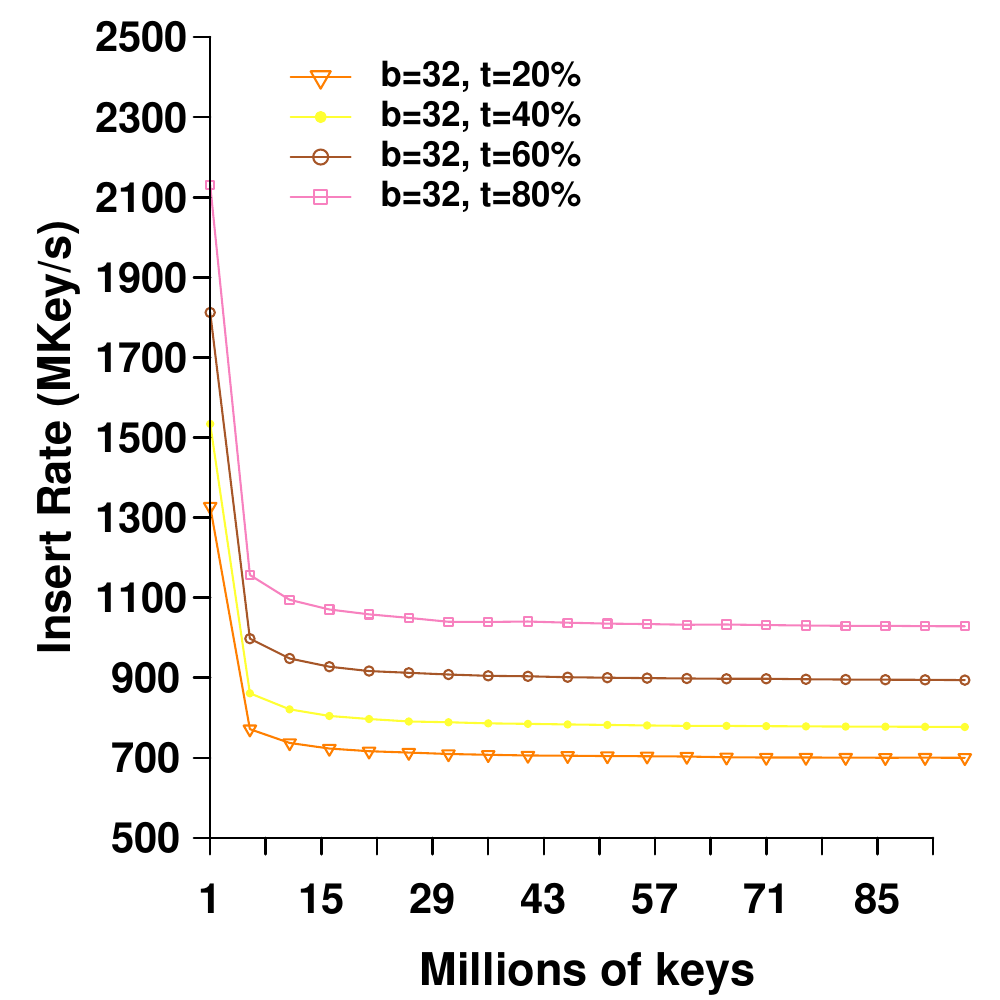}
                        &   \includegraphics[valign=m]{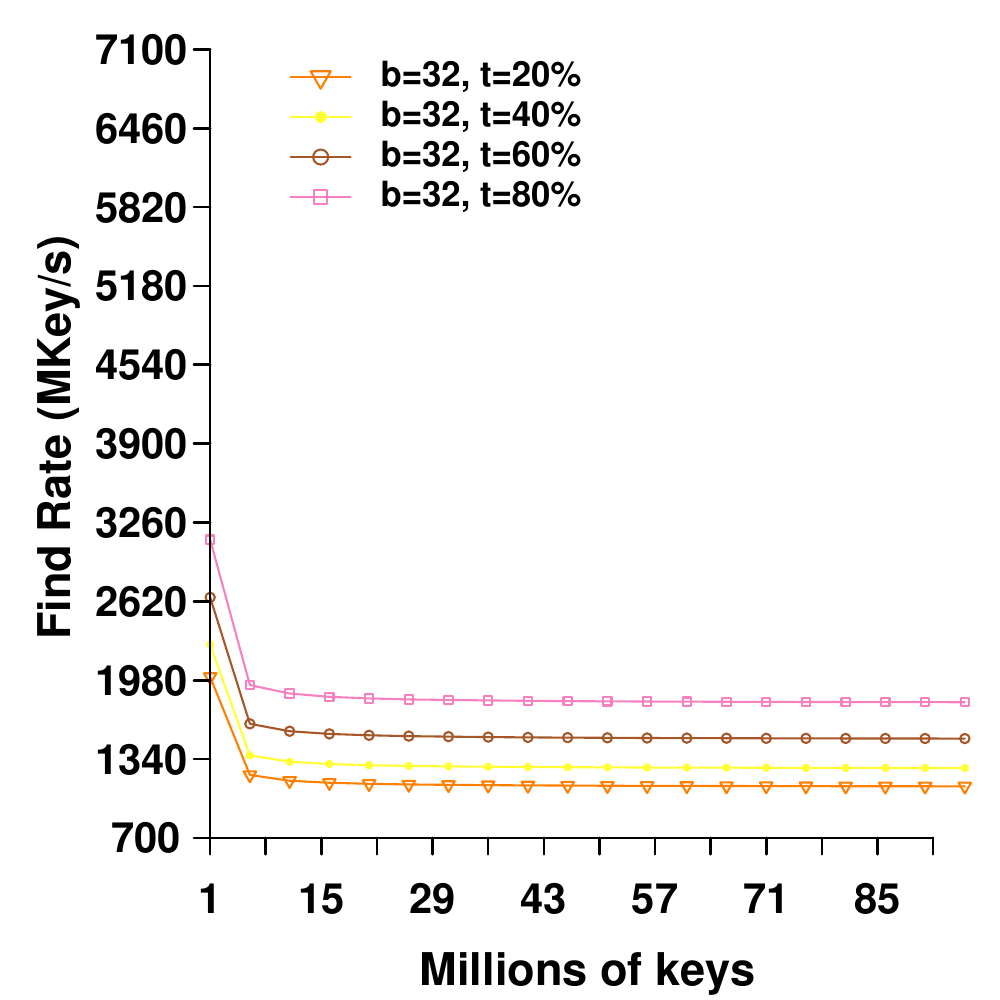}
                        &   \includegraphics[valign=m]{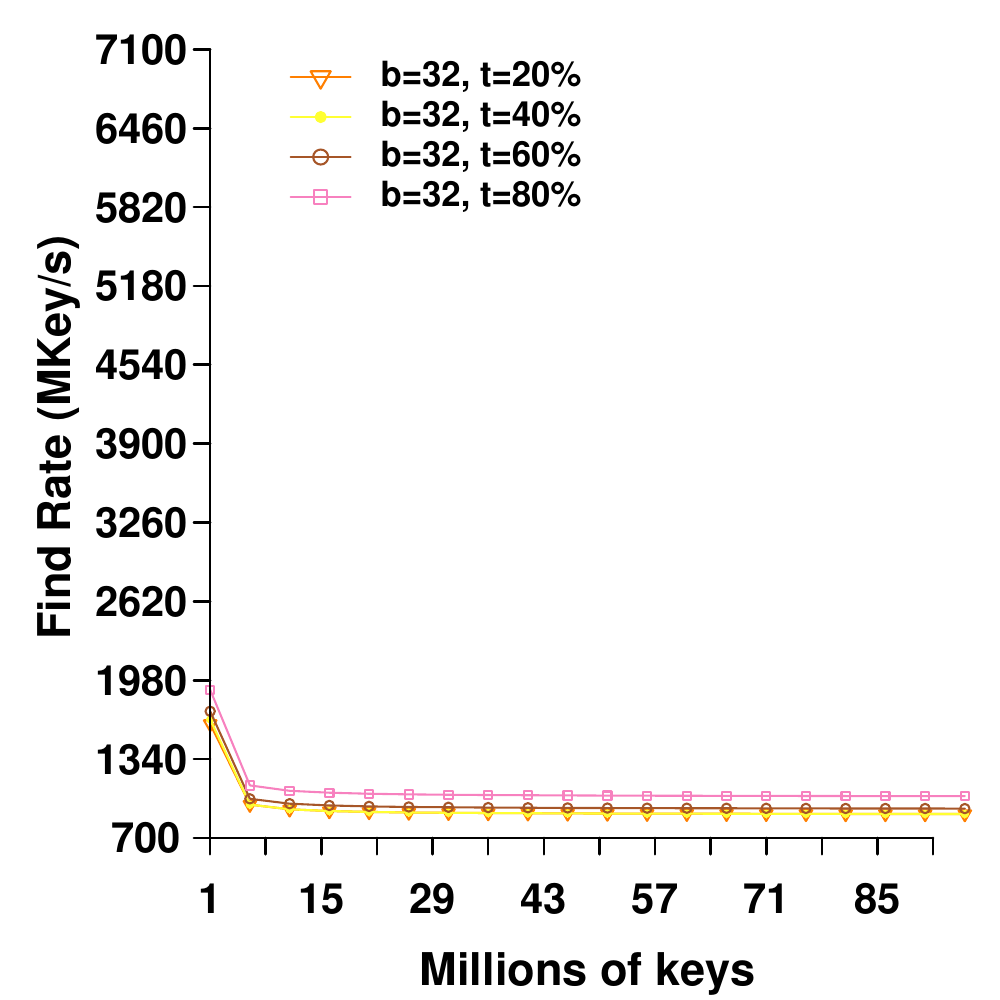}
                        &   \includegraphics[valign=m]{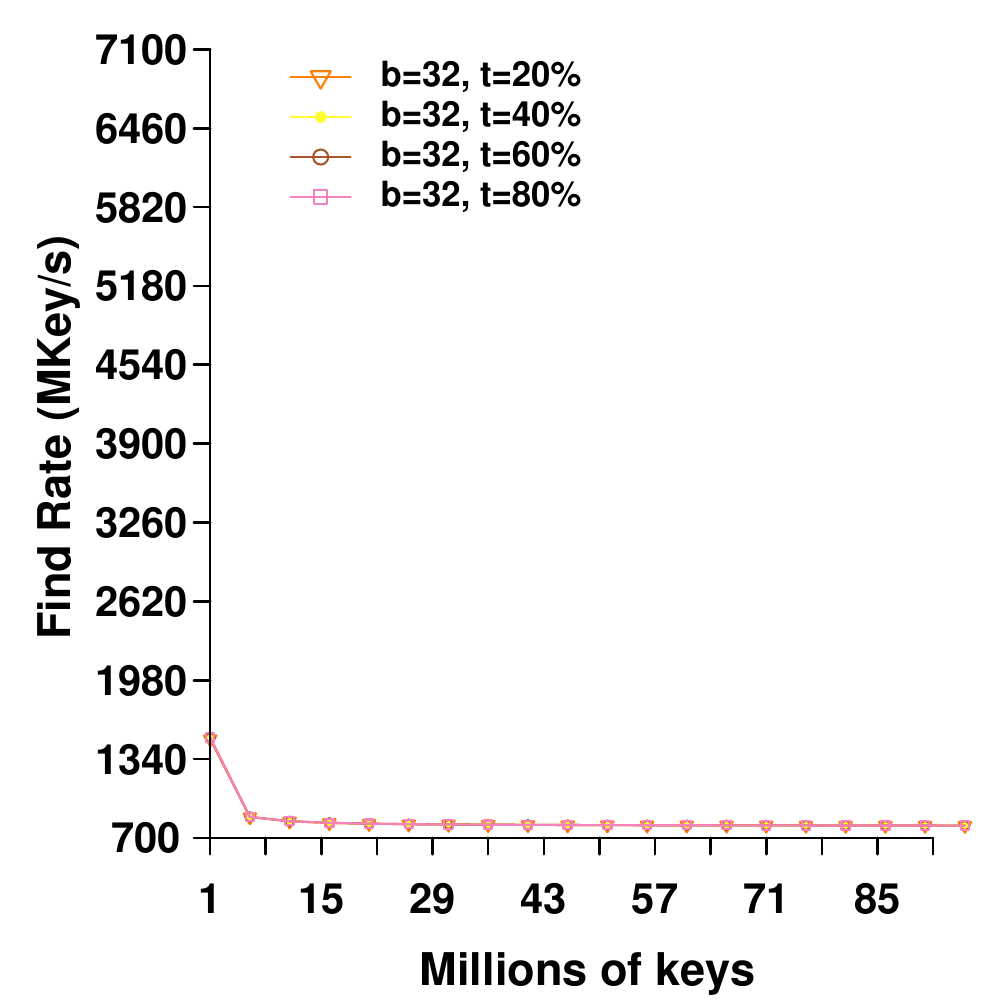}
\end{tabularx}
\caption{IHT throughput for insertion, 100\%, 50\%, and 0\% positive queries (from left to right).}
\label{fig:config_iht_fixed_load_factor}
\end{figure*}

\section{Conclusion and future work}

The number of probes an insertion (or query) operation performs is the primary performance limiter in a hash table given an efficient design and implementation. We efficiently implemented three different variants of static hash tables on the GPU, where the main difference between each hash table variant is the probing scheme.

For memory-bound problems such as building and querying a hash table, an efficient design fully utilizes the memory bandwidth. Bucketed hash tables where the bucket size matches the hardware cache line achieves a hash table design that fits the GPU\@. Our results show that resolving collisions using cuckoo hashing outperforms alternate techniques on the GPU\@. With around 1.43 average probes per key at load factors as high as 0.99, bucketed cuckoo hash tables are the best all-around choice among all hash tables we studied.

In future work, we would like to revisit the iceberg hashing technique. Specifically, we would like to achieve high load factors when the bucket size is most suitable for the GPU (i.e., $b=16$) by investigating different probing schemes after a bucket reaches its threshold. Moreover, to improve the performance of negative queries, we can use a quotient filter that stores information about keys that did not fit in the primary bucket.

\section*{Acknowledgments}

Thanks to Martin Dietzfelbinger, Lars Nyland, Nicolas Iskos, and Alex Conway for helpful technical suggestions. Thanks to Aleksy Ba\l{}azi{\'{n}}ski for pointing out listings errors.

The authors appreciate the research support of the National Science Foundation (awards~\# CCF-1637442 and~\# OAC-1740333), DARPA (AFRL awards \#~FA8650-18-2-7835 and \#~HR0011-18-3-0007), an Adobe Data Science Research Award, an NVIDIA AI Laboratory, and equipment donations from NVIDIA\@.

This material is based on research sponsored by the Air Force Research Lab (AFRL) and the Defense Advanced Research Projects Agency (DARPA). The U.S.\ Government is authorized to reproduce and distribute reprints for Governmental purposes notwithstanding any copyright notation thereon. The views and conclusions contained herein are those of the authors and should not be interpreted as necessarily representing the official policies or endorsements, either expressed or implied, of Air Force Research Lab (AFRL) and the Defense Advanced Research Projects Agency (DARPA) or the U.S.\ Government.
\bibliographystyle{IEEEtranS}
\bibliography{better-gpu-hash-tables}

\begin{thebibliography}{10}
\providecommand{\url}[1]{#1}
\csname url@samestyle\endcsname
\providecommand{\newblock}{\relax}
\providecommand{\bibinfo}[2]{#2}
\providecommand{\BIBentrySTDinterwordspacing}{\spaceskip=0pt\relax}
\providecommand{\BIBentryALTinterwordstretchfactor}{4}
\providecommand{\BIBentryALTinterwordspacing}{\spaceskip=\fontdimen2\font plus
\BIBentryALTinterwordstretchfactor\fontdimen3\font minus
  \fontdimen4\font\relax}
\providecommand{\BIBforeignlanguage}[2]{{%
\expandafter\ifx\csname l@#1\endcsname\relax
\typeout{** WARNING: IEEEtranS.bst: No hyphenation pattern has been}%
\typeout{** loaded for the language `#1'. Using the pattern for}%
\typeout{** the default language instead.}%
\else
\language=\csname l@#1\endcsname
\fi
#2}}
\providecommand{\BIBdecl}{\relax}
\BIBdecl

\bibitem{Alcantara:2009:RPH}
\BIBentryALTinterwordspacing
D.~A. Alcantara, A.~Sharf, F.~Abbasinejad, S.~Sengupta, M.~Mitzenmacher, J.~D.
  Owens, and N.~Amenta, ``Real-time parallel hashing on the {GPU},'' \emph{ACM
  Transactions on Graphics}, vol.~28, no.~5, pp. 154:1--154:9, Dec. 2009.
  [Online]. Available: \url{https://escholarship.org/uc/item/445536d6}
\BIBentrySTDinterwordspacing

\bibitem{Alcantara:2011:BAE}
D.~A. Alcantara, V.~Volkov, S.~Sengupta, M.~Mitzenmacher, J.~D. Owens, and
  N.~Amenta, ``Building an efficient hash table on the {GPU},'' in \emph{GPU
  Computing Gems}, W.~W. Hwu, Ed.\hskip 1em plus 0.5em minus 0.4em\relax Morgan
  Kaufmann, Oct. 2011, vol.~2, ch.~4, pp. 39--53.

\bibitem{Ashkiani:2018:ADH}
\BIBentryALTinterwordspacing
S.~Ashkiani, M.~Farach-Colton, and J.~D. Owens, ``A dynamic hash table for the
  {GPU},'' in \emph{Proceedings of the 32nd IEEE International Parallel and
  Distributed Processing Symposium}, ser. IPDPS 2018, May 2018, pp. 419--429.
  [Online]. Available: \url{https://escholarship.org/uc/item/2p48q0zg}
\BIBentrySTDinterwordspacing

\bibitem{Azar:1999:BA}
Y.~Azar, A.~Z. Broder, A.~R. Karlin, and E.~Upfal, ``Balanced allocations,''
  \emph{SIAM Journal on Computing}, vol.~29, no.~1, pp. 180--200, Jan. 1999.

\bibitem{Bell:2009:ISM}
N.~Bell and M.~Garland, ``Implementing sparse matrix-vector multiplication on
  throughput-oriented processors,'' in \emph{Proceedings of the 2009 ACM/IEEE
  Conference on Supercomputing}, ser. SC '09, Nov. 2009, pp. 18:1--18:11.

\bibitem{Breslow:2016:HTF}
A.~D. Breslow, D.~P. Zhang, J.~L. Greathouse, N.~Jayasena, and D.~M. Tullsen,
  ``{H}orton tables: Fast hash tables for in-memory data-intensive computing,''
  in \emph{Proceedings of the 2016 USENIX Conference on Usenix Annual Technical
  Conference}, ser. USENIX ATC '16.\hskip 1em plus 0.5em minus 0.4em\relax USA:
  USENIX Association, Jun. 2016, pp. 281--294.

\bibitem{Dietzfelbinger:2010:GMM}
M.~Dietzfelbinger, A.~Goerdt, M.~Mitzenmacher, A.~Montanari, R.~Pagh, and
  M.~Rink, ``Tight thresholds for cuckoo hashing via {XORSAT},'' in
  \emph{International Colloquium on Automata, Languages and Programming}, 2010,
  pp. 213--225.

\bibitem{Erlingsson:2006:ACA}
{\'U}.~Erlingsson, M.~Manasse, and F.~McSherry, ``A cool and practical
  alternative to traditional hash tables,'' in \emph{Proceedings of the 7th
  Workshop on Distributed Data and Structures}, ser. WDAS '06, Jan. 2006.

\bibitem{Fotakis:2005:SEH}
D.~Fotakis, R.~Pagh, P.~Sanders, and P.~Spirakis, ``Space efficient hash tables
  with worst case constant access time,'' \emph{Theory of Computing Systems},
  vol.~38, no.~2, pp. 229--248, 2005.

\bibitem{Harris:2017:CUDPP}
M.~Harris, J.~D. Owens, S.~Sengupta, Y.~Zhang, and A.~Davidson, ``{CUDPP}:
  {CUDA} data parallel primitives library,'' 2009--2017,
  \url{http://cudpp.github.io/}.

\bibitem{Jia:2018:DTN}
Z.~Jia, M.~Maggioni, B.~Staiger, and D.~P. Scarpazza, ``Dissecting the {NVIDIA}
  {V}olta {GPU} architecture via microbenchmarking,'' \emph{CoRR}, Apr. 2018,
  arXiv:cs.DC/1804.06826v1.

\bibitem{Junger:2018:WAL}
D.~J{\"u}nger, R.~Kobus, A.~M{\"u}ller, C.~Hundt, K.~Xu, W.~Liu, and
  B.~Schmidt, ``{WarpCore}: A library for fast hash tables on {GPUs},'' 2020.

\bibitem{Junger:2018:WMP}
D.~J{\"u}tnger, C.~Hundt, and B.~Schmidt, ``{WarpDrive}: Massively parallel
  hashing on multi-{GPU} nodes,'' in \emph{2018 IEEE International Parallel and
  Distributed Processing Symposium (IPDPS)}, 2018, pp. 441--450.

\bibitem{Lessley:2020:DHT}
B.~Lessley and H.~Childs, ``Data-parallel hashing techniques for {GPU}
  architectures,'' \emph{IEEE Transactions on Parallel and Distributed
  Systems}, vol.~31, no.~1, pp. 237--250, Jan. 2020.

\bibitem{Li:2021:DDH}
\BIBentryALTinterwordspacing
Y.~Li, Q.~Zhu, Z.~Lyu, Z.~Huang, and J.~Sun, ``Dy{Cuckoo}: Dynamic hash tables
  on {GPUs},'' in \emph{2021 IEEE 37th International Conference on Data
  Engineering (ICDE)}.\hskip 1em plus 0.5em minus 0.4em\relax IEEE Computer
  Society, Apr. 2021, pp. 744--755. [Online]. Available:
  \url{https://doi.ieeecomputersociety.org/10.1109/ICDE51399.2021.00070}
\BIBentrySTDinterwordspacing

\bibitem{Marsaglia:2003:XR}
G.~Marsaglia, ``Xorshift rngs,'' \emph{Journal of Statistical Software},
  vol.~8, no.~14, pp. 1--6, 2003.

\bibitem{Mitzenmacher:2001:TPO}
M.~Mitzenmacher, A.~W. Richa, and R.~Sitaraman, ``The power of two random
  choices: A survey of techniques and results,'' in \emph{Handbook of
  Randomized Computing}, ser. Combinatorial optimization, S.~Rajasekaran, P.~M.
  Pardalos, J.~H. Reif, and J.~Rolim, Eds.\hskip 1em plus 0.5em minus
  0.4em\relax Kluwer Academic Publishers, Jun. 2001, vol.~I, pp. 255--312.

\bibitem{Anon:2021:DBA}
{Omitted for Anonymity}, ``Dynamic balls-and-bins and iceberg hashing,''
  \emph{in progress}, 2021.

\bibitem{Pagh:2001:CUH}
R.~Pagh and F.~F. Rodler, ``Cuckoo hashing,'' in \emph{9th Annual European
  Symposium on Algorithms}, ser. Lecture Notes in Computer Science, vol.
  2161.\hskip 1em plus 0.5em minus 0.4em\relax Springer, Aug. 2001, pp.
  121--133.

\bibitem{Panigrahy:2005:EHW}
R.~Panigrahy, ``Efficient hashing with lookups in two memory accesses,'' in
  \emph{Proceedings of the Sixteenth Annual ACM-SIAM Symposium on Discrete
  Algorithms}, ser. SODA '05.\hskip 1em plus 0.5em minus 0.4em\relax USA:
  Society for Industrial and Applied Mathematics, 2005, p. 830–839.

\bibitem{Walzer:2020:RHF}
\BIBentryALTinterwordspacing
S.~Walzer, ``\BIBforeignlanguage{en}{Random hypergraphs for hashing-based data
  structures},'' Ph.D. dissertation, Technische Universit{\"a}t Ilmenau,
  Ilmenau, Nov 2020, dissertation, Technische Universit{\"a}t Ilmenau, 2020.
  [Online]. Available:
  \url{https://www.db-thueringen.de/receive/dbt_mods_00047127}
\BIBentrySTDinterwordspacing

\bibitem{Zhang:2015:MAC}
K.~Zhang, K.~Wang, Y.~Yuan, L.~Guo, R.~Lee, and X.~Zhang, ``{Mega-KV}: A case
  for {GPU}s to maximize the throughput of in-memory key-value stores,''
  \emph{Proceedings of the VLDB Endowment}, vol.~8, no.~11, pp. 1226--1237,
  Jul. 2015.

\end{thebibliography}
\end{document}